%% file: bmvc_review.tex
\documentclass{bmvc2k}

\usepackage{multirow}
\usepackage{booktabs}       
\usepackage{color}
\usepackage{colortbl}
\usepackage{graphicx} 
\usepackage{diagbox}
\usepackage{wrapfig}

\definecolor{Gray}{gray}{0.90}
\definecolor{white}{rgb}{1.0, 1.0, 1.0}

\definecolor{LightCyan}{RGB}{240, 224, 238}

\title{On Evaluating Adversarial Robustness of Volumetric Medical Segmentation Models }

\addauthor{Hashmat Shadab Malik}{hashmat.malik@mbzuai.ac.ae}{1}
\addauthor{Numan Saeed}{numan.saeed@mbzuai.ac.ae}{1}
\addauthor{Asif Hanif}{asif.hanif@mbzuai.ac.ae}{1}
\addauthor{Muzammal Naseer}{muzammal.naseer@ku.ac.ae}{4}
\addauthor{Mohammad Yaqub}{mohammad.yaqub@mbzuai.ac.ae}{ 1}
\addauthor{Salman Khan}{salman.khan@mbzuai.ac.ae}{ 1,2}
\addauthor{Fahad Shahbaz Khan}{fahad.khan@mbzuai.ac.ae}{ 1,3}

\addinstitution{
 Mohamed bin Zayed University of AI\\
}
\addinstitution{
 Australian National University
}
\addinstitution{
 Link\"oping University
}
\addinstitution{
 Center of Secure Cyber-Physical \\Security Systems, Khalifa University, Abu Dhabi, United Arab Emirates 
}
\runninghead{Malik et.al}{Adv. Robustness of Volumetric Medical Seg. Models}

\usepackage{amsmath,amsfonts,amssymb}

\begin{document}

\maketitle
\begin{abstract}

Volumetric medical segmentation models have achieved significant success on organ and tumor-based segmentation tasks in recent years.  However, their vulnerability to adversarial attacks remains largely unexplored, raising serious concerns regarding the real-world deployment of tools employing such models in the healthcare sector. This underscores the importance of investigating the robustness of existing models. In this context, our work aims to empirically examine the adversarial robustness across current volumetric segmentation architectures, encompassing Convolutional, Transformer, and Mamba-based models. We extend this investigation across four volumetric segmentation datasets, evaluating robustness under both \emph{white box} and \emph{black box} adversarial attacks. Overall, we observe that while both pixel and frequency-based attacks perform reasonably well under \emph{white box} setting, the latter performs significantly better under transfer-based \emph{black box} attacks. Across our experiments, we observe transformer-based models show higher robustness than convolution-based models with Mamba-based models being the most vulnerable. Additionally, we show that large-scale training of volumetric segmentation models improves the model's robustness against adversarial attacks. The code and robust models are available on \href{https://github.com/HashmatShadab/Robustness-of-Volumetric-Medical-Segmentation-Models}{Github}.

\end{abstract}

\section{Introduction}
\vspace{-0.5em}
\label{sec:intro}



The field of computer vision has witnessed remarkable progress in recent years, leading to substantial improvements across a diverse range of vision-based tasks spanning various domains. Despite this advancement, deep learning models are widely known to be susceptible to \emph{adversarial attacks} \cite{szegedy2013intriguing, goodfellow2014explaining,madry2017towards,carlini2017towards,su2019one,croce2020reliable}, which involves introducing carefully crafted, imperceptible perturbations to images, leading to incorrect response from the model.  Based on the amount of information available about the target model, adversarial attacks can be broadly classified into \emph{white-box} attacks, where the attacker has complete information and can directly craft an attack on the target model, and  the more practical setting of \emph{black-box} attacks, where access to the model is limited, and adversarial examples are crafted using query-based 
 \cite{ZOOgradest, decisonbased_Brendel2018, decison_based_Narodytska2017} and transfer-based approaches \cite{ dong2018boosting, naseer2019cross, malik2022adversarial}. Understanding the vulnerabilities of models against these attacks is crucial for deploying them in security-critical applications such as healthcare systems, as it offers insights into identifying blind spots in the models to ensure their robustness and reliability.

Several works \cite{shao2021adversarial, naseer2021intriguing, benz2021adversarial, bai2021transformers, pinto2022impartial} have extensively investigated the adversarial robustness of  Convolutional Neural Networks (CNNs) \cite{simonyan2014very, he2016deep} and Vision Transformers(ViTs) \cite{dosovitskiy2020image, touvron2021training}. However, these investigations have predominantly focused on models trained with natural images for the task of image classification, emphasizing the need to assess the robustness of vision-based models in the medical domain to ensure their safe deployment in healthcare systems. Within the medical domain, volumetric image segmentation stands out as an essential task for diagnoses, enabling the identification and delineation of organs or tumors within three-dimensional medical images such as Computed Tomography(CT) or Magnetic Resonance Imaging(MRI) scans \cite{landman2015miccai, bernard2018deep, andrearczyk2021overview, ma2021abdomenct}. While in recent years, the performance of volumetric segmentation models has improved significantly \cite{isensee2024nnu}, there has been limited works studying and evaluating the susceptibility of these models to adversarial attacks 
 \cite{daza2021towards, hanif2023frequency}. To address this gap, our study aims to establish the first comprehensive benchmark for evaluating the robustness of volumetric segmentation models against adversarial attacks.

For thoroughly evaluating the robustness of current volumetric segmentation models, we consider a wide range of architectures, encompassing Convolutional, Transformer, and the recently introduced Mamba-based models \cite{ma2024u}. We expand the scope of our analysis by evaluating these segmentation models across four diverse 3D segmentation datasets consisting of CT and MRI scans. We first evaluate the performance of these architectures under \emph{white-box} setting against different pixel and frequency-based attacks expanding on previous works which have done limited analysis under this setting \cite{daza2021towards, hanif2023frequency}. Further, we evaluate the models against transfer-based \emph{black-box} attacks by transferring the adversarial examples crafted on the surrogate model to unseen target models. To the best of our knowledge, no existing work has investigated the transferability of adversarial examples among different volumetric medical segmentation models, a practical setting that can be encountered in real-world scenarios. Furthermore, we expand our analysis by evaluating the current vision foundation model \cite{wang2023sam} on transfer-based attacks to assess its robustness capabilities, given its training on large-scale 3D medical datasets. Our contributions are as follows:


\begin{itemize}
    \setlength\itemsep{0.2em} 
    \setlength\parskip{0.2em} 
    \setlength\parsep{0.2em} 
    \item We provide the first comprehensive benchmark for analyzing the robustness of Convolution, Transformer, and Mamba-based volumetric medical image segmentation models against adversarial attacks.
    \item While all models show vulnerability to pixel and frequency-based attacks under \emph{white-box} setting, the latter shows significant transferability across models in the \emph{black-box} setting. Furthermore, transformer-based models in general exhibit higher robustness against adversarial attacks, while Mamba-based models are more vulnerable.
    \item Vision foundational models trained on large-scale datasets exhibit better robustness against \emph{black-box} attacks.
\end{itemize}

\section{Background and Related Work}

\noindent \textbf{Notations and Terminologies.} A clean  volumetric image is represented as $x \in \mathbb{R}^{H\times W\times D}$, with its binary segmentation mask $y \in \{0,1\}^{C \times H\times W\times D}$, where $C$ signifies the number of classes. A volumetric segmentation model $\mathcal{F}$, predicts the segmentation mask of the input images, expressed as $\mathcal{F}: \mathcal{X} \mapsto \mathcal{Y}$, where $\mathcal{X}$ and $\mathcal{Y}$ denote the sets of input images and corresponding segmentation masks, respectively. An adversarial attack crafted on model $\mathcal{F}$ involves manipulating the input image $x$, typically in the pixel or frequency domains to generate the adversarial image $x'$. The adversarial manipulation $\delta=(x'-x)$ while imperceptible, leads to an incorrect response from the model $\mathcal{F}(x') \neq \mathcal{F}(x)$.

\noindent \textbf{Adversarial Attacks.} The vulnerability of deep learning models to adversarial examples \cite{szegedy2013intriguing, goodfellow2014explaining, kurakin2018adversarial} has prompted the research community to investigate the robustness of current models against various forms of adversarial attacks \cite{goodfellow2014explaining, madry2017towards,carlini2017towards, su2019one,croce2020reliable}. Several works have delved into analyzing and comparing the adversarial robustness of CNN and transformer-based classifier models trained on natural images \cite{shao2021adversarial, naseer2021intriguing, benz2021adversarial, bai2021transformers, pinto2022impartial}. A few works in the natural domain have focused on designing adversarial attacks tailored for 2D segmentation models \cite{gu2022segpgd, agnihotri2023cospgd, jia2023transegpgd}.
These methods propose a dynamic pixel-level loss, enabling the attack to direct attention to regions of the image where the model maintains accurate predictions during the attack iterations. Focusing on the task of volumetric medical segmentation, limited works \cite{daza2021towards, hanif2023frequency} have delved into understanding the robustness of medical segmentation models. In \cite{daza2021towards}, adversarial attacks proposed for classification tasks are extended to target volumetric medical segmentation models, while \cite{hanif2023frequency} proposes a frequency domain attack to fool the medical segmentation models by dropping imperceptible information in the frequency domain of the input image. However, both works provide a limited examination of robustness, confining the scope to just \emph{white-box} attacks across a restricted range of models and datasets. In contrast, our work aims to provide a comprehensive empirical evaluation of current volumetric medical segmentation architectures across adversarial attacks encompassing both \emph{white-box} and \emph{black-box} settings.


\noindent \textbf{Volumetric Medical Segmentation Models.} Convolutional Neural Networks (CNNs) and Transformers are two popular architectures used in volumetric medical image segmentation. CNNs, such as U-Net\cite{unet} and SegResNet \cite{myronenko20193d}, are particularly good at extracting hierarchical image features. The shared weights in CNN-based models across the features help to capture local-level information, making them well-suited for capturing translational invariances. In contrast to CNNs, transformer-based models \cite{dosovitskiy2020image, liu2021swin} attention mechanism captures global information, increasing the effective receptive field of the model \cite{luo2016understanding}.  To exploit the complementary strengths of both, hybrid-based models like TransUNet \cite{chen2021transunet}, UNETR \cite{hatamizadeh2022unetr}, and SwinUNETR \cite{hatamizadeh2022swin} have been proposed to incorporate both CNNs and Transformers into the network. While Transformer models excel at modeling long-range dependencies via their attention mechanism, a significant drawback is their quadratic computational scaling with input size, making them resource-intensive for  3D medical segmentation. Recently, state space sequence models (SSMs) \cite{gu2021efficiently, gu2023mamba} proposed in the natural language domain have been adapted for vision tasks \cite{liu2024vmamba, zhu2024vision, li2024videomamba, chen2024video}, providing the capability to handle long-range dependencies while maintaining a linear computational cost. In the medical segmentation domain, \cite{ma2024u} introduced UMamba, a network merging CNNs' local feature extraction with SSMs' efficient long-range dependency handling capabilities. Based on the above model architectures, our work provides an empirical robustness evaluation of  CNNs, transformers, and recently introduced state space models against adversarial attacks in the context of volumetric medical segmentation. Furthermore, given the recent trend of enhancing the generalization capabilities of vision models through large-scale training, we extend our analysis to examine the robustness of SAM-Med3D \cite{wang2023sam}, a foundational model for volumetric segmentation trained on large-scale segmentation datasets.

\section{Research Scope}


\noindent \textbf{Research Goal.} In recent years, volumetric medical segmentation models have achieved significant success on organ and tumor-based segmentation tasks. However, the vulnerability of these models to adversarial attacks raises serious concerns about their real-world deployment in the healthcare sector. This vulnerability raises doubts among clinicians regarding the reliability of tools employing such models and underscores the importance of investigating and enhancing their robustness. In this context, our work aims to \emph{empirically} investigate the adversarial robustness across different volumetric segmentation architectures, encompassing Convolutional, Transformer, and Mamba-based models. We extend this investigation across four volumetric segmentation datasets, assessing the models' robustness under both \emph{white-box} and \emph{black-box} adversarial attack scenarios. Our work does not claim to provide theoretical reasoning behind the robustness behavior across models, instead as one of the first works in studying the vulnerability of volumetric segmentation models, we provide detailed empirical evaluations across different adversarial settings. Furthermore, we conduct frequency-based analyses to gain further insights into the models' behavior to adversarial attacks and examine the impact of large-scale training on the model's robustness. While the scope of this work is limited, we hope this work paves the way for future research focusing on the robustness of medical segmentation models.

\noindent \textbf{Datasets.} We utilize four 3D segmentation datasets: \texttt{BTCV} (\textit{$30$ abdominal CT scans from liver cancer patients with $13$ organ annotations}) \cite{landman2015miccai}, \texttt{ACDC} (\textit{$150$ MRI images of cardiac abnormalities with heart organ annotations}) \cite{bernard2018deep}, \texttt{Hecktor} (\textit{$524$ CT/PET scans of head and neck cancer patients annotated for primary and nodal tumor}) \cite{andrearczyk2021overview}, and AbdomenCT-1k (\textit{$1112$ abdominal CT scans from diverse medical centers with annotations for abdominal region}) \cite{ma2021abdomenct}. For more details about the datasets, refer to Appendix \ref{appendix:datasets}.

\noindent \textbf{Models.} In our experiments, we consider CNN models (UNet \cite{unet} and SegResNet \cite{myronenko20193d}), Transformer models (UNETR \cite{hatamizadeh2022unetr} and Swin-UNETR \cite{hatamizadeh2022swin}), and recently introduced Visual State Space models (UMamba-B and UMamba-E \cite{ma2024u}).  All models are trained from scratch on the four mentioned datasets with the input size of $96\times96\times96$, following the training methodology outlined in \cite{hanif2023frequency}. Further, we consider SAM-Med3D \cite{wang2023sam}, a vision foundation model for volumetric medical images trained on large-scale segmentation datasets.

\noindent \textbf{Adversarial Attacks and Metrics.} We consider two categories of adversarial attacks: pixel-based attacks, which include Fast Gradient Sign Method (\texttt{FGSM}) \cite{goodfellow2014explaining}, Projected Gradient Descent (PGD) \cite{madry2017towards}, and Cosine-PGD (\texttt{CosPGD}) \cite{agnihotri2023cospgd}, and frequency-based attacks, for which we consider Volumetric Adversarial Frequency Attack (\texttt{VAFA}) \cite{hanif2023frequency}. \texttt{CosPGD}, originally proposed for 2D segmentation tasks, is adapted to work in 3D segmentation scenarios, whereas \texttt{VAFA} operates directly on volumetric data. All the above methods are based on minimizing the Dice Similarity Score (DSC) \cite{sudre2017generalised} during the attack optimization. We also report model performance on additive Gaussian Noise (\texttt{GN}). For evaluating segmentation task, we report the Dice Similarity Coefficient (DSC) and the mean $95\%$ Hausdorff Distance (HD95) score on clean samples. To asses the Attack Success Rate (ASR), we introduce two metrics, ASR-D and ASR-H, which measure the average performance decline of models under adversarial attacks:
\begin{equation}
\footnotesize
\texttt{ASR-D} = \frac{1}{\vert \mathcal{D}_{\text{test}}\vert}\sum_{x,y\in\mathcal{D}_{\text{test}}} \bigg\vert \mathrm{DSC}\big(\mathcal{F}(x), y\big) - \mathrm{DSC}\big(\mathcal{F}(x^{\prime}), y\big) \bigg\vert
\end{equation}
\begin{equation}
\footnotesize
\texttt{ASR-H} = \frac{1}{\vert \mathcal{D}_{\text{test}}\vert} \sum_{x,y\in\mathcal{D}_{\text{test}}} \bigg\vert \mathrm{HD95}\big(\mathcal{F}(x^{\prime}), y\big) - \mathrm{HD95}\big(\mathcal{F}(x), y\big) \bigg\vert
\end{equation}


\section{Experimental Results}

\subsection{Robustness against White-Box Attacks} 
\label{sec:wbox}
\begin{table}[!t]
\centering\small
   \setlength{\tabcolsep}{3.8pt}
   \scalebox{0.63}[0.63]{
    \begin{tabular}{c|c|cc|cc|cc|cc|cc|cc|cc}
        \toprule
        \rowcolor{LightCyan} 
        Dataset & Attack  & \multicolumn{2}{c|}{\textbf{UNet}} & \multicolumn{2}{c|}{\textbf{SegResNet}}  & \multicolumn{2}{c|}{\textbf{UNETR}}  & \multicolumn{2}{c|}{\textbf{SwinUNETR}}  &  \multicolumn{2}{c|}{\textbf{UMamba-B}} & \multicolumn{2}{c|}{\textbf{UMamba-E}} & \multicolumn{2}{c}{\textbf{Average}} \\
                \rowcolor{LightCyan} 
          &  & ~$\texttt{ASR-D}\hspace{-0.3em}$~ & $\texttt{ASR-H}\hspace{-0.3em}$~  &  $\texttt{ASR-D}\hspace{-0.3em}$~ & ~$\texttt{ASR-H}\hspace{-0.3em}$~ & $\texttt{ASR-D}\hspace{-0.3em}$~  &  $\texttt{ASR-H}\hspace{-0.3em}$ & $\texttt{ASR-D}\hspace{-0.3em}$~  &  $\texttt{ASR-H}\hspace{-0.3em}$ & $\texttt{ASR-D}\hspace{-0.3em}$~  &  $\texttt{ASR-H}\hspace{-0.3em}$ & $\texttt{ASR-D}\hspace{-0.3em}$~  &  $\texttt{ASR-H}\hspace{-0.3em}$ &  $\texttt{ASR-D}\hspace{-0.3em}$ &  $\texttt{ASR-H}\hspace{-0.3em}$\\
 \midrule
\rowcolor{Gray} \multirow{4}{*}{\rotatebox[origin=c]{0}{\parbox[c]{1.8cm}{\centering\texttt{BTCV}}}} & Clean  & 75.72 & 9.15 & 80.84 & 8.14 & 72.53 & 15.08 & 78.07 & 10.01 & 78.37 & 8.12 & 77.06 & 11.11 & - & - \\
& \texttt{GN}  & 0.32 & 1.16 & 0.50 & 0.30 & 0.20 & -0.63 & 0.63 & -0.28 & 0.68 & 1.32 & 1.72 & 10.69 & 0.675 & 2.093\\
& \texttt{FGSM}  & 26.23 & 39.08 & 20.87 & 29.22 & 21.55 & 29.78 & 19.27 & 30.84 & 19.74 & 27.33 & 23.47 & 34.43 & 21.86 &	31.78\\
& \texttt{PGD}  & 43.66 & 80.30 & 60.73 & 91.99 & 24.52 & 38.70 & 23.08 & 54.42 & 48.29 & 89.92 & 47.55 & 93.40 & 41.31	& 74.79\\
& \texttt{CosPGD}  & 43.21 & 66.74 & 61.16 & 18.60 & 25.07 & 36.06 & 25.14 & 48.45 & 47.41 & 74.63 & 45.96 & 68.58 & 41.33 &	52.18\\
& \texttt{VAFA}  & 56.23 & 77.60 & 58.84 & 64.02 & 45.85 & 43.75 & 53.77 & 52.37 & 61.72 & 70.92 & 60.61 & 80.24 & 56.17 & 64.82\\
    \midrule
    \rowcolor{Gray} \multirow{6}{*}{\rotatebox[origin=c]{0}{\parbox[c]{1.8cm}{\centering\texttt{ACDC}}}} & Clean  & 85.52 & 5.75 & 89.65 & 2.56 & 76.37 & 16.31 & 84.19 & 7.93 & 88.22 & 6.01 & 80.91 & 8.48 & - & -\\
& \texttt{GN}  & 0.51 & -0.08 & 0.39 &0.33 & 1.49 & 2.18 & 1.13 & 2.76 & 1.27 & 0.67 & 2.80 & 3.19 & 1.265 & 1.508 \\
& \texttt{FGSM}  & 29.97 & 16.24 & 23.07 & 9.32 & 43.21 & 13.46 & 28.08 & 13.90 & 15.67 & 15.18 & 29.26 & 15.74 & 28.21	& 13.97\\
& \texttt{PGD}  & 64.10 & 33.39 & 69.23 & 34.78 & 54.06 & 19.13 & 62.50 & 29.06 & 63.27 & 28.65 & 55.90 & 23.95 & 61.51	& 28.16\\
& \texttt{CosPGD}  & 62.90 & 31.09 & 67.32 & 36.05 & 53.29 & 18.49 & 61.69 & 28.27 & 62.68 & 25.31 & 56.47 & 23.07 & 60.73	& 27.05\\
& \texttt{VAFA}  & 35.67 & 21.69 & 35.18 & 20.82 & 26.98 & 6.32 & 35.75 & 17.64 & 34.84 & 19.21 & 29.30 & 17.24 & 32.95 &	17.15\\
    \midrule
   \rowcolor{Gray} \multirow{4}{*}{\rotatebox[origin=c]{0}{\parbox[c]{1.8cm}{\centering\texttt{Hecktor}}}} & Clean  & 73.91 & 11.36 & 74.73 & 11.08 & 72.36 & 14.61 & 71.61 & 22.07 & 73.49 & 10.89 & 72.19 & 13.29 & - & - \\
& \texttt{GN}  & 1.80 & 1.61 & 1.76 & -0.87 & 1.09 & 0.25 & 1.97 & 2.70 & 1.57 & 1.24 & 5.63 & -0.42 & 2.303 & 0.751\\
& \texttt{FGSM}  & 30.13 & 26.43 & 27.03 & 21.16 & 27.47 & 20.47 & 27.79 & 30.88 & 23.33 & 19.23 & 25.04 & 18.36 & 26.80	& 22.76\\
& \texttt{PGD}  & 39.84 & 67.25 & 41.71 & 70.09 & 37.43 & 59.97 & 37.49 & 56.78 & 39.18 & 67.50 & 37.90 & 64.96 & 38.93	& 64.43\\
& \texttt{CosPGD}  & 39.69 & 67.26 & 41.32 & 70.20 & 37.47 & 59.23 & 37.40 & 55.55 & 39.08 & 67.55 & 37.82 & 64.56 & 38.80	& 64.06\\
& \texttt{VAFA}  & 29.61 & 30.83 & 27.18 & 22.00 & 24.35 & 19.21 & 27.17 & 20.07 & 24.22 & 25.10 & 27.38 & 23.55 & 26.65	& 23.46\\

    \midrule
   \rowcolor{Gray} \multirow{6}{*}{\rotatebox[origin=c]{0}{\parbox[c]{1.8cm}{\centering\texttt{Abdomen-CT}}}} & Clean  &  76.79 & 19.72 & 80.89 & 13.30 & 71.35 & 27.73 & 79.33 & 25.79 & 81.08 & 15.51 & 78.05 & 18.31 & - & - \\
& \texttt{GN}  & 0.66 & 2.25 & 3.30 & 4.86 & 0.22 & 6.64 & 0.38 & 3.30 & 1.75 & 1.21 & 1.78 & -0.15 & 1.348 & 3.018 \\
& \texttt{FGSM}  & 29.94 & 40.47 & 30.16 & 35.77 & 22.87 & 31.12 & 25.87 & 45.24 & 29.72 & 38.38 & 29.24 & 33.82 & 27.97	& 37.47\\
& \texttt{PGD}  & 47.65 & 71.13 & 63.47 & 93.43 & 26.52 & 48.35 & 32.38 & 63.09 & 62.88 & 96.28 & 54.46 & 78.53 & 47.89	& 75.14\\
& \texttt{CosPGD}  & 45.99 & 57.75 & 62.01 & 88.95 & 26.71 & 46.61 & 32.54 & 57.13 & 62.09 & 86.38 & 53.06 & 63.70 & 47.07	& 66.75\\
& \texttt{VAFA}  & 55.87 & 68.47 & 63.21 & 71.61 & 49.01 & 60.35 & 53.90 & 50.62 & 66.09 & 76.58 & 64.17 & 74.60 & 58.71	& 67.04\\
\midrule
\texttt{Average} & - & 42.54&49.73&47.03&48.63&34.15&34.44&36.49&40.89&43.76&51.76&42.35&48.67 & - & -\\
\midrule
\end{tabular}
}
\caption{\small Performance of models against \emph{white box} attacks across different datasets is reported. }
\label{tab:wbox}
\end{table}

\begin{figure}[!t]
\begin{minipage}{\linewidth}
    \centering
    \includegraphics[  trim= 2mm 0mm 0mm 0mm, clip, width=0.24\textwidth]{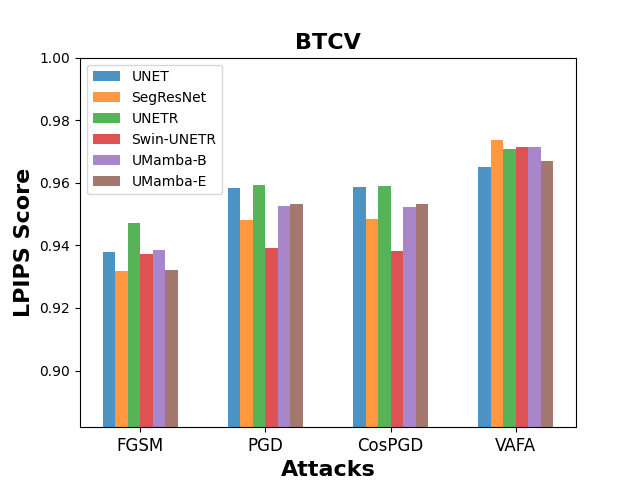}
    \includegraphics[  trim= 2mm 0mm 0mm 0mm, clip, width=0.24\textwidth]{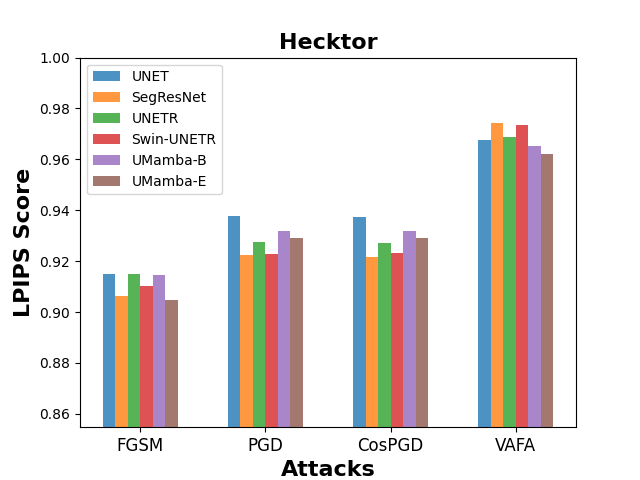}
        \includegraphics[ trim= 2mm 0mm 0mm 0mm, clip, width=0.24\textwidth]{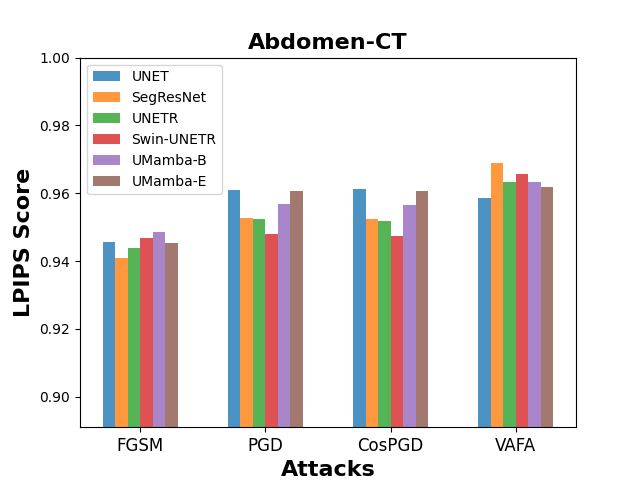}
    \includegraphics[ trim= 2mm 0mm 0mm 0mm, clip, width=0.24\textwidth]{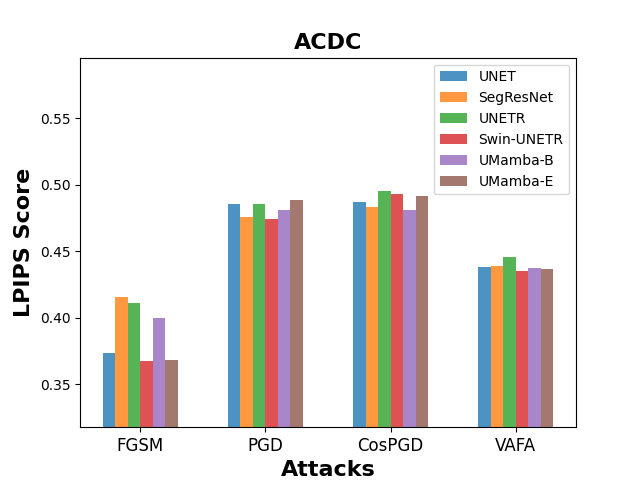}
\end{minipage}
\caption{\small LPIPS scores for adversarial examples crafted on different segmentation models. }
\label{fig: LPIPS_score}
\end{figure}

In this section, we examine the robustness of segmentation models in white-box settings. 
For pixel-based attacks, we consider \texttt{FGSM}  \cite{szegedy2013intriguing}, \texttt{PGD} \cite{madry2017towards} and \texttt{CosPGD} \cite{agnihotri2023cospgd} and use perturbation budget $\epsilon=\frac{8}{255}$ and set $20$ optimization steps for iterative attacks.
For frequency-based attack \texttt{VAFA} \cite{hanif2023frequency}, we generate adversarial examples with $q_{max}=30$, employing a patch size of $32\times32\times32$, while keeping other parameters consistent with the default settings as outlined in \cite{hanif2023frequency}. Adversarial examples are generated on the validation sets of \texttt{BTCV}, \texttt{ACDC}, \texttt{Hecktor}, and \texttt{Abdomen-CT} datasets by attacking the models trained on the respective datasets. In Table \ref{tab:wbox}, we show segmentation model performance based on attack success rates for DSC and HD95 metrics. We observe that for the \texttt{BTCV} and \texttt{Abdomen-CT} datasets, which are utilized for the organ segmentation task, \texttt{VAFA} achieves the highest ASR-D among all attacks. On average, across all the models, \texttt{VAFA} demonstrates an increase of $14.84\%$ and $11.64\%$ over the best pixel-based attacks for the \texttt{BTCV} and \texttt{Abdomen-CT} datasets, respectively. However, for \texttt{ACDC} and \texttt{Hecktor} datasets, which are used for MRI-based organ segmentation and CT-based tumor segmentation, we observe iterative pixel-based attacks perform much better than other attacks, especially for the \texttt{ACDC} dataset (an increase of $28.56\%$ over \texttt{VAFA}). When examining the average attack success rate across all datasets, we note that CNN-based models exhibit the least robustness, with SegResNet having the highest average ASR-D of $47.03\%$. Following closely are UNet and UMamba-base models. In contrast, transformer-based models demonstrate the highest level of robustness, with UNETR exhibiting an average ASR-D of $34.15\%$.

We further look into the perceptual quality of adversarial images crafted of different models across different datasets. In Figure \ref{fig: LPIPS_score}, we report LPIPs score \cite{zhang2018unreasonable} to compute the similarity between the adversarial and clean volumetric image. Our results reveal a general trend: pixel-based attacks tend to reduce the LPIPS score more than the frequency-based \texttt{VAFA} attack. Specifically, the single-step \texttt{FGSM} attack demonstrates the most significant decrease in score. Notably, \texttt{VAFA} achieves higher similarity scores with respect to the clean images, suggesting the potential for generating even stronger adversarial examples by increasing the bound $q_{max}$ while maintaining a comparable LPIPS score compared to other attacks. However, this trend deviates on the \texttt{ACDC} dataset, where all attacks lead to a notable drop in the LPIPS score. Iterative pixel-based attacks yield the highest LPIPs score, followed by \texttt{VAFA} and \texttt{FGSM} attacks.  Further analyses on \emph{white-box} attacks with  perturbation budgets: $\epsilon=\frac{4}{255}$ for pixel-based attacks and $q_{max} \in \{10,20\}$ in Appendix \ref{appendix: wb_results_eps_4}.

\subsection{Robustness against Black-Box Attacks}
\label{sec:bbox}
\begin{figure}[!t]
    \centering
    \includegraphics[width=\linewidth]{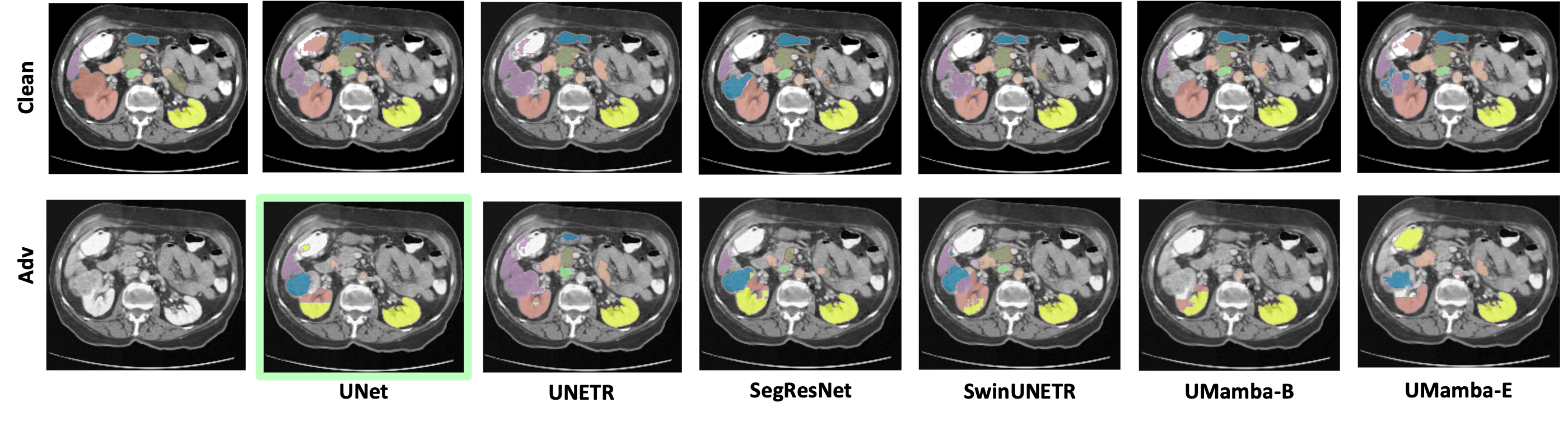}
    \caption{\small Comparing multi-organ segmentation across various models under transfer-based \emph{black box} attacks, where adversarial examples are generated on UNet and transferred to other unseen models.}
    \label{fig:vis}
\end{figure}


In this section, we evaluate and compare the robustness of different models against transfer-based \emph{black-box} attacks.  These attacks exploit the \emph{transferability} property of adversarial examples i.e., adversarial examples crafted on a surrogate model transfer to unknown target models. For our evaluation, we evaluate the transferability of all segmentation models by using them interchangeably as both surrogate and target models. In Table \ref{tab: bbox btcv}, \ref{tab: bb abdomen}, \ref{tab: bb acdc} and \ref{tab: bb hecktor} we report results on transferability of models on \texttt{BTCV}, \texttt{Abdomen-CT}, \texttt{ACDC}, and \texttt{Hecktor} dataset. On all the datasets except \texttt{Hecktor}, we observe that frequency-based attacks demonstrate significant transferability across all models compared to pixel-based attacks. Additionally, among pixel-based attacks, the single-step \texttt{FGSM} attack performs better, suggesting that adversarial examples crafted by iterative pixel-based attacks tend to overfit on the surrogate models, resulting in lower transferability\cite{dong2018boosting}. This is evident as pixel-based iterative attacks outperform the \texttt{FGSM} attack on the surrogate model. While one-step gradient methods do not overfit on the surrogate model\cite{kurakin2016adversarial}, they tend to have a low success rate, making them ineffective for \emph{black box} attacks. Frequency-based attacks alleviate this trade-off between white-box attacks and transferability, providing a high level of success on \emph{white box} attacks while also being highly transferable. However, we note a general lack of transferability among all attacks on the \texttt{Hecktor} dataset (Table \ref{tab: bb hecktor}). Interestingly,  compared to iterative pixel and frequency-based attacks, single-step \texttt{FGSM} attack demonstrate better performance across all the models.

\begin{table}[!t]
\centering\small
  \setlength{\tabcolsep}{3.8pt}
   \scalebox{0.63}[0.63]{
    \begin{tabular}{l|c|cc|cc|cc|cc|cc|cc|cc}
        \toprule
        \rowcolor{LightCyan} 
         Target $\rightarrow$ & Attack  & \multicolumn{2}{c|}{\textbf{UNet}} & \multicolumn{2}{c|}{\textbf{SegResNet}}  & \multicolumn{2}{c|}{\textbf{UNETR}}  & \multicolumn{2}{c|}{\textbf{SwinUNETR}}  &  \multicolumn{2}{c|}{\textbf{UMamba-B}} & \multicolumn{2}{c|}{\textbf{UMamba-E}}  & \multicolumn{2}{c}{\textbf{Average}}\\
        \rowcolor{LightCyan} 
          Surrogate $\downarrow$ &  & ~$\texttt{ASR-D}\hspace{-0.3em}$~ & $\texttt{ASR-H}\hspace{-0.3em}$~  &  $\texttt{ASR-D}\hspace{-0.3em}$~ & ~$\texttt{ASR-H}\hspace{-0.3em}$~ & $\texttt{ASR-D}\hspace{-0.3em}$~  &  $\texttt{ASR-H}\hspace{-0.3em}$ & $\texttt{ASR-D}\hspace{-0.3em}$~  &  $\texttt{ASR-H}\hspace{-0.3em}$ & $\texttt{ASR-D}\hspace{-0.3em}$~  &  $\texttt{ASR-H}\hspace{-0.3em}$ & $\texttt{ASR-D}\hspace{-0.3em}$~  &  $\texttt{ASR-H}\hspace{-0.3em}$ &
          $\texttt{ASR-D}\hspace{-0.3em}$~  &  $\texttt{ASR-H}\hspace{-0.3em}$ \\
          \midrule
\rowcolor{Gray}  & Clean  & 75.72 & 9.15 & 80.84 & 8.14 & 72.53 & 15.08 & 78.07 & 10.01 & 78.37 & 8.12 & 77.06 & 11.11 & - & - \\
\rowcolor{Gray} & \texttt{GN}  & 0.32 & 1.16 & 0.50 & 0.30 & 0.20 & -0.63 & 0.63 & -0.28 & 0.68 & 1.32 & 1.72 & 10.69 & 0.675 &	2.093\\
\midrule
\multirow{4}{*}{\rotatebox[origin=c]{0}{\parbox[c]{1.6cm}{\centering\texttt{UNet}}}} 
& \texttt{FGSM}  & \cellcolor{green!25}26.23 & \cellcolor{green!25}39.08 & 2.86 & 1.35 & 2.02 & 0.91 & 3.10 & 2.25 & 3.73 & 1.91 & 5.22 & 14.84 & 3.386	& 4.250\\
& \texttt{PGD}  & \cellcolor{green!25}43.66 & \cellcolor{green!25}80.30 & 1.06 & 0.59 &  0.51 & 0.20 & 1.18 & 0.68 & 1.45 & 0.90 & 2.32 & 2.75 & 1.304	& 1.020\\
& \texttt{CosPGD}  & \cellcolor{green!25}43.21 & \cellcolor{green!25}66.74 & 1.06 & 0.57 &  0.54 & -0.14 & 1.20 & 1.01 & 1.54 & 2.24 & 2.27 & 5.39 & 1.322 &	1.810\\
& \texttt{VAFA}  & \cellcolor{green!25}56.23 & \cellcolor{green!25}77.60 & 46.45 & 45.17 &  27.04 & 19.60 & 36.59 & 26.99 & 49.37 & 49.13 & 57.68 & 60.84 & 43.42	& 40.35\\
\midrule
\multirow{4}{*}{\rotatebox[origin=c]{0}{\parbox[c]{1.6cm}{\centering\texttt{SegResNet}}}} & \texttt{FGSM}  & 4.02 & 2.86 & \cellcolor{green!25} 20.87 & \cellcolor{green!25}29.22 & 2.60 & 0.26 & 5.59 & 1.80 & 9.64 & 6.65 & 9.11 & 11.63 & 6.192	& 4.640\\
& \texttt{PGD}  & 2.38 & 4.79 & \cellcolor{green!25}60.73 & \cellcolor{green!25}91.99 & 1.79 & 0.07 & 3.12 & 2.28 & 7.58 & 6.13 & 6.98 & 10.56 & 4.370	& 4.766\\
& \texttt{CosPGD}  & 2.43 & 4.61 & \cellcolor{green!25}61.16 & \cellcolor{green!25}88.60 & 1.78 & 0.04 & 3.03 & 1.74 & 7.87 & 6.04 & 7.27 & 10.20 & 4.476 &	4.526\\
& \texttt{VAFA}  & 39.29 & 35.28 & \cellcolor{green!25}58.84 & \cellcolor{green!25}64.02 & 22.37 & 13.11 & 32.53 & 26.82 & 49.98 & 47.75 & 57.53 & 58.34 & 40.34	& 36.26\\
\midrule
\multirow{4}{*}{\rotatebox[origin=c]{0}{\parbox[c]{1.6cm}{\centering\texttt{UNETR}}}} & \texttt{FGSM}  & 4.83 & 2.92 & 4.71 & 2.37 & \cellcolor{green!25}21.55 & \cellcolor{green!25}29.78 & 6.59 & 2.96 & 5.90 & 2.23 & 5.85 & 3.94 & 5.576	& 2.884\\
& \texttt{PGD}  & 3.48 & 4.34 & 2.38 & 1.43 & \cellcolor{green!25}{24.52} & \cellcolor{green!25}{38.70} & 3.43 & 2.61 & 3.75 & 2.17 & 4.33 & 4.18 & 3.474	& 2.946 \\
& \texttt{CosPGD}  & 3.41 & 4.46 & 2.11 & 1.15 & \cellcolor{green!25}{25.07} & \cellcolor{green!25}{36.06} & 3.24 & 3.43 & 3.59 & 1.68 & 4.25 & 4.32 & 3.320	& 3.008\\
& \texttt{VAFA}  & 35.45 & 26.09 & 38.29 & 26.70 & \cellcolor{green!25}{45.85} & \cellcolor{green!25}{43.75} & 36.17 & 29.53 & 42.21 & 35.85 & 45.59 & 34.34 & 39.54 & 30.50\\
\midrule
\multirow{4}{*}{\rotatebox[origin=c]{0}{\parbox[c]{1.6cm}{\centering\texttt{SwinUNETR}}}} & \texttt{FGSM}  & 4.25 & 3.43 & 5.75 & 2.72 & 3.59 & 0.66 & \cellcolor{green!25} 19.27 & \cellcolor{green!25}30.83 & 6.49 & 3.34 & 6.91 & 8.01 & 5.398	& 3.632\\
& \texttt{PGD}  & 3.44 & 4.53 & 2.92 & 1.47 & 2.36 & 2.09 & \cellcolor{green!25} 23.98 & \cellcolor{green!25} 54.42 & 3.75 & 3.44 & 4.87 & 9.59 & 3.468	& 4.224\\
& \texttt{CosPGD}  & 3.32 & 5.44 & 2.97 & 1.57 & 2.33 & 3.05 & \cellcolor{green!25} 25.14 & \cellcolor{green!25} 48.45 & 3.65 & 3.91 & 4.87 & 10.01 & 3.428	& 4.796\\
& \texttt{VAFA}  & 46.50 & 44.96 & 50.54 & 36.06 & 33.84 & 23.14 & \cellcolor{green!25} 53.77 & \cellcolor{green!25}52.37 & 52.68 & 41.47 & 55.10 & 45.35 & 47.73	& 38.20\\
\midrule
\multirow{4}{*}{\rotatebox[origin=c]{0}{\parbox[c]{1.6cm}{\centering\texttt{UMamba-B}}}} & \texttt{FGSM}  & 4.33 & 4.99 & 8.88 & 5.15 & 2.40 & 0.89 & 5.44 & 2.72 & \cellcolor{green!25}19.74 & \cellcolor{green!25}27.33 & 11.31 & 12.02 & 6.472 &	5.154\\
& \texttt{PGD}  & 2.45 & 2.87 & 4.73 & 4.45 & 1.47 & 0.53 & 2.48 & 2.30 & \cellcolor{green!25}48.29 & \cellcolor{green!25}89.92 & 9.34 & 12.52 & 4.094 &	4.534\\
& \texttt{CosPGD}  & 2.42 & 3.77 & 4.43 & 4.64 & 1.42 & -0.11 & 2.33 & 1.71 & \cellcolor{green!25}47.41 & \cellcolor{green!25}74.63 & 8.80 & 14.33 & 3.880	& 4.868\\
& \texttt{VAFA}  & 40.54 & 33.63 & 50.39 & 41.58 & 25.60 & 19.34 & 35.18 & 28.18 & \cellcolor{green!25}61.72 & \cellcolor{green!25}70.92 & 60.05 & 62.62& 42.35 &	37.07\\
\midrule
\multirow{4}{*}{\rotatebox[origin=c]{0}{\parbox[c]{1.6cm}{\centering\texttt{UMamba-E}}}} & \texttt{FGSM}  & 3.13 & 3.18 & 5.15 & 3.18 & 2.04 & 1.19 & 3.59 & 1.01 & 7.01 & 6.46 & \cellcolor{green!25}23.47 & \cellcolor{green!25}34.43 & 4.184	& 3.004\\
& \texttt{PGD}  & 1.33 & 4.51 & 1.75 & 1.41 & 0.84 & -0.04 & 1.16 & 1.22 & 2.79 & 2.62 & \cellcolor{green!25}47.55 & \cellcolor{green!25}93.40 & 1.574	& 1.944\\
& \texttt{CosPGD}  & 1.31 & 4.58 & 1.70 & 1.80 & 0.78 & 0.30 & 1.12 & 1.23 & 2.54 & 3.40 & \cellcolor{green!25}45.96 & \cellcolor{green!25}68.58 & 1.490	& 2.262\\
& \texttt{VAFA}  & 31.64 & 27.63 & 38.37 & 28.60 & 18.92 & 10.49 & 23.66 & 15.98 & 45.47 & 39.01 & \cellcolor{green!25}60.61 & \cellcolor{green!25}80.24 & 31.61	& 24.34\\
\midrule
\end{tabular}
}
\caption{\small  Performance of models against transfer-based \emph{black box} attacks on  \texttt{BTCV} dataset. }
\label{tab: bbox btcv}
\end{table}

\begin{table}[!t]
\centering\small
  \setlength{\tabcolsep}{3.8pt}
   \scalebox{0.63}[0.63]{
    \begin{tabular}{l|c|cc|cc|cc|cc|cc|cc|cc}
        \toprule
        \rowcolor{LightCyan} 
        Target $\rightarrow$ & Attack  & \multicolumn{2}{c|}{\textbf{UNet}} & \multicolumn{2}{c|}{\textbf{SegResNet}}  & \multicolumn{2}{c|}{\textbf{UNETR}}  & \multicolumn{2}{c|}{\textbf{SwinUNETR}}  &  \multicolumn{2}{c|}{\textbf{UMamba-B}} & \multicolumn{2}{c|}{\textbf{UMamba-E}} & \multicolumn{2}{c}{\textbf{Average}} \\
                \rowcolor{LightCyan} 
          Surrogate $\downarrow$ &  & ~$\texttt{ASR-D}\hspace{-0.3em}$~ & $\texttt{ASR-H}\hspace{-0.3em}$~  &  $\texttt{ASR-D}\hspace{-0.3em}$~ & ~$\texttt{ASR-H}\hspace{-0.3em}$~ & $\texttt{ASR-D}\hspace{-0.3em}$~  &  $\texttt{ASR-H}\hspace{-0.3em}$ & $\texttt{ASR-D}\hspace{-0.3em}$~  &  $\texttt{ASR-H}\hspace{-0.3em}$ & $\texttt{ASR-D}\hspace{-0.3em}$~  &  $\texttt{ASR-H}\hspace{-0.3em}$ & $\texttt{ASR-D}\hspace{-0.3em}$~  &  $\texttt{ASR-H}\hspace{-0.3em}$ &
          $\texttt{ASR-D}\hspace{-0.3em}$~  &  $\texttt{ASR-H}\hspace{-0.3em}$ \\
          \midrule
\rowcolor{Gray}  & Clean  & 76.79 & 19.72 & 80.89 & 13.30 & 71.35 & 27.73 & 79.33 & 25.79 & 81.08 & 15.51 & 78.05 & 18.31 & - & - \\
\rowcolor{Gray} & \texttt{GN}  & 0.66 & 2.25 & 3.30 & 4.86 & 0.22 & 6.64 & 0.38 & 3.30 & 1.75 & 1.27 & 1.78 & -0.15 & 1.348	& 3.028\\
\midrule
\multirow{4}{*}{\rotatebox[origin=c]{0}{\parbox[c]{1.6cm}{\centering\texttt{UNet}}}} 
& \texttt{FGSM}  & \cellcolor{green!25}29.94 & \cellcolor{green!25}40.47 & 8.99 & 5.20  & 2.45 & 7.27 & 4.34 & 5.95 & 8.53 & 7.17 & 7.81 & 6.12 & 6.424	& 6.340\\
& \texttt{PGD}  & \cellcolor{green!25}47.65 & \cellcolor{green!25}71.13  &  4.54 & 2.26 & 1.16 & 5.19 & 1.60 & 2.98 & 3.86 & 3.78 & 3.58 & 1.84 & 2.948	& 3.210\\
& \texttt{CosPGD}  & \cellcolor{green!25}45.99 & \cellcolor{green!25}57.75  & 4.45 & 2.36 & 1.11 & 6.14 & 1.47 & 2.73 & 3.65 & 2.76 & 3.47 & 1.94 & 2.830 &	3.190\\
& \texttt{VAFA}  & \cellcolor{green!25}55.87 & \cellcolor{green!25}68.47  &  54.27 & 53.09 & 32.99 & 35.84 & 40.58 & 31.49 & 58.72 & 55.74 & 59.16 & 56.80 & 49.14 &	46.59\\
\midrule
\multirow{4}{*}{\rotatebox[origin=c]{0}{\parbox[c]{1.6cm}{\centering\texttt{SegResNet}}}} & \texttt{FGSM}  & 6.29 & 3.87 & \cellcolor{green!25}30.16 & \cellcolor{green!25}35.77 & 2.64 & 6.93 & 5.04 & 2.17 & 16.17 & 13.63 & 14.11 & 8.41 & 8.850	& 7.002\\
& \texttt{PGD}  & 3.03 & 3.05 & \cellcolor{green!25}63.47 & \cellcolor{green!25}93.43 & 1.45 & 4.85 & 2.43 & 3.04 & 10.21 & 8.06 & 8.24 & 6.49 & 5.070	& 5.100\\
& \texttt{CosPGD}  & 2.72 & 4.35 & \cellcolor{green!25}62.01 & \cellcolor{green!25}88.95 & 1.42 & 4.53 & 2.33 & 3.38 & 9.49 & 7.62 & 7.69 & 5.250 & 4.730	& 5.030\\
& \texttt{VAFA}  & 38.32 & 31.64 & \cellcolor{green!25} 63.21 & \cellcolor{green!25} 71.61 & 28.02 & 26.86 & 34.33 & 22.06 & 58.65 & 48.33 & 58.30 & 46.32 & 43.52 &	35.04\\
    \midrule
    \multirow{4}{*}{\rotatebox[origin=c]{0}{\parbox[c]{1.6cm}{\centering\texttt{UNETR}}}} & \texttt{FGSM}  & 6.41 & 3.83 & 9.21 & 3.36 & \cellcolor{green!25}22.87 & \cellcolor{green!25}31.12 & 7.79 & 5.69 & 9.38 & 9.68 & 8.77 & 3.11 & 8.312	& 5.134 \\
& \texttt{PGD}  & 4.22 & 3.27 & 6.04 & 3.89 & \cellcolor{green!25}{26.52} & \cellcolor{green!25}{48.35} & 4.05 & 5.17 & 6.26 & 3.33 & 7.23 & 5.50 & 5.560	& 4.232\\
& \texttt{CosPGD}  & 4.12 & 3.01 & 5.78 & 3.75 & \cellcolor{green!25}{26.71} & \cellcolor{green!25}{46.61} & 3.87 & 4.82 & 6.26 & 2.96 & 7.05 & 4.74 & 5.416	& 3.856\\
& \texttt{VAFA}  & 37.11 & 28.88 & 48.26 & 42.75 & \cellcolor{green!25}{49.01} & \cellcolor{green!25}{60.35} & 41.22 & 28.70 & 51.48 & 38.72 & 51.28 & 40.31 & 45.87	& 35.87\\
    \midrule
    \multirow{4}{*}{\rotatebox[origin=c]{0}{\parbox[c]{1.6cm}{\centering\texttt{SwinUNETR}}}} & \texttt{FGSM}  & 6.80 & 4.72 & 11.50 & 5.18 & 4.54 & 7.32 & \cellcolor{green!25} 25.87 & \cellcolor{green!25}45.24 & 10.87 & 7.94 & 9.93 & 5.54 & 8.728	& 6.140\\
& \texttt{PGD}  & 4.12 & 4.42 & 7.70 & 4.90 & 2.58 & 10.62 & \cellcolor{green!25}32.38 & \cellcolor{green!25}63.09 & 6.69 & 7.92 & 6.65 & 5.92 & 5.548	& 6.760\\
& \texttt{CosPGD}  & 3.91 & 3.95 & 7.30 & 5.04 & 2.50 & 10.36 & \cellcolor{green!25}32.54 & \cellcolor{green!25}57.13 & 6.49 & 6.79 & 6.38 & 5.60 & 5.322	& 6.350\\
& \texttt{VAFA}  & 46.32 & 40.91 & 53.38 & 47.21 & 38.25 & 36.63 & \cellcolor{green!25}53.90 & \cellcolor{green!25}50.62 & 56.97 & 50.12 & 55.99 & 47.08 & 50.18	& 44.39\\
\midrule
    \multirow{4}{*}{\rotatebox[origin=c]{0}{\parbox[c]{1.6cm}{\centering\texttt{UMamba-B}}}} & \texttt{FGSM}  & 6.19 & 5.19 & 15.83 & 10.68 & 2.66 & 8.21 & 5.08 & 5.61 & \cellcolor{green!25}29.72 & \cellcolor{green!25}38.38 & 15.86 & 11.70 & 9.124	& 8.278\\
& \texttt{PGD}  & 3.30 & 2.45 & 11.14 & 8.54 & 1.42 & 4.98 & 2.44 & 2.64 & \cellcolor{green!25}62.88 & \cellcolor{green!25}96.28 & 12.33 & 10.23 & 6.130	& 5.770\\
& \texttt{CosPGD}  & 3.33 & 3.27 & 10.54 & 7.64 & 1.41 & 5.35 & 2.41 & 2.90 & \cellcolor{green!25}62.09 & \cellcolor{green!25}86.38 & 11.69 & 11.73 & 5.880	& 6.180\\
& \texttt{VAFA}  & 38.30 & 35.75 & 56.69 & 52.14 & 27.55 & 30.64 & 34.72 & 26.39 & \cellcolor{green!25}66.09 & \cellcolor{green!25}76.58 & 59.59 & 51.33 & 43.37	& 39.25\\
\midrule
    \multirow{4}{*}{\rotatebox[origin=c]{0}{\parbox[c]{1.6cm}{\centering\texttt{UMamba-E}}}} & \texttt{FGSM}  & 5.86 & 3.76 & 12.62 & 7.57 & 2.66 & 7.34 & 4.48 & 3.73 & 15.42 & 14.84 & \cellcolor{green!25}29.24 & \cellcolor{green!25}33.82 & 8.208	& 7.448\\
& \texttt{PGD}  & 2.59 & 1.99 & 7.07 & 4.28 & 1.10 & 4.39 & 1.78 & 3.10 & 9.90 & 7.60 & \cellcolor{green!25}54.46 & \cellcolor{green!25}78.53 & 4.488	& 4.270\\
& \texttt{CosPGD}  & 2.38 & 1.44 & 6.74 & 3.76 & 1.05 & 4.83 & 1.68 & 2.99 & 9.22 & 7.39 & \cellcolor{green!25}53.06 & \cellcolor{green!25}63.70 & 4.214	& 4.080\\
& \texttt{VAFA}  & 37.38 & 33.43 & 53.50 & 47.84 & 28.03 & 30.90 & 32.80 & 25.18 & 57.77 & 57.00 & \cellcolor{green!25}64.17 & \cellcolor{green!25}74.60 & 41.90	& 38.87\\
\midrule 
\end{tabular}
}
\caption{\small Performance of models against transfer-based \emph{black box} attacks on  \texttt{Abdomen-CT} dataset.}
\label{tab: bb abdomen}
\end{table}

To evaluate the effectiveness of surrogate models in crafting transferable adversarial examples, we report the average attack success rate across the target models. We observe that SwinUNETR surrogate models tend to craft the most transferable adversarial examples, achieving the highest average ASR-D of $47.73\%$, $50.18\%$, and $16.11\%$ on \texttt{BTCV}, \texttt{Abdomen-CT}, and \texttt{Hecktor} datasets, respectively. However, on \texttt{ACDC}, it achieves a score of $27.06\%$, slightly lower than the $27.81\%$ obtained by SegResNet. This effectiveness of SwinUNETR as a surrogate model can be attributed to its hybrid design, incorporating attributes of both CNN and transformer architectures, which contribute to the generalizability of adversarial examples across different architectures. Furthermore, upon averaging the ASR-D achieved by target models on the most transferable adversarial examples crafted by each surrogate model, we observe that the transformer-based model UNETR tend to be more robust. Similar to our \emph{white box} analysis, UNETR achieves the lowest average ASR-D of $25.54\%$, $30.96\%$, $5.49\%$ on \texttt{BTCV}, \texttt{Abdomen-CT}, and \texttt{Hecktor}, respectively. However, on \texttt{ACDC}, SegResNet with a score of $16.52\%$ is more robust. Furthermore, consistent with Table \ref{tab:wbox}, we note that Mamba-based models generally exhibit greater vulnerability to transfer-based attacks compared to their counterparts. In Figure \ref{fig:vis}, we show a qualitative comparison of transfer-based \emph{black box} attack. Detailed analysis of \emph{black box} setting is in Appendix \ref{appendix: bb_results_eps_4}.




\begin{table}[!t]
\centering\small
  \setlength{\tabcolsep}{3.8pt}
   \scalebox{0.63}[0.63]{
    \begin{tabular}{l|c|cc|cc|cc|cc|cc|cc|cc}
        \toprule
        \rowcolor{LightCyan} 
        Target $\rightarrow$ & Attack  & \multicolumn{2}{c|}{\textbf{UNet}} & \multicolumn{2}{c|}{\textbf{SegResNet}}  & \multicolumn{2}{c|}{\textbf{UNETR}}  & \multicolumn{2}{c|}{\textbf{SwinUNETR}}  &  \multicolumn{2}{c|}{\textbf{UMamba-B}} & \multicolumn{2}{c|}{\textbf{UMamba-E}}  & \multicolumn{2}{c}{\textbf{Average}}\\
                \rowcolor{LightCyan} 
          Surrogate $\downarrow$ &  & ~$\texttt{ASR-D}\hspace{-0.3em}$~ & $\texttt{ASR-H}\hspace{-0.3em}$~  &  $\texttt{ASR-D}\hspace{-0.3em}$~ & ~$\texttt{ASR-H}\hspace{-0.3em}$~ & $\texttt{ASR-D}\hspace{-0.3em}$~  &  $\texttt{ASR-H}\hspace{-0.3em}$ & $\texttt{ASR-D}\hspace{-0.3em}$~  &  $\texttt{ASR-H}\hspace{-0.3em}$ & $\texttt{ASR-D}\hspace{-0.3em}$~  &  $\texttt{ASR-H}\hspace{-0.3em}$ & $\texttt{ASR-D}\hspace{-0.3em}$~  &  $\texttt{ASR-H}\hspace{-0.3em}$ & $\texttt{ASR-D}\hspace{-0.3em}$~  &  $\texttt{ASR-H}\hspace{-0.3em}$\\
          \midrule
\rowcolor{Gray}  & Clean  & 85.52 & 5.75 & 89.65 & 2.56 & 76.37 & 16.31 & 84.19 & 7.93 & 88.22 & 6.01 & 80.91 & 8.48 & - & - \\
\rowcolor{Gray} & \texttt{GN}  & 0.51 & -0.08 & 0.39 & 0.33 & 1.49 & 2.18 & 1.13 & 2.76 & 1.27 & 0.67 & 2.80 & 3.19 & 1.265	& 1.508\\
\midrule
\multirow{4}{*}{\rotatebox[origin=c]{0}{\parbox[c]{1.6cm}{\centering\texttt{UNet}}}} 
& \texttt{FGSM}  & \cellcolor{green!25}29.97 & \cellcolor{green!25}16.24 & 3.00 & 1.99 & 3.05 & 2.56 & 3.07 & 3.29 & 3.64 & 5.18 & 5.95 & 4.94 & 3.742	& 3.590\\
& \texttt{PGD}  & \cellcolor{green!25}64.10 & \cellcolor{green!25}33.39 & 1.60 & 2.65 & 1.70 & 2.65 & 1.62 & 2.66 & 1.90 & 4.07 & 3.09 & 2.34 & 1.982	& 2.870\\
& \texttt{CosPGD}  & \cellcolor{green!25}62.90 & \cellcolor{green!25}31.09 & 1.47 & 2.56 & 2.01 & 2.42 & 1.27 & 2.19 & 1.72 & 3.91 & 3.58 & 3.49 & 2.010	& 2.910\\
& \texttt{VAFA}  & \cellcolor{green!25}35.67 & \cellcolor{green!25}21.69 & 15.50 & 15.61 & 20.28 & 5.46 & 25.47 & 15.02 & 27.95 & 18.45 & 27.23 & 17.50 & 23.29	& 14.41\\
\midrule
\multirow{4}{*}{\rotatebox[origin=c]{0}{\parbox[c]{1.6cm}{\centering\texttt{SegResNet}}}} & \texttt{FGSM}  & 1.89 & 0.57 & \cellcolor{green!25}23.07 & \cellcolor{green!25}9.32 & 3.34 & 3.41 & 3.78 & 4.82 & 7.07 & 13.94 & 6.34 & 5.98 & 4.484	& 5.744\\
& \texttt{PGD}  & 1.14 & 0.59 & \cellcolor{green!25}69.23 & \cellcolor{green!25}34.78 & 1.78 & 1.31 & 1.92 & 1.91 & 2.50 & 4.47 & 4.52 & 2.86 & 2.370	& 2.230\\
& \texttt{CosPGD}  & 0.98 & 0.33 & \cellcolor{green!25}67.32 & \cellcolor{green!25}36.05 & 1.58 & 1.74 & 1.75 & 1.65 & 2.53 & 3.57 & 4.50 & 2.84 & 2.270 &	2.030\\
& \texttt{VAFA}  & 33.69 & 21.47 & \cellcolor{green!25}35.18 & \cellcolor{green!25}20.82 & 21.41 & 5.29 & 27.19 & 16.27 & 28.98 & 18.26 & 27.79 & 16.90 & 27.81	& 15.64\\
\midrule
\multirow{4}{*}{\rotatebox[origin=c]{0}{\parbox[c]{1.6cm}{\centering\texttt{UNETR}}}} & \texttt{FGSM}  & 5.37 & 7.03 & 6.32 & 3.92 & \cellcolor{green!25}43.21 & \cellcolor{green!25}13.46 & 8.40 & 8.81 & 10.74 & 19.92 & 8.03 & 6.45 & 7.772	& 9.226\\
& \texttt{PGD}  & 2.27 & 7.94 & 4.11 & 4.63 & \cellcolor{green!25}{54.06} & \cellcolor{green!25}{19.13} & 4.07 & 7.26 & 7.55 & 16.47 & 12.54 & 8.50 & 6.108	& 8.960\\
& \texttt{CosPGD}  & 2.54 & 7.73 & 3.57 & 4.53 & \cellcolor{green!25}{53.29} & \cellcolor{green!25}{18.49} & 4.07 & 7.48 & 6.85 & 15.10 & 10.39 & 9.03 & 5.484	& 8.774\\
& \texttt{VAFA}  & 32.85 & 20.91 & 15.47 & 13.22 & \cellcolor{green!25}{26.98} & \cellcolor{green!25}{6.32} & 28.29 & 16.54 & 29.95 & 18.84 & 28.75 & 16.61 & 27.06	& 17.22\\
\midrule
\multirow{4}{*}{\rotatebox[origin=c]{0}{\parbox[c]{1.6cm}{\centering\texttt{SwinUNETR}}}} & \texttt{FGSM}  & 2.41 & 1.86 & 8.31 & 2.55 & 4.46 & 1.84 & \cellcolor{green!25}28.08 & \cellcolor{green!25}13.90 & 7.53 & 9.60 & 6.19 & 3.26 & 5.780	& 3.822\\
& \texttt{PGD}  & 1.48 & 1.74 & 5.28 & 4.19 & 4.64 & 2.91 & \cellcolor{green!25}62.50 & \cellcolor{green!25}29.06 & 5.80 & 9.22 & 8.76 & 4.29 & 5.192	& 4.470\\
& \texttt{CosPGD}  & 1.60 & 2.00 & 5.27 & 4.69 & 4.46 & 3.93 & \cellcolor{green!25}61.69 & \cellcolor{green!25}28.27 & 4.88 & 13.97 & 10.42 & 4.75 & 5.326	& 5.870\\
& \texttt{VAFA}  & 34.02 & 21.19 & 19.47 & 14.60 & 22.87 & 6.94 & \cellcolor{green!25}35.75 & \cellcolor{green!25} 17.64 & 30.54 & 19.00 & 28.39 & 16.79 & 27.06	& 15.70\\
\midrule
\multirow{4}{*}{\rotatebox[origin=c]{0}{\parbox[c]{1.6cm}{\centering\texttt{UMamba-B}}}} & \texttt{FGSM}  & 3.97 & 8.78 & 6.38 & 2.55 & 4.54 & 3.68 & 4.58 & 5.81 & \cellcolor{green!25}15.67 & \cellcolor{green!25}15.18 & 9.37 & 6.38 & 5.768	& 5.440\\
& \texttt{PGD}  & 2.70 & 7.06 & 7.16 & 3.62 & 4.23 & 2.82 & 4.24 & 4.40 & \cellcolor{green!25}63.27 & \cellcolor{green!25}28.65 & 11.25 & 6.81 & 5.920	& 4.940\\
& \texttt{CosPGD}  & 2.94 & 4.51 & 6.96 & 4.81 & 3.87 & 2.47 & 4.05 & 4.04 & \cellcolor{green!25}62.68 & \cellcolor{green!25}25.31 & 11.98 & 6.38 & 5.960 &	4.440\\
& \texttt{VAFA}  & 34.06 & 21.60 & 18.29 & 15.23 & 22.19 & 6.51 & 27.05 & 5.46 & \cellcolor{green!25}34.84 & \cellcolor{green!25}19.21 & 29.08 & 17.79 & 26.13	& 13.32\\
\midrule
\multirow{4}{*}{\rotatebox[origin=c]{0}{\parbox[c]{1.6cm}{\centering\texttt{UMamba-E}}}} & \texttt{FGSM}  & 3.08 & 7.66 & 4.69 & 2.19 & 5.14 & 4.28 & 2.37 & 4.23 & 6.25 & 9.76 & \cellcolor{green!25}29.26 & \cellcolor{green!25}15.74 & 4.306	& 5.624\\
& \texttt{PGD}  & 0.86 & 0.36 & 1.94 & 0.87 & 1.86 & 1.85 & 1.24 & 1.24 & 2.10 & 1.45 & \cellcolor{green!25}55.90 & \cellcolor{green!25}23.95 & 1.600	& 1.150\\
& \texttt{CosPGD}  & 0.73 & 0.57 & 2.34 & 1.28 & 1.51 & 1.27 & 1.18 & 1.59 & 2.03 & 1.13 & \cellcolor{green!25}56.47 & \cellcolor{green!25}23.07 & 1.558	& 1.170\\
& \texttt{VAFA}  & 30.51 & 21.33 & 14.08 & 13.21 & 18.97 & 4.49 & 23.97 & 15.92 & 26.96 & 18.53 & \cellcolor{green!25}29.30 & \cellcolor{green!25}17.24 & 22.90	& 14.70\\
\midrule
\end{tabular}
}
\caption{\small Performance of models against transfer-based \emph{black box} attacks on  \texttt{ACDC} dataset. }
\label{tab: bb acdc}
\end{table}

\begin{table}[!h]
\centering\small
  \setlength{\tabcolsep}{3.8pt}
   \scalebox{0.63}[0.63]{
    \begin{tabular}{l|c|cc|cc|cc|cc|cc|cc|cc}
        \toprule
        \rowcolor{LightCyan} 
        Target $\rightarrow$ & Attack  & \multicolumn{2}{c|}{\textbf{UNet}} & \multicolumn{2}{c|}{\textbf{SegResNet}}  & \multicolumn{2}{c|}{\textbf{UNETR}}  & \multicolumn{2}{c|}{\textbf{SwinUNETR}}  &  \multicolumn{2}{c|}{\textbf{UMamba-B}} & \multicolumn{2}{c|}{\textbf{UMamba-E}}  & \multicolumn{2}{c}{\textbf{Average}}\\
                \rowcolor{LightCyan} 
          Surrogate $\downarrow$ &  & ~$\texttt{ASR-D}\hspace{-0.3em}$~ & $\texttt{ASR-H}\hspace{-0.3em}$~  &  $\texttt{ASR-D}\hspace{-0.3em}$~ & ~$\texttt{ASR-H}\hspace{-0.3em}$~ & $\texttt{ASR-D}\hspace{-0.3em}$~  &  $\texttt{ASR-H}\hspace{-0.3em}$ & $\texttt{ASR-D}\hspace{-0.3em}$~  &  $\texttt{ASR-H}\hspace{-0.3em}$ & $\texttt{ASR-D}\hspace{-0.3em}$~  &  $\texttt{ASR-H}\hspace{-0.3em}$ & $\texttt{ASR-D}\hspace{-0.3em}$~  &  $\texttt{ASR-H}\hspace{-0.3em}$ &
          $\texttt{ASR-D}\hspace{-0.3em}$~  &  $\texttt{ASR-H}\hspace{-0.3em}$ \\
          \midrule
\rowcolor{Gray}  & Clean  & 73.91 & 11.36 & 74.73 & 11.08 & 72.36 & 14.61 &71.61 & 22.07 & 73.50 & 10.89 & 72.19 & 13.29 & - & - \\
\rowcolor{Gray} & \texttt{GN}  & 1.80 & 1.61 & 1.47 & -1.00 & 1.08 & -0.10 &  2.00 & 1.58 & 1.38 & 3.07 & 5.61 & 0.16 & 2.223	& 0.887\\
\midrule
\multirow{4}{*}{\rotatebox[origin=c]{0}{\parbox[c]{1.6cm}{\centering\texttt{UNet}}}} 
& \texttt{FGSM}  & \cellcolor{green!25} 30.13 & \cellcolor{green!25}26.43 & 15.62 & 7.15 & 5.87 & 2.62 & 12.85 & 8.54 & 14.24 & 7.74 & 14.72 & 5.79 & 12.66	& 6.370\\
& \texttt{PGD}  & \cellcolor{green!25} 39.84 & \cellcolor{green!25}67.25 & 6.49 & 2.72 &  2.49 & 2.93 & 6.62 & 5.26 & 6.12 & 3.73 & 6.64 & 4.51 & 5.672	& 3.830\\
& \texttt{CosPGD}  & \cellcolor{green!25} 39.69 & \cellcolor{green!25}67.26 & 5.25 & 1.84 &  2.34 & 2.22 & 5.50 & 3.93 & 4.92 & 2.22 & 5.98 & 2.81 & 4.798	& 2.600\\
& \texttt{VAFA}  & \cellcolor{green!25} 29.61 & \cellcolor{green!25}30.83 & 5.92 & 3.72 &  9.26 & 3.76 & 6.51 & -0.02 & 8.51 & 7.04 & 11.82 & 8.56 & 8.400	& 4.610\\
\midrule
\multirow{4}{*}{\rotatebox[origin=c]{0}{\parbox[c]{1.6cm}{\centering\texttt{SegResNet}}}} & \texttt{FGSM}  & 13.29 & 10.07 & \cellcolor{green!25} 27.03 & \cellcolor{green!25} 21.16 & 4.85 & 2.56 & 14.60 & 8.13 & 19.58 & 16.23 & 17.06 & 7.79 & 13.87	& 8.956\\
& \texttt{PGD}  & 12.06 & 9.53 & \cellcolor{green!25} 41.71 & \cellcolor{green!25} 70.09 & 3.77 & 2.52 & 12.98 & 7.00 & 22.29 & 20.59 & 17.66 & 12.26 & 13.75	& 10.38\\
& \texttt{CosPGD}  & 11.92 & 8.53 & \cellcolor{green!25} 41.32 & \cellcolor{green!25} 70.20 & 3.55 & 2.02 & 12.87 & 7.12 & 23.25 & 21.36 & 18.04 & 9.99 & 13.93	& 9.800\\
& \texttt{VAFA}  & 4.84 & 6.41 & \cellcolor{green!25} 27.18 & \cellcolor{green!25} 22.00 & 6.12 & 3.10 & 6.06 & -0.22 & 8.27 & 6.90 & 9.97 & 10.25 & 7.050 &	5.290\\
\midrule
\multirow{4}{*}{\rotatebox[origin=c]{0}{\parbox[c]{1.6cm}{\centering\texttt{UNETR}}}} & \texttt{FGSM}  & 15.75 & 7.91 & 15.72 & 4.52 & \cellcolor{green!25}27.47 & \cellcolor{green!25}20.47 & 17.32 & 9.20 & 15.62 & 6.22 & 15.24& 6.33 & 15.93	& 6.836 \\
& \texttt{PGD}  & 15.58 & 8.99 & 13.71 & 5.81 & \cellcolor{green!25}{37.43} & \cellcolor{green!25}{59.97} & 16.06 & 10.68 & 13.12 & 8.51 & 13.21 & 4.560 & 14.33 &	7.710\\
& \texttt{CosPGD}  & 14.89 & 9.82 & 12.76 & 6.21 & \cellcolor{green!25}{37.47} & \cellcolor{green!25}{59.23} & 15.48 & 10.22 & 11.95 & 8.75 & 12.24 & 6.82 & 13.46	& 8.364\\
& \texttt{VAFA}  & 9.18 & 8.00 & 5.05 & 2.73 & \cellcolor{green!25}{24.35} & \cellcolor{green!25}{19.21} & 13.76 & 5.46 & 8.12 & 8.30 & 13.60 & 7.99 & 9.940 &	6.496\\
\midrule
\multirow{4}{*}{\rotatebox[origin=c]{0}{\parbox[c]{1.6cm}{\centering\texttt{SwinUNETR}}}} & \texttt{FGSM}  & 15.95 & 10.46 & 19.14 & 7.89 & 7.70 & 4.78 & \cellcolor{green!25}27.79 & \cellcolor{green!25}30.88 & 18.96 & 13.82 & 18.80 & 8.19 & 16.11	& 9.028 \\
& \texttt{PGD}  & 18.66 & 14.51 & 18.64 & 9.08 & 6.07 & 5.22 & \cellcolor{green!25}37.49 & \cellcolor{green!25}56.78 & 18.36 & 12.83 & 18.01 & 10.31 & 15.94	& 10.39\\
& \texttt{CosPGD}  & 18.08 & 12.52 & 18.12 & 10.27 & 5.83 & 4.82 & \cellcolor{green!25}37.40 & \cellcolor{green!25}55.55 & 18.13 & 14.29 & 17.49 & 9.79 & 15.53 & 	10.34\\
& \texttt{VAFA}  & 8.33 & 8.40 & 6.35 & 4.38 & 10.39 & 5.64 & \cellcolor{green!25}27.17 & \cellcolor{green!25}20.07 & 7.57 & 5.64 & 13.63 & 7.39 & 9.250 &	6.290\\
\midrule
\multirow{4}{*}{\rotatebox[origin=c]{0}{\parbox[c]{1.6cm}{\centering\texttt{UMamba-B}}}} & \texttt{FGSM}  & 13.57 & 7.27 & 19.23 & 7.70 & 5.13 & 2.12 & 15.32 & 10.95 & \cellcolor{green!25}23.34 & \cellcolor{green!25}19.23 & 17.68 & 6.57 & 14.18	& 6.922\\
& \texttt{PGD}  & 12.08 & 7.98 & 25.08 & 28.91 & 3.29 & 2.27 & 13.49 & 9.23 & \cellcolor{green!25}39.19 & \cellcolor{green!25}67.50 & 21.16 & 17.41 & 15.02	& 13.16\\
& \texttt{CosPGD}  & 12.04 & 8.98 & 26.16 & 29.73 & 3.24 & 3.00 & 12.56 & 8.95 & \cellcolor{green!25}39.09 & \cellcolor{green!25}67.55 & 20.80 & 17.87 & 14.96	& 13.71\\
& \texttt{VAFA}  & 7.30 & 9.50 & 9.74 & 5.24 & 8.25 & 3.89 & 7.08 & -0.54 & \cellcolor{green!25}24.23 & \cellcolor{green!25}25.10 & 13.67 & 8.840 & 9.210 &	5.390 \\
\midrule
\multirow{4}{*}{\rotatebox[origin=c]{0}{\parbox[c]{1.6cm}{\centering\texttt{UMamba-E}}}} & \texttt{FGSM}  & 11.27 & 5.31 & 15.44 & 5.47 & 3.93 & 2.49 & 11.27 & 7.22 & 16.01 & 11.39 & \cellcolor{green!25}25.04 & \cellcolor{green!25}18.36 & 11.58	& 6.376\\
& \texttt{PGD}  & 4.61 & 3.59 & 5.96 & 1.14 & 1.77 & 0.92 & 4.69 & 3.07 & 6.75 & 5.84 & \cellcolor{green!25}37.90 & \cellcolor{green!25}64.96 & 4.756	& 2.910\\
& \texttt{CosPGD}  & 4.36 & 2.50 & 5.24 & 1.80 & 1.69 & 0.46 & 4.27 & 3.97 & 6.07 & 6.10 & \cellcolor{green!25}37.82 & \cellcolor{green!25}64.56 & 4.326	& 2.970\\
& \texttt{VAFA}  & 5.82 & 9.15 & 4.89 & 2.82 & 8.10 & 4.51 & 5.75 & 2.10 & 7.47 & 9.23 & \cellcolor{green!25} 27.38 & \cellcolor{green!25}23.55 & 6.410	& 5.560\\
\midrule
\end{tabular}
}
\caption{\small Performance of models against transfer-based \emph{black box} attacks on  \texttt{Hecktor} dataset.}
\label{tab: bb hecktor}
\end{table}


\subsection{Frequency Analysis} 
\label{sec:freq_analysis}
\begin{figure}[!t]
\begin{minipage}{\linewidth}
    \centering
    \includegraphics[ width=0.24\textwidth]{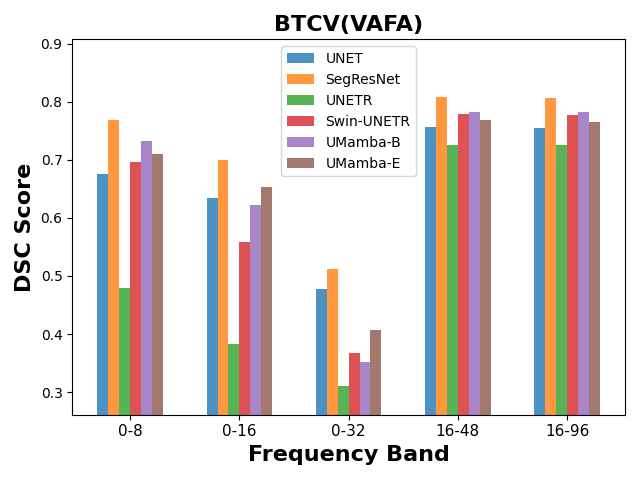}
    \includegraphics[  width=0.24\textwidth]{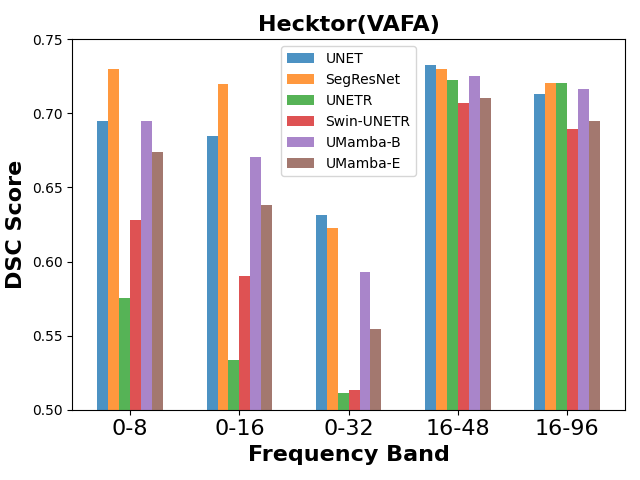}
        \includegraphics[  width=0.24\textwidth]{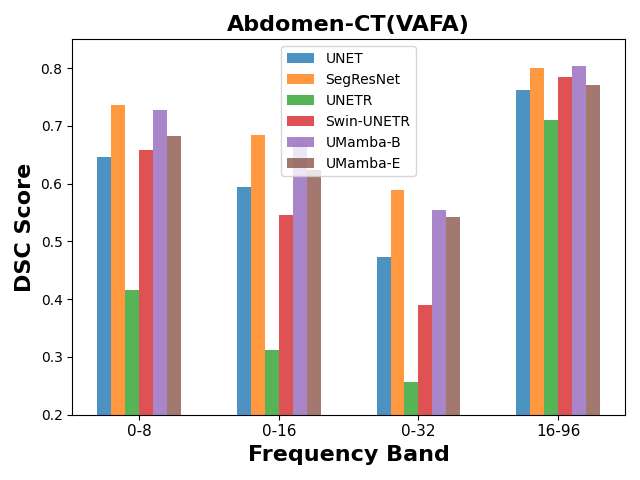}
    \includegraphics[  width=0.24\textwidth]{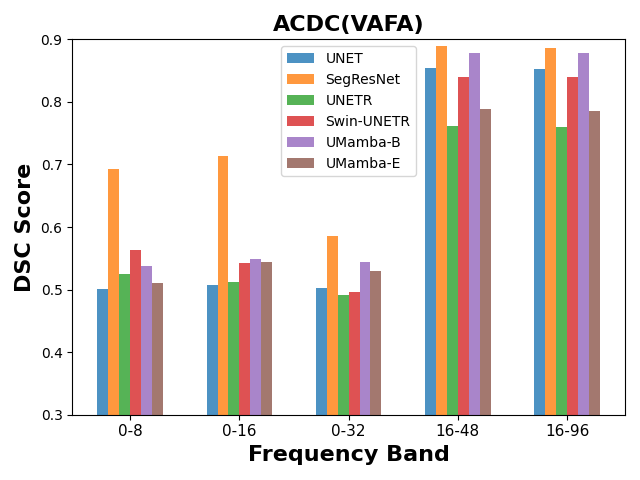}

\end{minipage}

    \caption{\small \textbf{Frequency Analysis (\texttt{VAFA}):} Low frequency components of adversarial perturbation cause significant performance degradation(DSC score is reported). }
    \label{fig:freq_analysis}
\end{figure}

Given the overall effectiveness of \texttt{VAFA} in achieving higher transferability of adversarial examples compared to pixel-based attacks, in this section we delve deeper into the frequency analysis of \texttt{VAFA} to study which frequency components lead to drop in performance of models. Following \cite{shao2021adversarial}, we implement an adversarial attack incorporating a frequency filter $M$, restricting perturbations to specific frequency domains. The filter operation is defined as {\footnotesize
$x'_{\text{freq}} = \text{IDCT}(\text{DCT}(x'-x) \odot M) + x$
}, where \texttt{DCT} and \texttt{IDCT} denote Discrete Cosine Transform and its inverse, respectively. Similar to \cite{shao2021adversarial}, using the filter $M$, we extract 3D cubes of varying size $n$ from the top left corner as part of the low-frequency components $(0,n)$ where $n \in \{8,16,32\}$. Similarly, mid-frequency $(16-48)$ and high-frequency $(16-96)$ components are also extracted. Our results as shown in Figure \ref{fig:freq_analysis}, show that across all models, the performance drop in adversarial perturbations crafted by \texttt{VAFA} is attributed to their low-frequency components.
This finding deviates from previous studies in the natural domain \cite{shao2021adversarial,bai2021transformers}, which demonstrate that high-frequency components are linked to a decrease in performance. Our results align with recent works\cite{maiya2021frequency, li2024exploiting} which suggest that the potency of adversarial examples cannot be solely attributed to high-frequency components. Instead, it depends on various factors including the model architecture and the characteristics of the dataset. We provide further frequency analysis across different attacks in the Appendix \ref{appendix: freq_analysis}.

\subsection{Robustness of SAM-Med3D against Black-box Attacks}
\begin{figure}[!t]
\begin{minipage}{\linewidth}
    \centering
    \includegraphics[ width=0.24\textwidth]{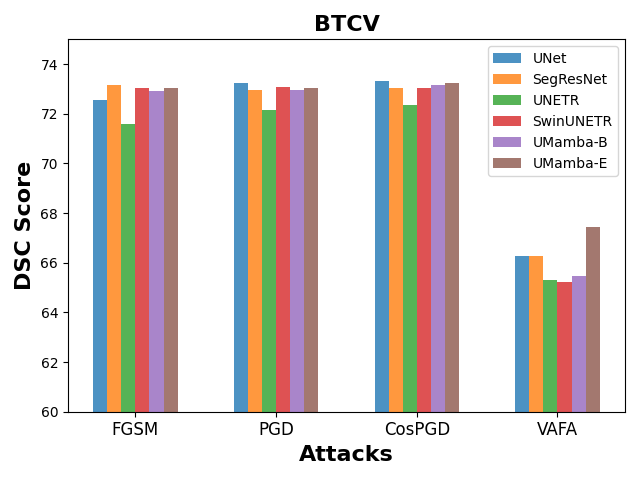}
    \includegraphics[  width=0.24\textwidth]{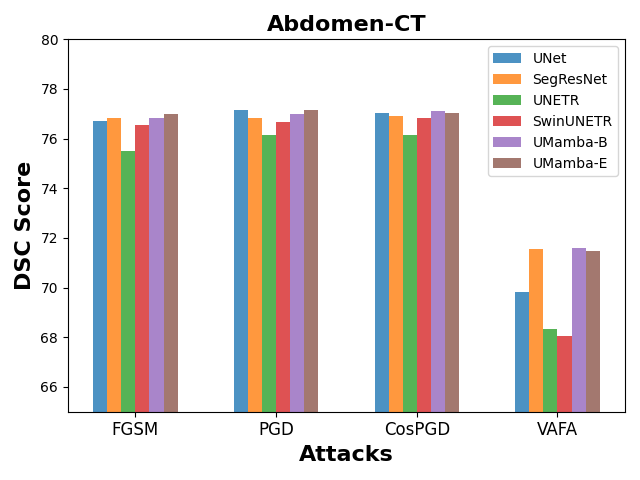}
        \includegraphics[  width=0.24\textwidth]{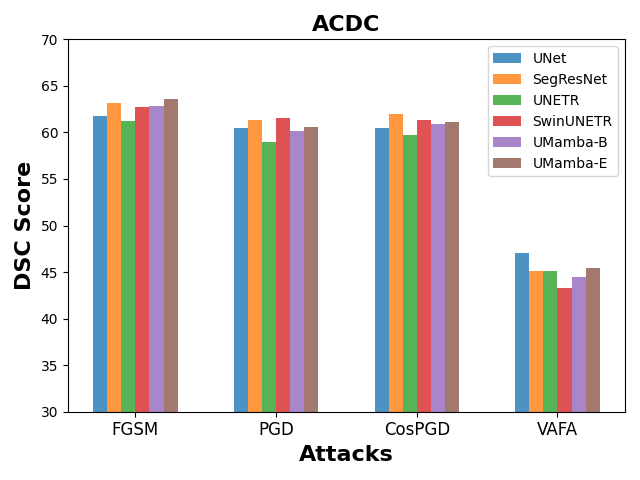}
    \includegraphics[  width=0.24\textwidth]{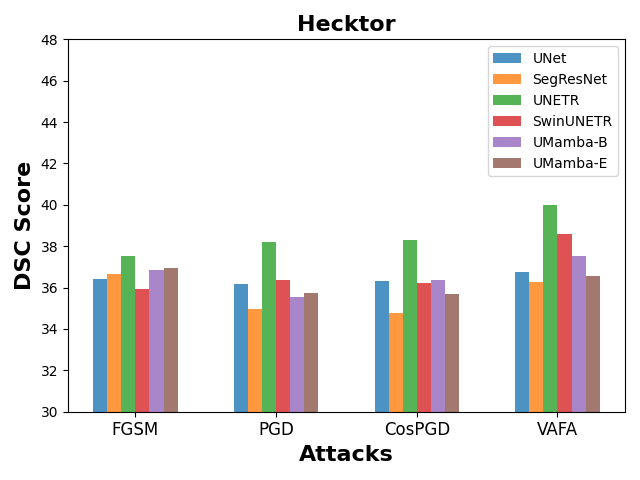}

\end{minipage}

    \caption{\small Evaluating SAM-Med3D against transfer-based \emph{black box} attacks.}
    \label{fig:medsam}
\end{figure}

We assess the robustness of the SAM-Med3D model  against adversarial examples crafted by different surrogate models. In Figure \ref{fig:medsam}, we report the DSC score obtain by SAM-Med3D on different datasets.
 Similar to  Section \ref{sec:bbox}, we observe that adversarial examples crafted using pixel-based attacks do not transfer to SAM-Med3D, while adversarial examples crafted using \texttt{VAFA} exhibit greater transferability. However, in general, we observe that the drop in performance of SAM-Med3D against transfer-based attacks is low compared to models trained on individual datasets(see Section \ref{sec:bbox}), which can be attributed to the robust generalization capabilities of the model, which are attained through effective training on large-scale datasets. We report detailed results on SAM-Med3D in Appendix \ref{appendix: bb_results_eps_4}.

\section{Conclusion}
This work presents the first comprehensive empirical study on the robustness of current volumetric medical segmentation models against adversarial attacks. Our results indicate that current pixel-based attacks while successful under \emph{white box} settings show limited transferabilty to unknown target models. In contrast, frequency-based attacks result in highly transferable adversarial examples. We show that low-frequency changes to images in frequency-based attacks are responsible for the attack success. Across different models, we observe transformer-based models are more robust than CNNS and Mamba-based models under different adversarial settings. Moreover, we highlight the enhanced robustness of vision foundational models trained on large-scale datasets. Our study highlights the importance of evaluating the robustness of vision-based models in the medical field. We hope this will foster future efforts to improve the robustness of these models.

\bibliography{egbib}

\newpage
\appendix

\noindent\begin{huge} \textbf{Appendix} \vspace{4mm} \end{huge}

We offer further insights into our study through various sections in the appendix. Firstly, we provide additional details regarding the datasets utilized in Section \ref{appendix:datasets} and elaborate on the training recipe for the models, as discussed in Section \ref{appendix:training}. In Section \ref{appendix: wb_results_eps_4}, we delve deeper into the \emph{white box} analysis, providing ablations across all datasets, examining pixel and frequency-based attacks at different perturbation budgets. Likewise, Section \ref{appendix: freq_analysis} expands on the frequency analysis initially presented in Section \ref{sec:freq_analysis}, covering different adversarial attacks. Lastly, in Section \ref{appendix: bb_results_eps_4}, we present results on transfer-based \emph{black box} attacks across all datasets, considering various perturbation strengths, and offer further insights derived from these findings.
\textbf{Our well-documented code and pretrained weights will be made publicly available.}


\section{Datasets}
\label{appendix:datasets}
Medical imaging data sets encompass a range of imaging techniques, including Positron Emission Tomography (PET), Computed Tomography (CT), and Magnetic Resonance Imaging (MRI). PET scans effectively show the metabolic or biochemical activity within the body, and CT imaging offers high-resolution images of the body's internal structure. Similarly, MRIs effectively differentiate between soft tissues without the use of ionizing radiation. These imaging modalities acquire comprehensive and complementary information about the body's organs, functions, and tumors. Thus, to conduct a comprehensive benchmarking analysis of the model's robustness and susceptibility to adversarial attacks, we utilize four different segmentation datasets: \texttt{BTCV}, \texttt{ACDC}, \texttt{Hecktor}, and \texttt{Abdomen-CT}, which consist of medical images from CT and MRI modalities encompassing different tumor and organ segmentations. 

\noindent \textbf{BTCV:} The \texttt{BTCV} dataset consists of 30 abdominal CT scans from metastatic liver cancer patients acquired from a single medical center. Each CT scan is manually annotated for 13 abdominal organs (Spleen, Right Kidney, Left Kideny, Gallbladder, Esophagus, Liver, Stomach, Aorta, IVC, Portal and Splenic Veins, Pancreas, Right adrenal gland, and Left adrenal gland). The CT scan size is $512 \times 512$ pixels, the number of slices ranges from 80 to 225, and the slice thickness ranges from 1 to 6 mm.

\noindent \textbf{ACDC:} The Automated Cardiac Diagnosis Challenge (\texttt{ACDC}) dataset consists of 150 MRI images from patients with cardiac abnormalities acquired from a single medical center. Each MRI scan is manually annotated for different heart organs, such as the left ventricle (LV), right ventricle (RV), and myocardium (MYO). The number of MRI slices ranges from 28 to 40, and the slice thickness ranges from 5 to 8 mm. The spatial resolution goes from 1.37 to 1.68 mm$^2/$pixel. 

\noindent \textbf{HECKTOR:}The \texttt{Hecktor} dataset consists of 524 CT/PET scans of head and neck cancer patients collected from seven medical centers. It was manually annotated for primary gross tumor volumes (GTVp) and nodal gross tumor volumes (GTVn). The CT scan size ranges from $128 \times 128$ to $512 \times 512$, the number of slices ranges from 67 to 736, and the slice thickness ranges from 1 to 2.8 mm.

\noindent \textbf{AbdomenCT-1k:} The AbdomenCT-1k dataset consists of 1112 abdominal CT scans from 12 medical centers, including multi-phase, multi-vendor, and multi-disease cases. All the scans' annotations for the liver, kidney, spleen, and pancreas are provided. The CT scan size has resolutions of $512 \times 512 $pixels with varying voxel sizes and slice thicknesses between 1.25 to 5 mm.

\section{Training Details }
\label{appendix:training}
For  \texttt{BTCV}, \texttt{Abdomen-CT}, and \texttt{ACDC} all models trained  for 5000 epochs, while for \texttt{Hecktor} we use only 500 epochs. A batch size of 3 and a learning rate (\texttt{lr}) of $1e-4$ with the \texttt{warmup\_cosine} scheduler is used. During training, images are normalized to the range of $[0,1]$, and a 3D random crop of size $96\times96\times96$ is selected as an input to the segmentation model. Augmentations include \texttt{RandomFlip} (for all three spatial dimensions), \texttt{RandomRotate90}, \texttt{RandomScaleIntensity}, and \texttt{RandomShiftIntensity}. During inference, we employed a sliding window approach, dividing the input volume of arbitrary size into 3D sliding windows of size $96\times96\times96$ with a 50\% overlap. The predictions of overlapping voxels were combined using a Gaussian kernel.

\section{Robustness against White-box Attacks }
\label{appendix: wb_results_eps_4}
\begin{figure}[!t]
\begin{minipage}{\linewidth}
    \centering
    \includegraphics[ width=0.24\textwidth]{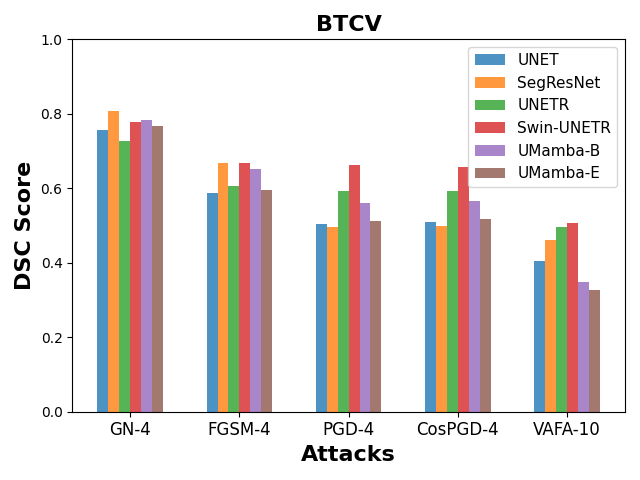}
    \includegraphics[  width=0.24\textwidth]{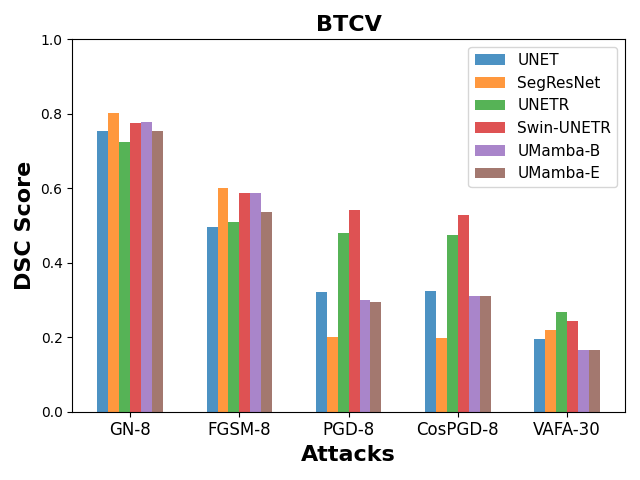}
        \includegraphics[  width=0.24\textwidth]{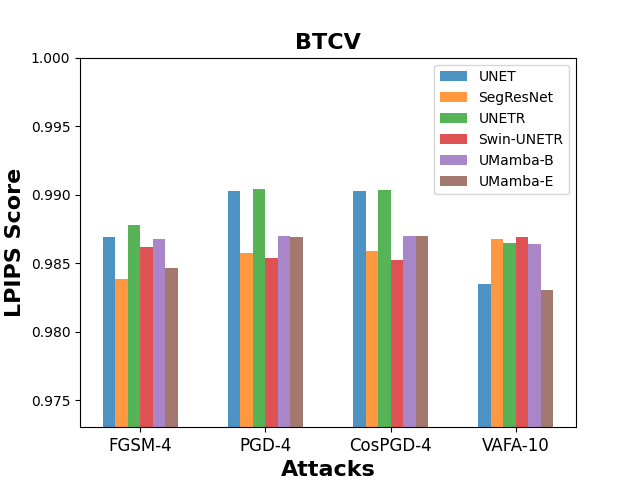}
    \includegraphics[  width=0.24\textwidth]{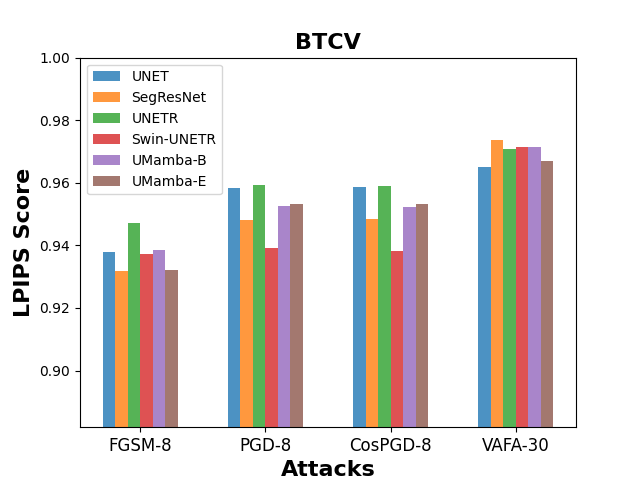}
\end{minipage}
\begin{minipage}{\linewidth}
    \centering
    \includegraphics[ width=0.24\textwidth]{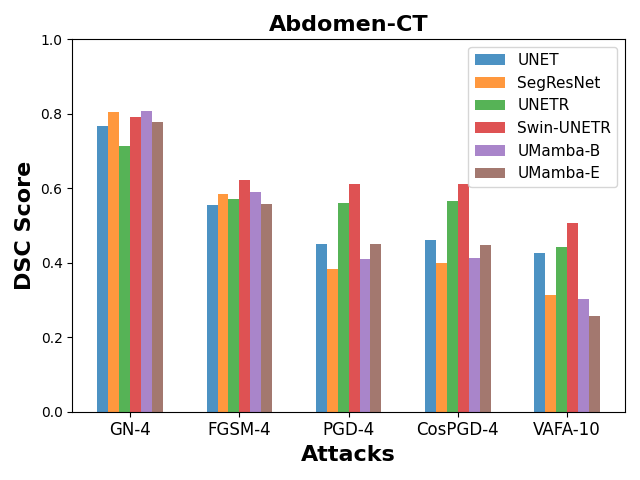}
    \includegraphics[  width=0.24\textwidth]{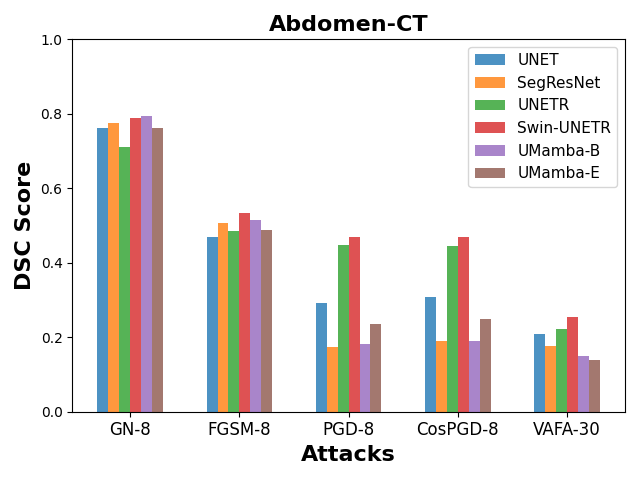}
        \includegraphics[  width=0.24\textwidth]{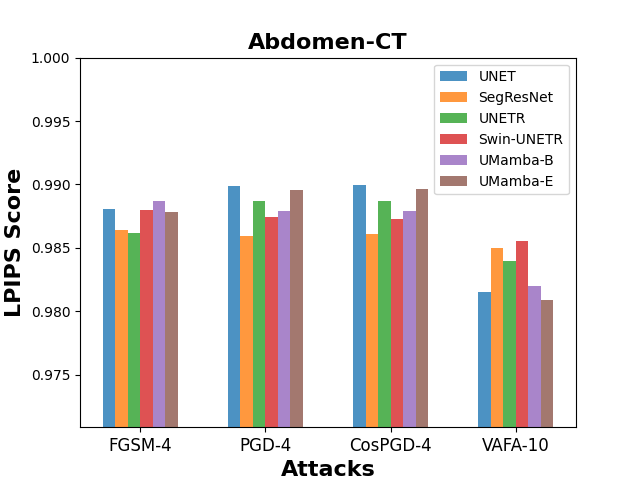}
    \includegraphics[  width=0.24\textwidth]{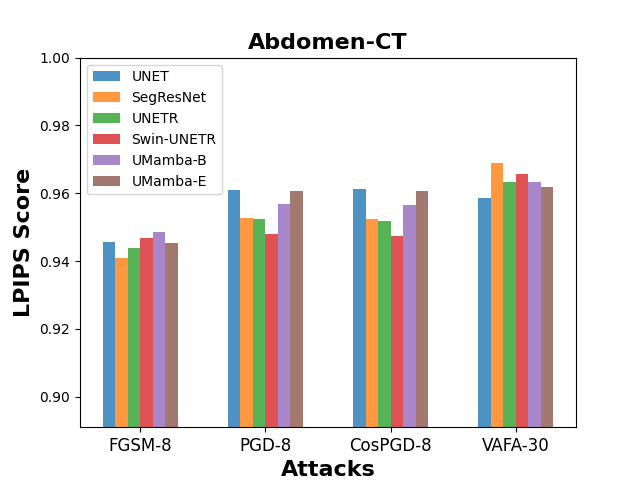}
\end{minipage}
\begin{minipage}{\linewidth}
    \centering
    \includegraphics[ width=0.24\textwidth]{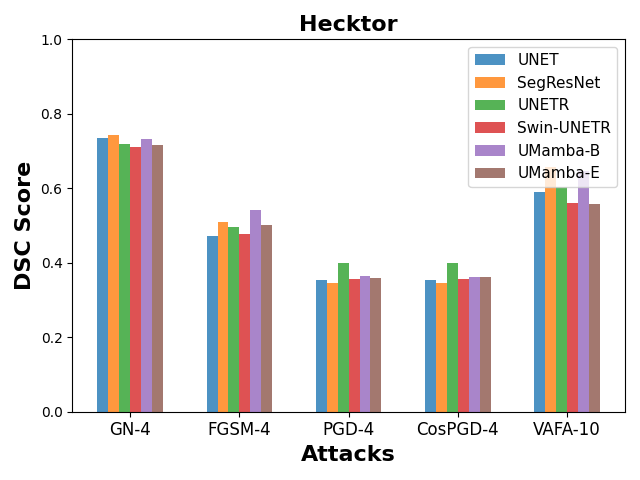}
    \includegraphics[  width=0.24\textwidth]{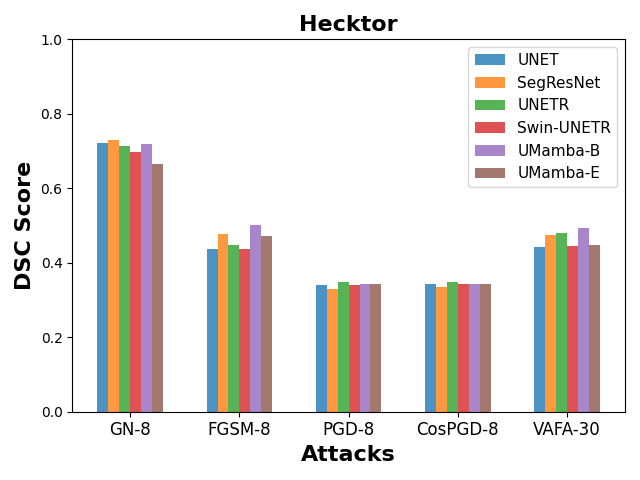}
        \includegraphics[  width=0.24\textwidth]{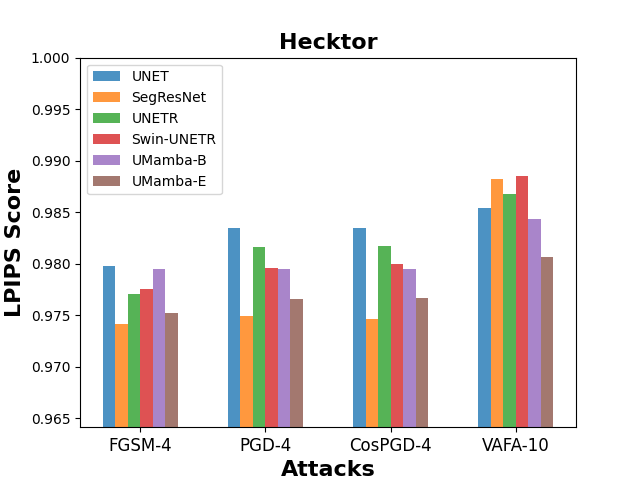}
    \includegraphics[  width=0.24\textwidth]{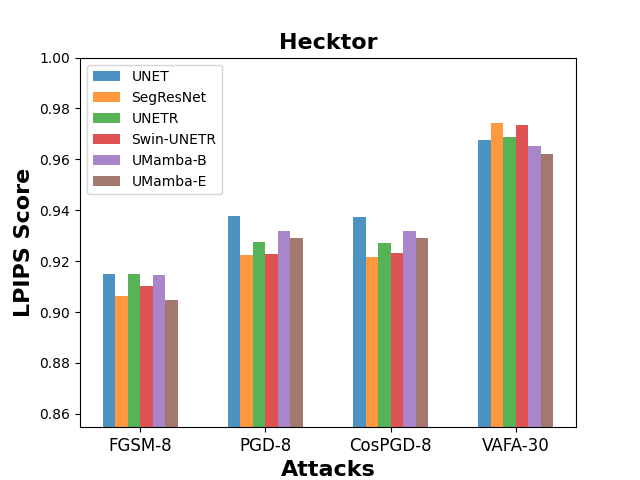}
\end{minipage}
\begin{minipage}{\linewidth}
    \centering
    \includegraphics[ width=0.24\textwidth]{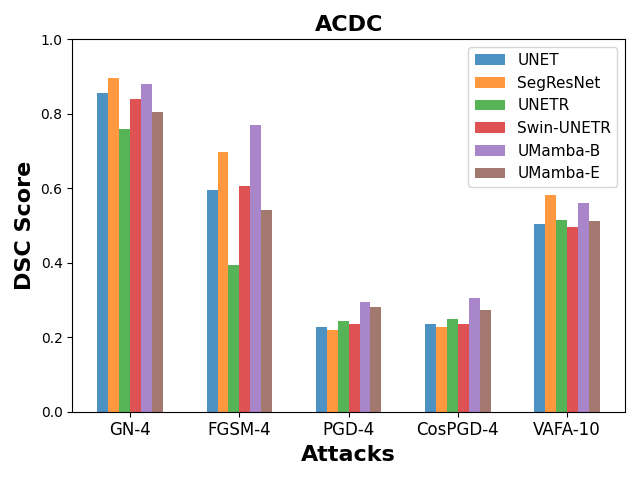}
    \includegraphics[  width=0.24\textwidth]{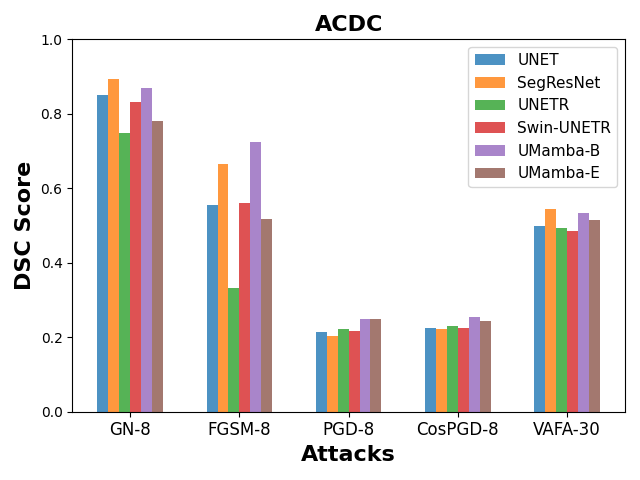}
        \includegraphics[  width=0.24\textwidth]{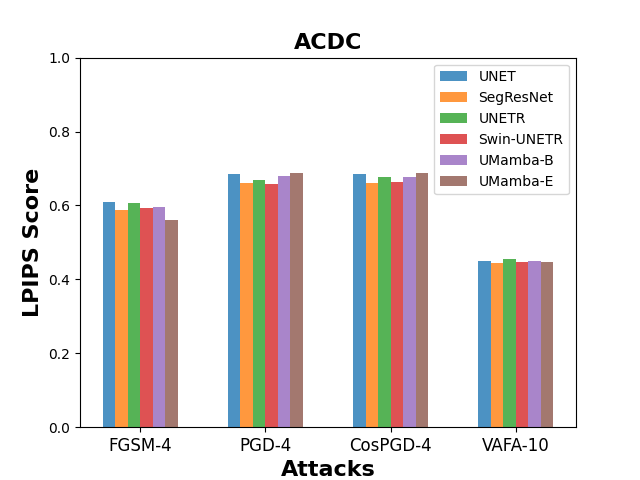}
    \includegraphics[  width=0.24\textwidth]{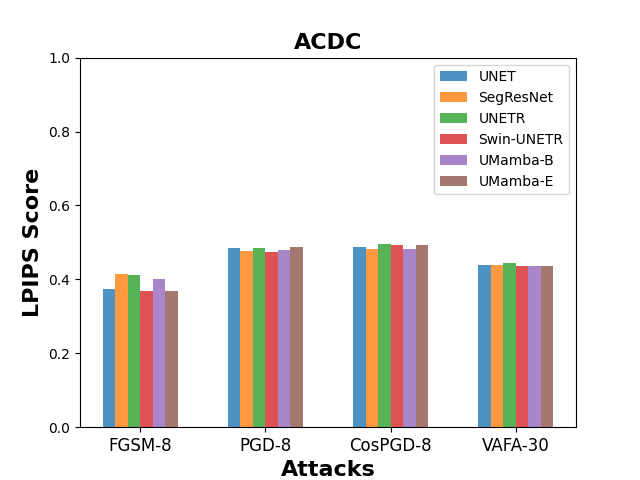}
\end{minipage}

    \caption{\small \textbf{White Box Attack Ablation:} Evaluating robustness of volumetric segmentation models on \emph{white box} attacks. For pixel-based attacks results are reported for $\epsilon=\frac{4}{255}$ and $\epsilon=\frac{8}{255}$ indicated by attack names followed by the suffixes $-4$ or $-8$, respectively. Regarding frequency-based attack \texttt{VAFA}, the results are reported with a constraint on $q_{\text{max}}$ set to $10$ and $30$, denoted as \texttt{VAFA-10} and \texttt{VAFA-30}, respectively. DSC score \emph{(lower is better)} and LPIPS score \emph{(higher is better)} are reported on the generated adversarial examples.}
    \label{fig:wb_appendix}
\end{figure}

\begin{figure}[!t]
\begin{minipage}{\linewidth}
    \centering
    \includegraphics[ width=0.24\textwidth]{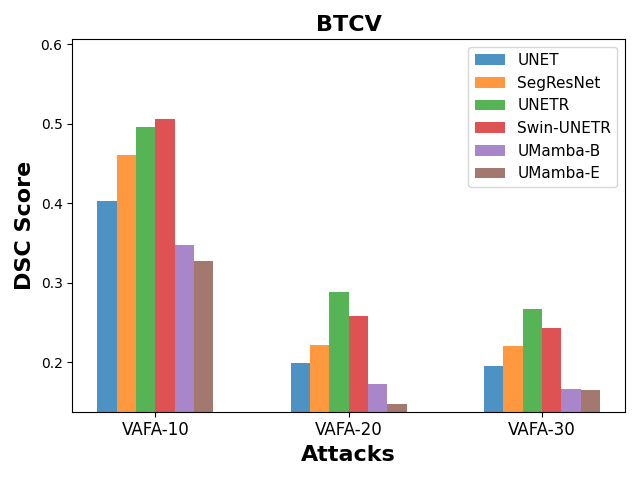}
    \includegraphics[  width=0.24\textwidth]{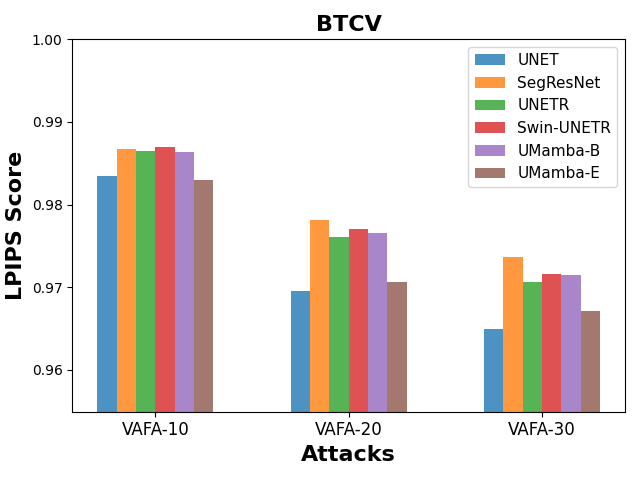}
       \includegraphics[ width=0.24\textwidth]{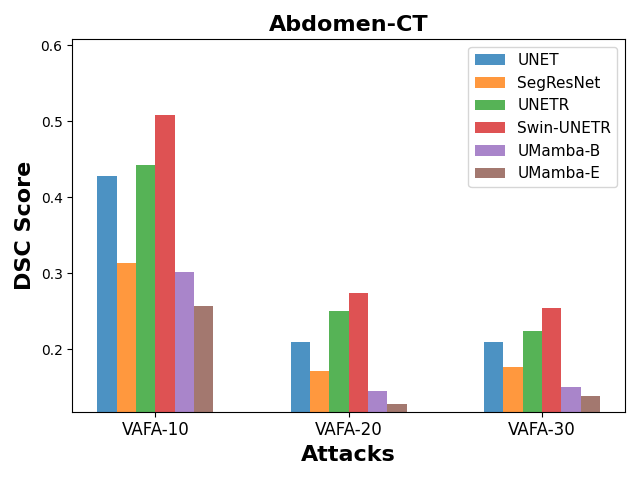}
    \includegraphics[  width=0.24\textwidth]{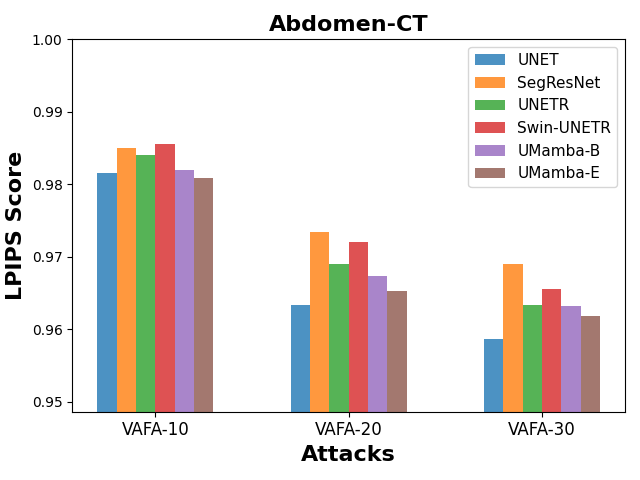}
\end{minipage}
\begin{minipage}{\linewidth}
    \centering
          \includegraphics[ width=0.24\textwidth]{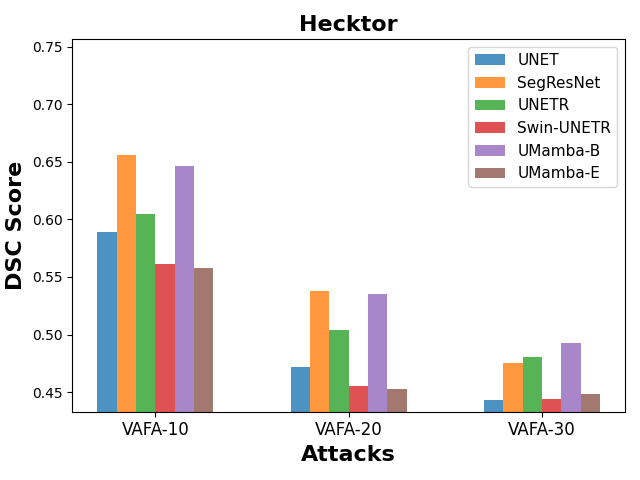}
    \includegraphics[  width=0.24\textwidth]{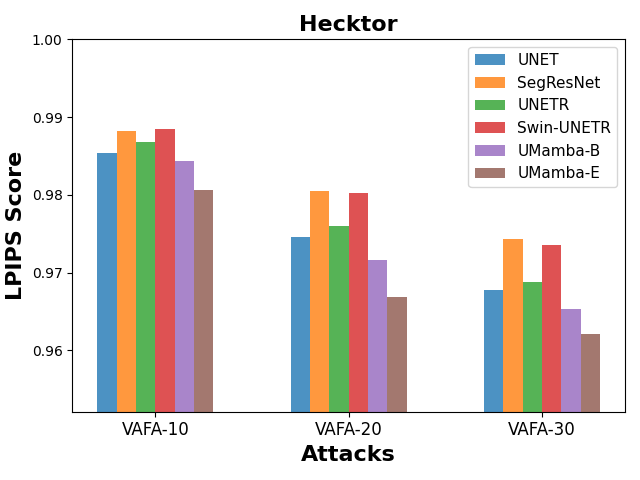}
       \includegraphics[ width=0.24\textwidth]{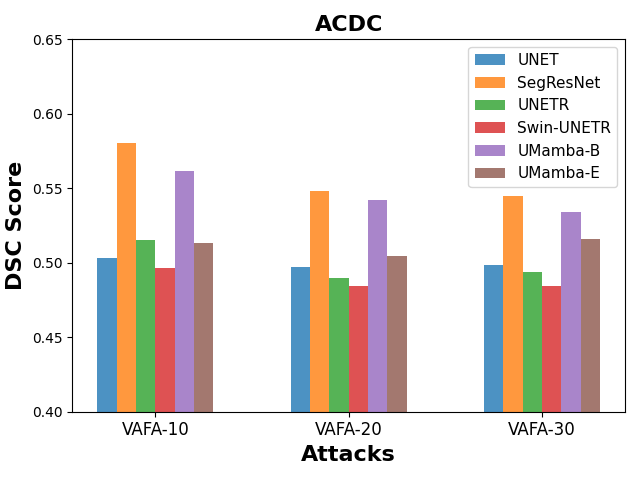}
    \includegraphics[  width=0.24\textwidth]{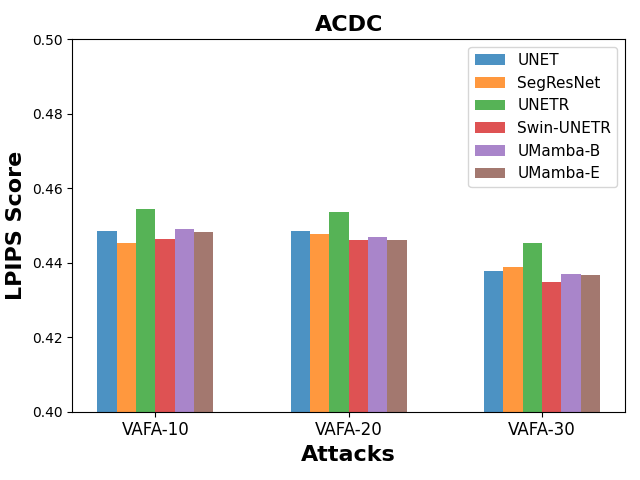}
\end{minipage}

    \caption{\small  \textbf{Frequency Attack Ablation:} Examining the robustness of volumetric segmentation models \texttt{VAFA} based \emph{white box} attack. Adversarial examples are generated at $q_{max} \in \{10, 20, 30\}$, represented as \texttt{VAFA-10}, \texttt{VAFA-20}, and \texttt{VAFA-30}, respectively. DSC score \emph{(lower is better)} and LPIPS score \emph{(higher is better)} are reported on the generated adversarial examples.  }
    \label{fig:wb_vafa_appendix}
\end{figure}

In Figure \ref{fig:wb_appendix}, we report robustness of the volumetric segmentation models on \emph{white box} attacks across \texttt{BTCV}, \texttt{Abdomen-CT}, \texttt{Hecktor}, and \texttt{ACDC} datasets. For pixel-based attacks we craft adversarial examples at $l_{\infty}$ perturbation budget $\epsilon \in \{\frac{4}{255}, \frac{8}{255}\}$ for \texttt{FGSM}, \texttt{PGD}, 
 and \texttt{CosPGD}. For frequency-based attack \texttt{VAFA} we craft adversarial examples with $q_{max} \in \{10, 30\}$. We report DSC and LPIPS score on generated adversarial examples. Similar to our results in Table \ref{tab:wbox} in the main paper, 
we observe that \texttt{VAFA} causes the most drop in DSC score of models on \texttt{BTCV} and \texttt{Abdomen-CT} dataset, while iterative pixel-based attacks \texttt{PGD} and \texttt{CosPGD} cause the most drop on \texttt{Hecktor} and \texttt{ACDC} dataset. Furthermore, we provide ablation across \texttt{VAFA} attack with varying $q_{max} \in \{10, 20, 30\}$ in Figure \ref{fig:wb_vafa_appendix}. With an increase in $q_{max}$, we observe a decrease in the DSC score across all the datasets. Consequently, we also observe a drop in the LPIPS score of the generated adversarial examples. 

\section{Adversarial Examples}
\label{appendix: adv_examples}

\begin{figure}[h]
    \centering
    \includegraphics[width=\linewidth]{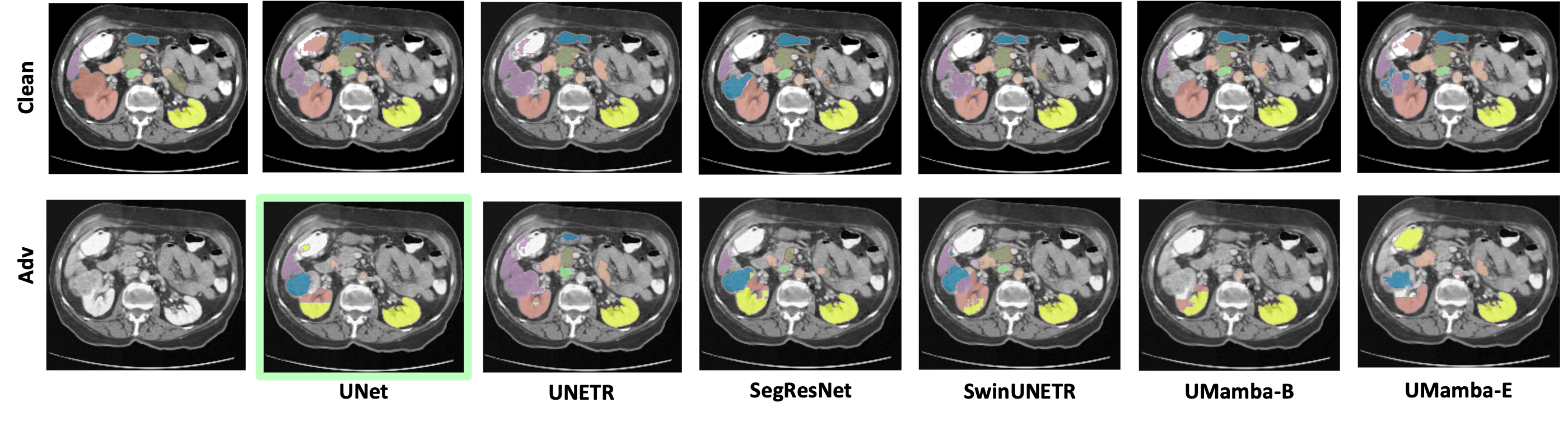}
    \caption{\small Comparing multi-organ segmentation across various models under transfer-based \emph{black box} attacks, where adversarial examples are generated on UNet and transferred to other unseen models.}
    \label{fig:vis_app}
\end{figure}

In Figure \ref{fig:vis_app}, adversarial example using \texttt{VAFA} is crafted on UNet model trained on \texttt{Abdomen-CT} dataset. Segmentation prediction of all the volumetric segmentation models is shown for the clean sample and the adversarial one. We can clearly observe that the adversarial example causes the predictions to change across all models.

\section{Frequency Analysis of White-Box Attacks }
\label{appendix: freq_analysis}

\begin{figure}[!t]
\begin{minipage}{\linewidth}
    \centering
    \includegraphics[ trim= 30mm 25mm 30mm 10mm, clip, width=0.19\textwidth]{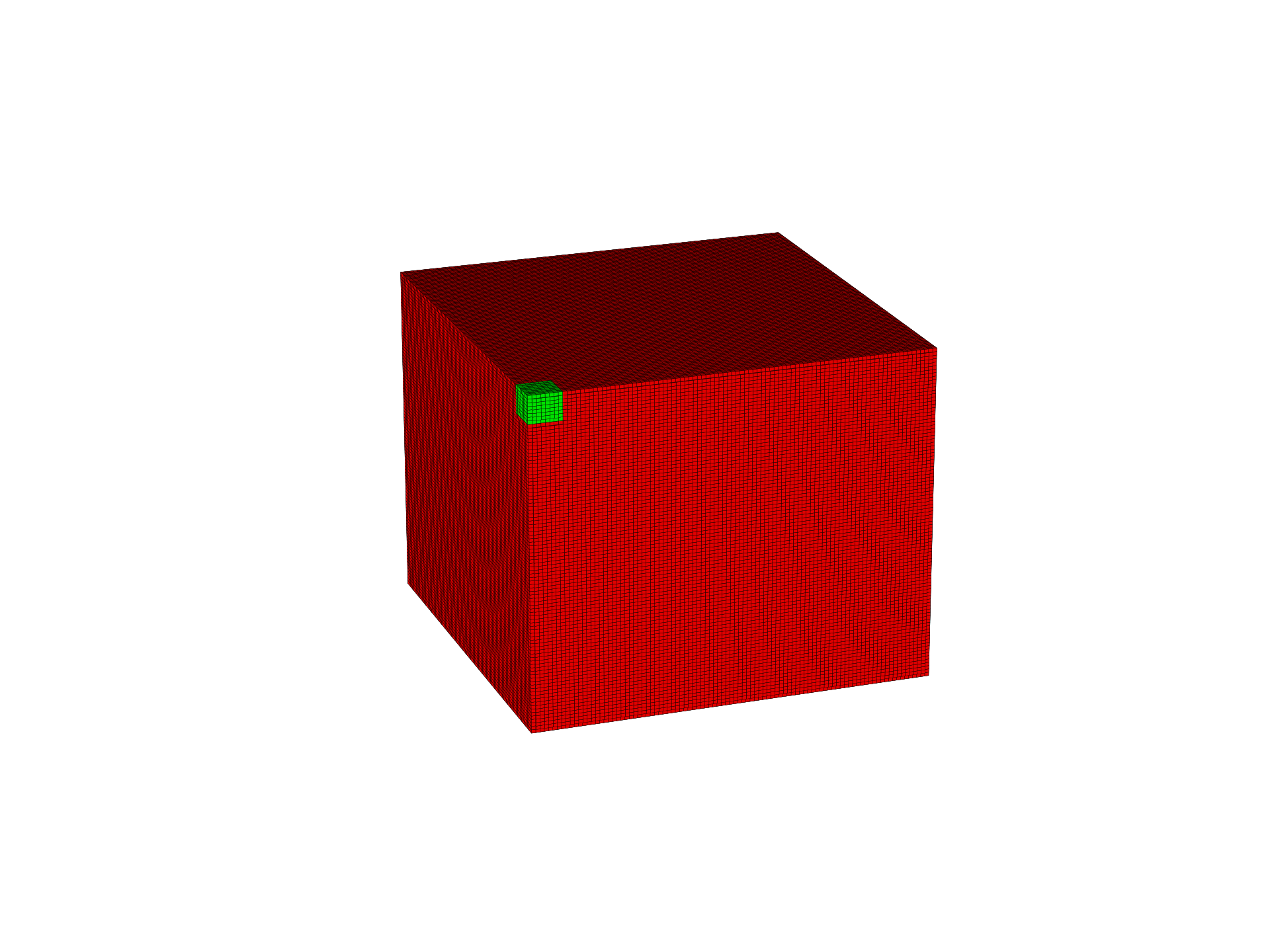}
    \includegraphics[ trim= 30mm 25mm 30mm 10mm, clip,  width=0.19\textwidth]{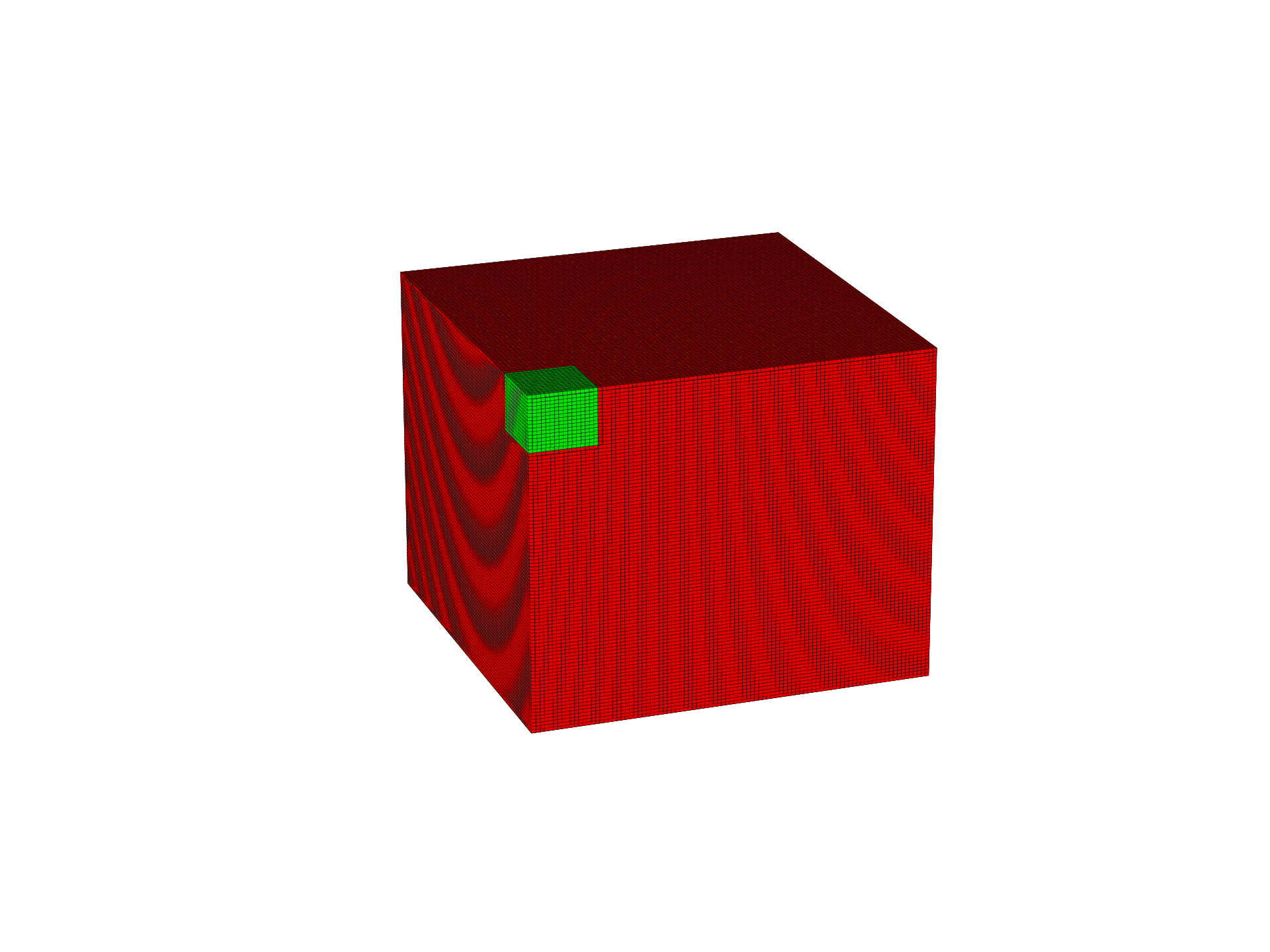}
       \includegraphics[trim= 30mm 25mm 30mm 10mm, clip,  width=0.19\textwidth]{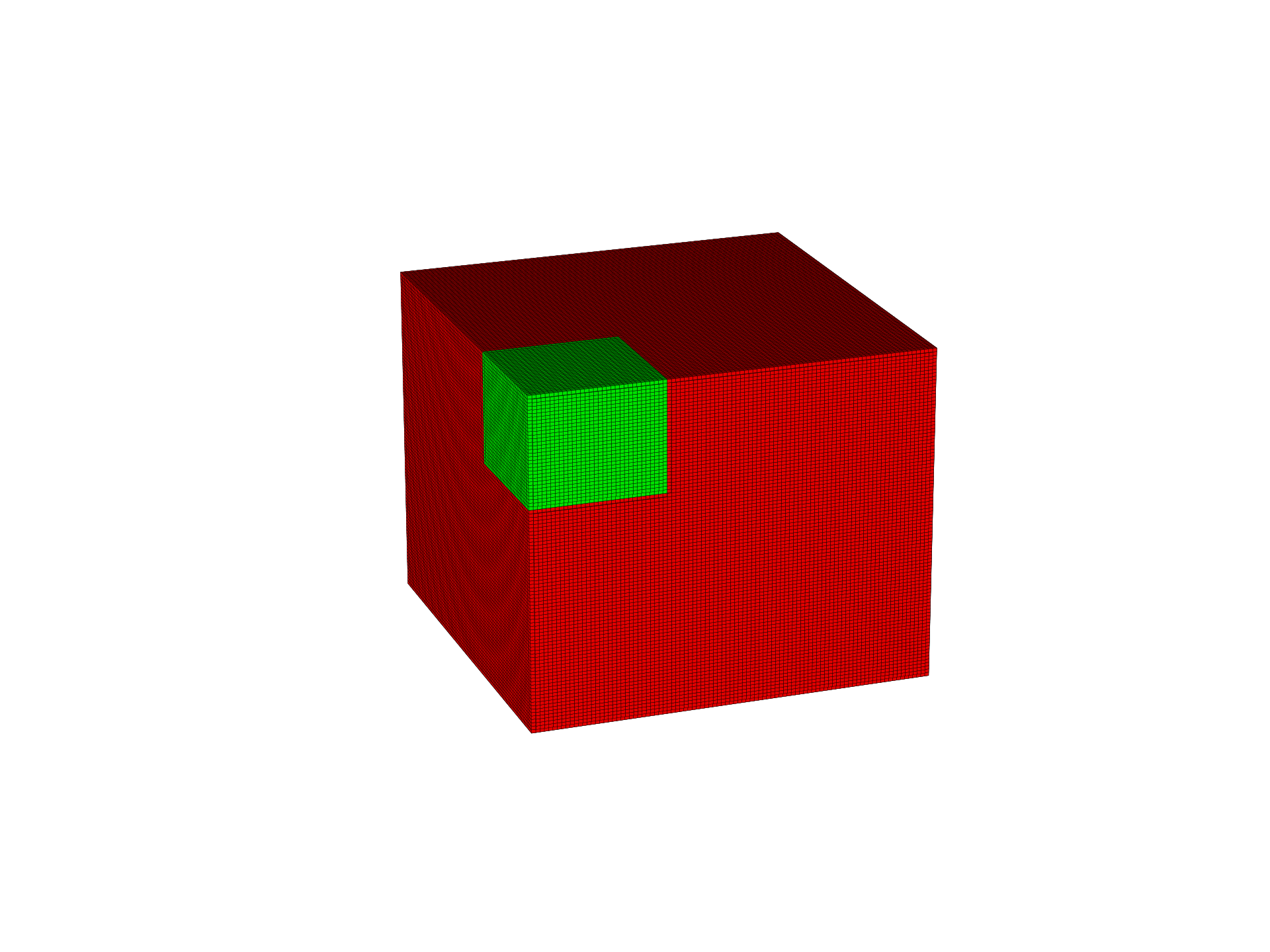}
       \includegraphics[ trim= 30mm 25mm 30mm 10mm, clip,  width=0.19\textwidth]{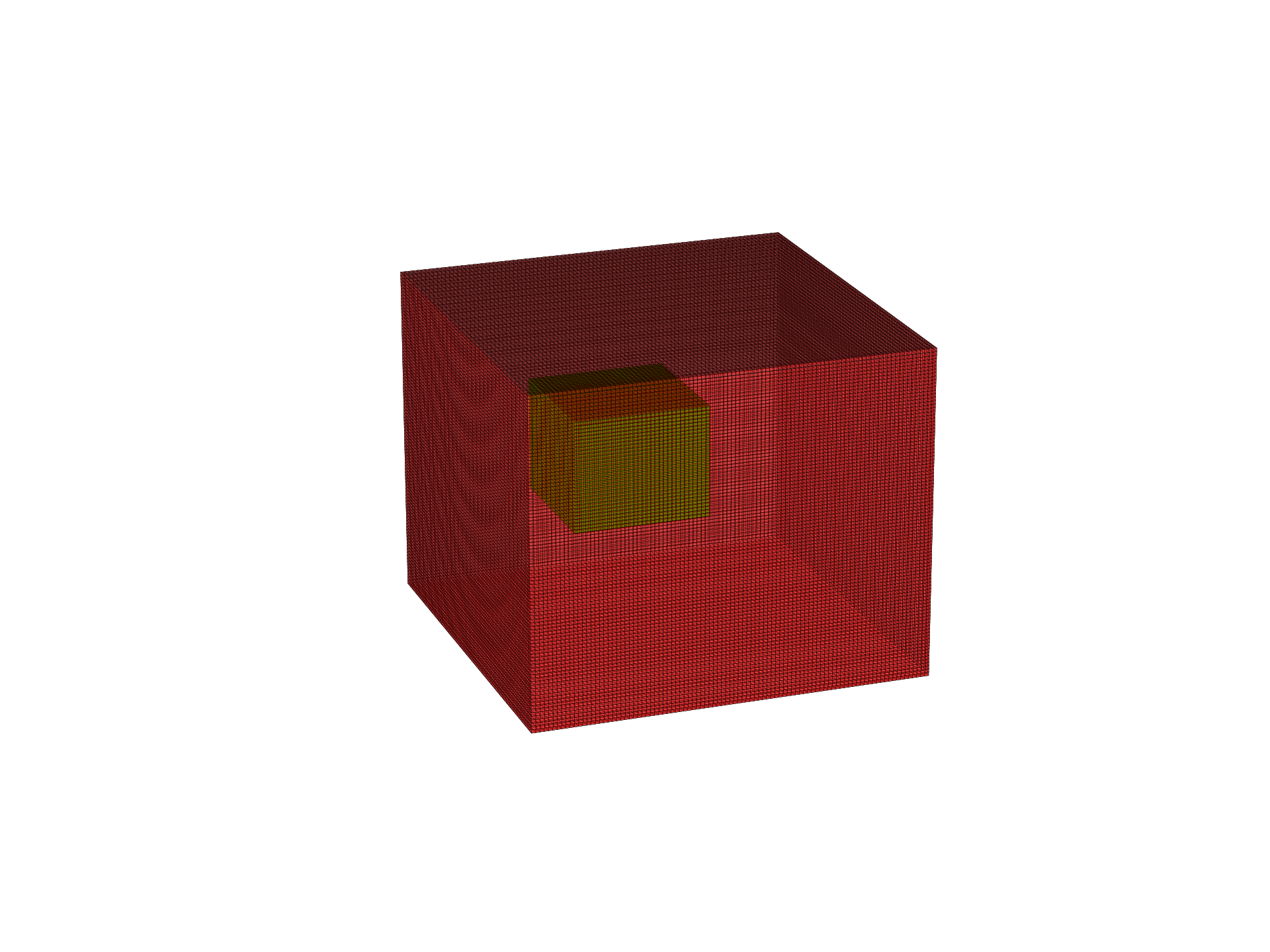}
    \includegraphics[ trim= 30mm 25mm 30mm 10mm, clip,  width=0.19\textwidth]{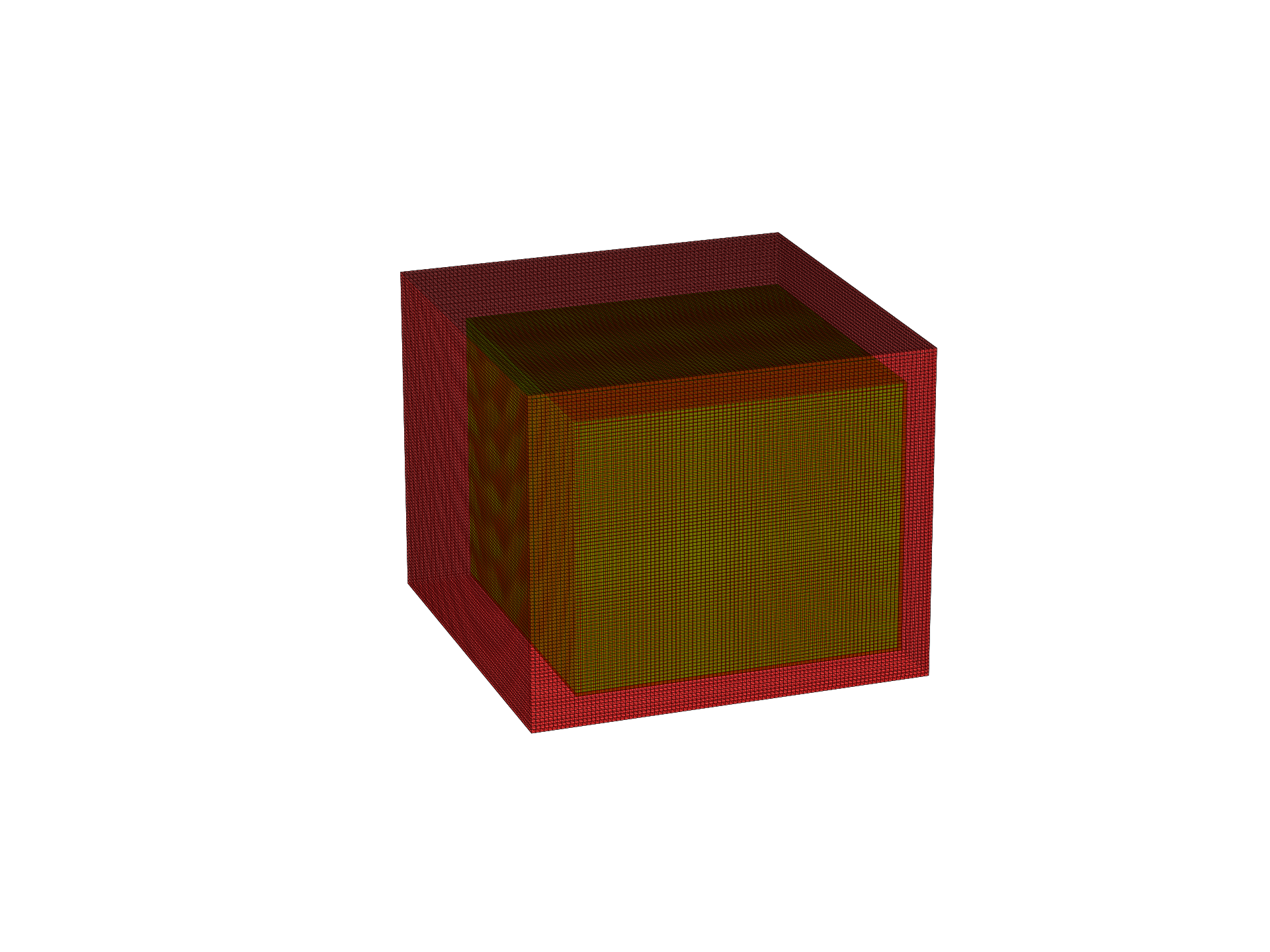}
\end{minipage}
    \caption{\small  \textbf{Frequency Analysis Filters:} The frequencies associated with the red section are eliminated, while those linked to the green section are allowed to pass. These filters are labeled as $(0-8)$, $(0-16)$, $(0-32)$, $(16-48)$, and $(16-96)$ \emph{(from right to left)}.  }
    \label{fig:freq_filter_appendix}
\end{figure}

\begin{figure}[!t]
\begin{minipage}{\linewidth}
    \centering
    \includegraphics[ width=0.24\textwidth]{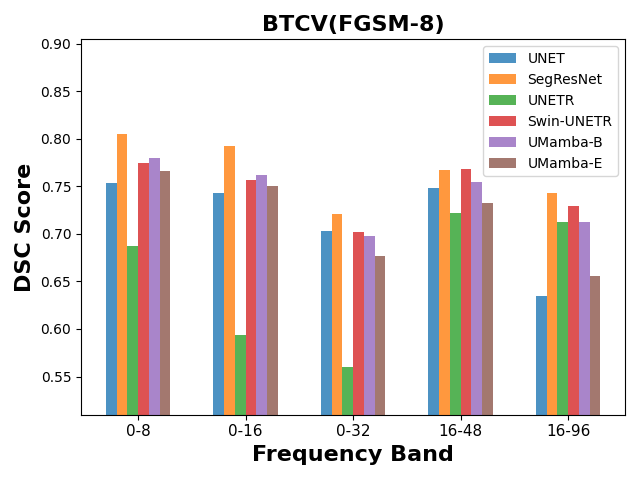}
    \includegraphics[  width=0.24\textwidth]{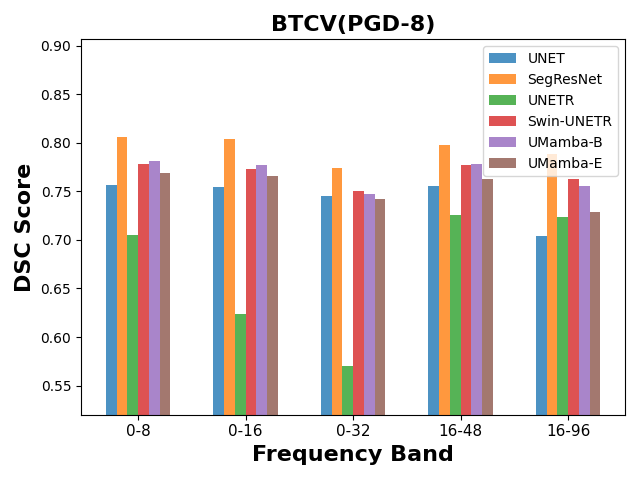}
       \includegraphics[ width=0.24\textwidth]{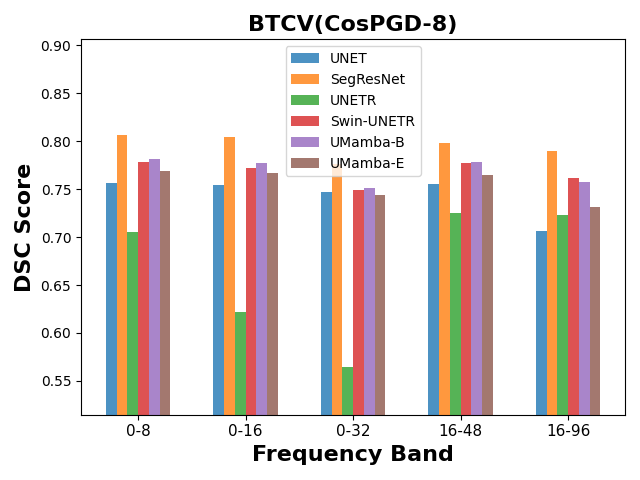}
    \includegraphics[  width=0.24\textwidth]{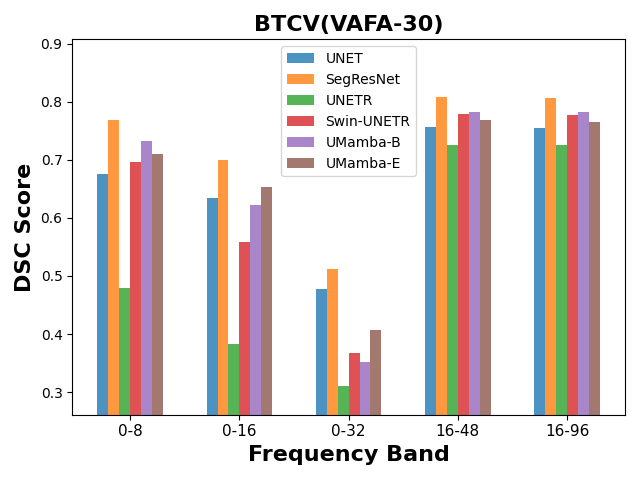}
\end{minipage}
\begin{minipage}{\linewidth}
    \centering
    \includegraphics[ width=0.24\textwidth]{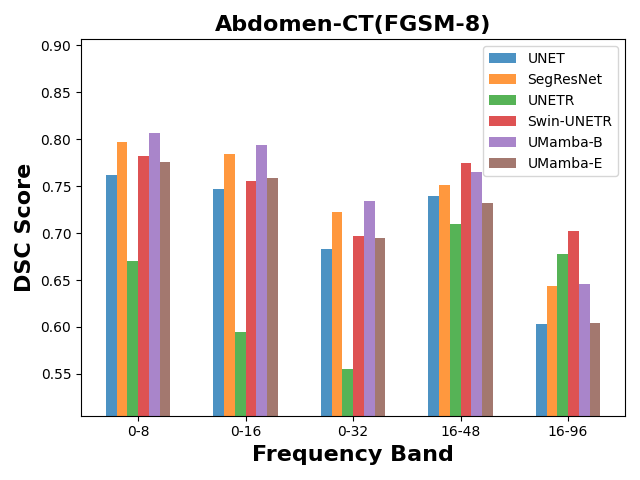}
    \includegraphics[  width=0.24\textwidth]{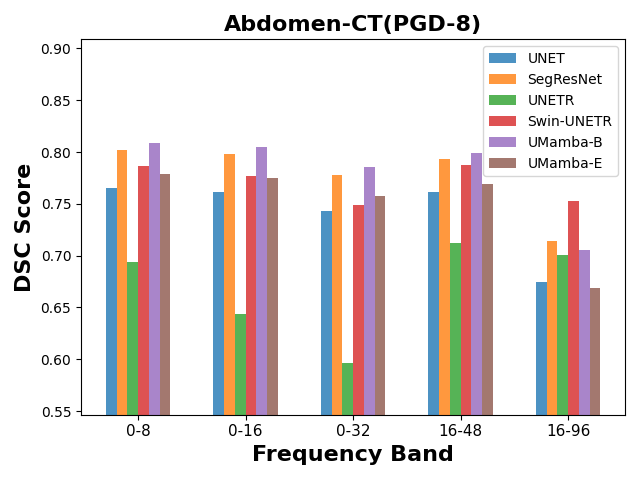}
       \includegraphics[ width=0.24\textwidth]{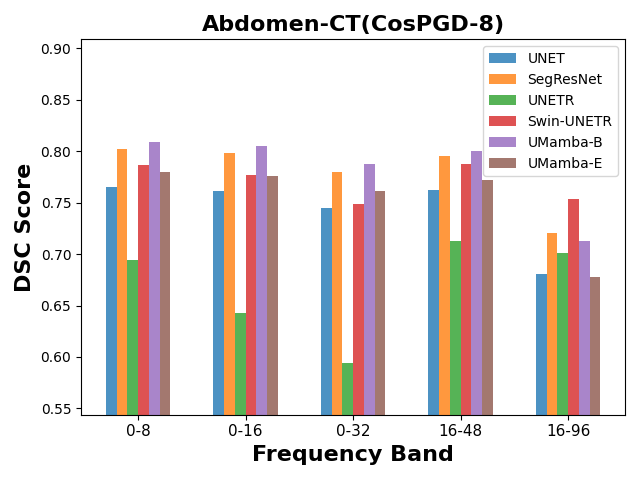}
    \includegraphics[  width=0.24\textwidth]{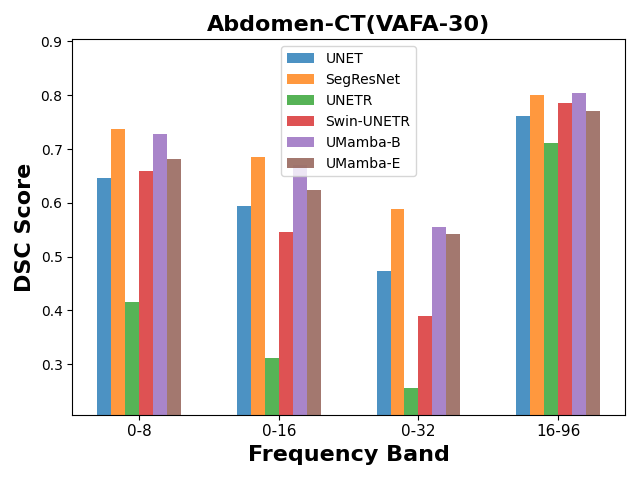}
\end{minipage}
\begin{minipage}{\linewidth}
    \centering
    \includegraphics[ width=0.24\textwidth]{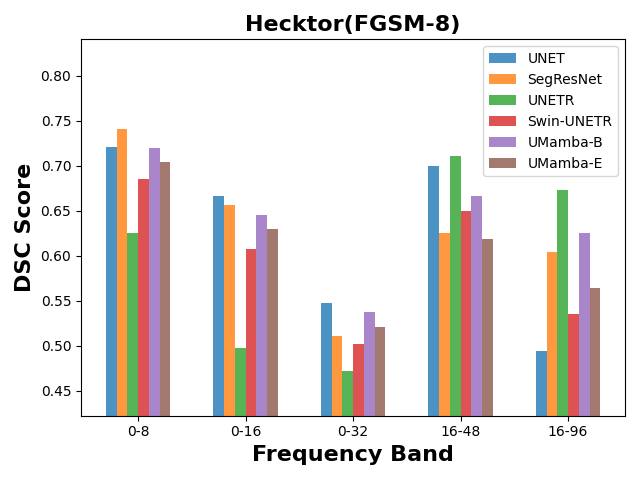}
    \includegraphics[  width=0.24\textwidth]{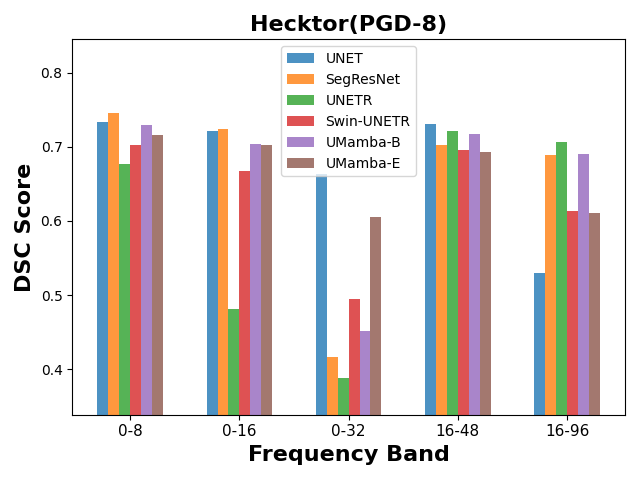}
       \includegraphics[ width=0.24\textwidth]{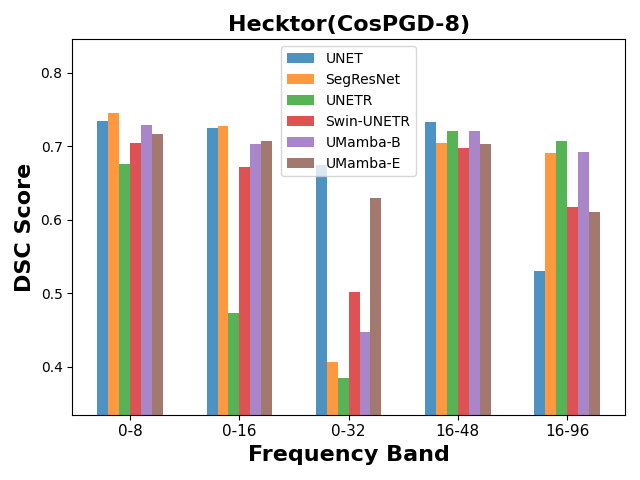}
    \includegraphics[  width=0.24\textwidth]{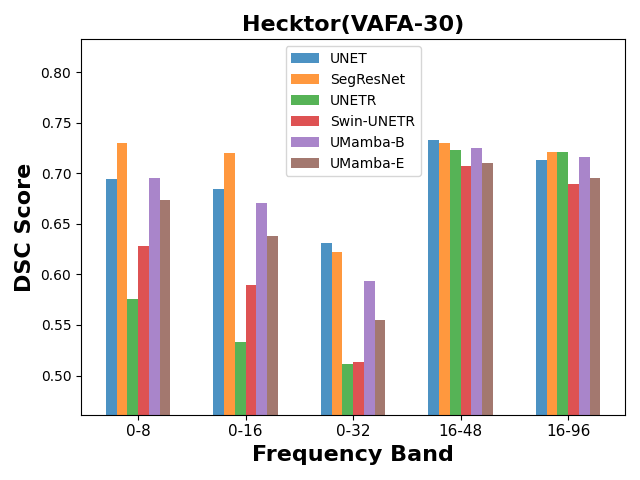}
\end{minipage}
\begin{minipage}{\linewidth}
    \centering
    \includegraphics[ width=0.24\textwidth]{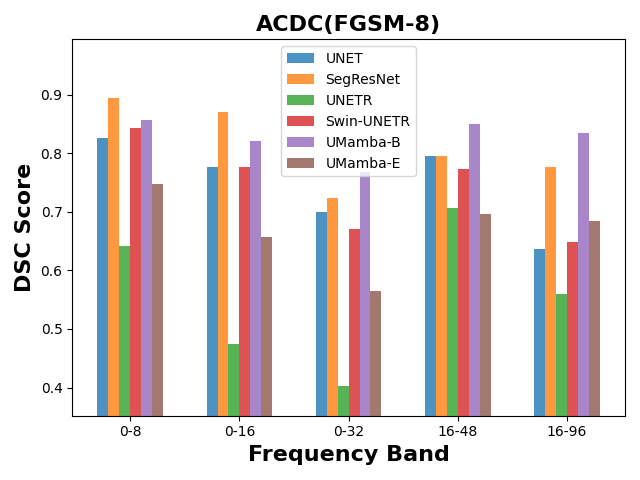}
    \includegraphics[  width=0.24\textwidth]{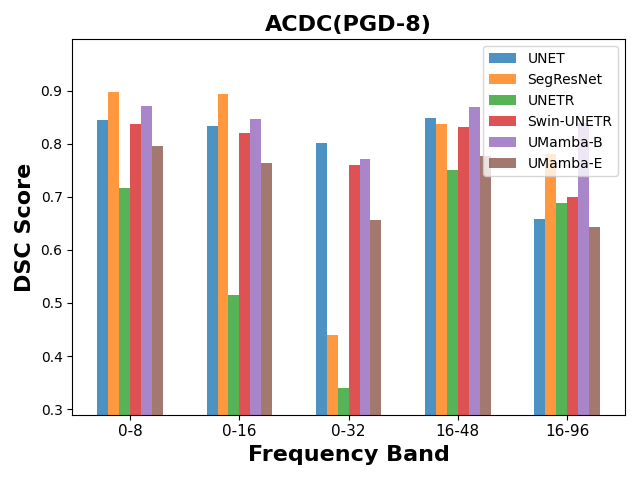}
       \includegraphics[ width=0.24\textwidth]{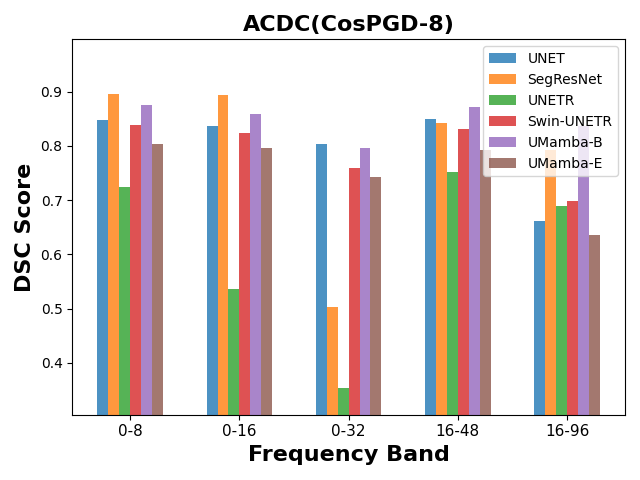}
    \includegraphics[  width=0.24\textwidth]{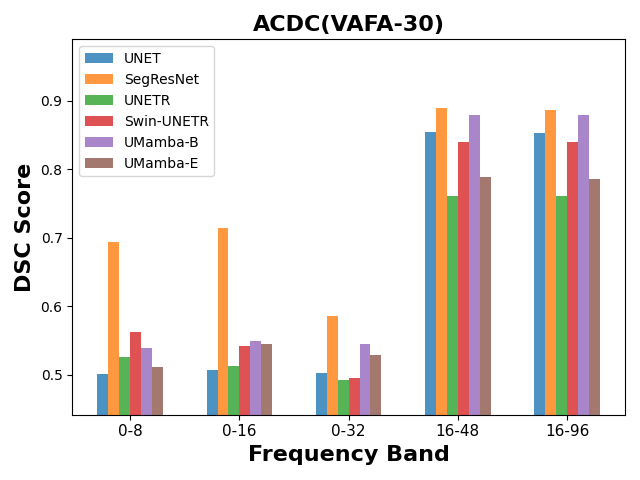}
\end{minipage}

    \caption{\small \textbf{Frequency Analysis on FGSM, PGD, CosPGD, and VAFA}: We report the performance drop on models in \emph{white box} settings across adversarial attacks. We restrict the adversarial perturbations to be added to the image within different frequency ranges.  For pixel-based attacks results are reported for $\epsilon=\frac{8}{255}$ indicated by attack names followed by the suffixes $-8$, respectively. Regarding frequency-based attack \texttt{VAFA}, the results are reported with a constraint on $q_{\text{max}}$ set to  $30$, denoted as \texttt{VAFA-30}. DSC score \emph{(lower is better)} is  reported on the generated adversarial examples.}
    \label{fig:freq_analysis_appendix}
\end{figure}

In this section, we delve deeper into the frequency analysis of adversarial attacks to study which frequency components lead to drop in performance of models. Following \cite{shao2021adversarial}, we implement an adversarial attack incorporating a frequency filter $M$, restricting perturbations to specific frequency domains. The filter operation is defined as {\footnotesize
$x'_{\text{freq}} = \text{IDCT}(\text{DCT}(x'-x) \odot M) + x$
}, where \texttt{DCT} and \texttt{IDCT} denote Discrete Cosine Transform and its inverse, respectively. Similar to \cite{shao2021adversarial}, using the filter $M$, we extract 3D cubes of varying size $n$ from the top left corner as part of the low-frequency components $(0,n)$ where $n \in \{8,16,32\}$. Similarly, mid-frequency $(16-48)$ and high-frequency $(16-96)$ components are also extracted. See Figure \ref{fig:freq_filter_appendix} for the design of filters. While in Figure \ref{fig:freq_analysis} of the main paper, we provide frequency analysis on \texttt{VAFA}, which shows the best transferability across target models. Figure \ref{fig:freq_analysis_appendix} expands this analysis to encompass all the adversarial attacks employed in our experiments.
For pixel-based attacks, we report results at $\epsilon=\frac{8}{255}$ and for \texttt{VAFA} at $q_{max}=30$. In the case of the \texttt{VAFA} attack, which demonstrates significant \emph{transferability} to the target models in \emph{black box} setting, we note that the low-frequency components of the adversarial examples predominantly contribute to the performance decline across surrogate volumetric segmentation models. While a similar trend is observed with pixel-based attacks, it is not as pronounced as with \texttt{VAFA}. For instance, when analyzing the \texttt{Abdomen-CT} and \texttt{ACDC} datasets, we find that the high-frequency components of adversarial examples generated by pixel-based attacks also result in a noticeable performance decrease across models. However, as discussed in Section \ref{sec:bbox}, these adversarial examples produced by pixel-based attacks exhibit very limited transferability.

\section{Robustness against Black-box Attacks }
In Tables \ref{tab: bbox btcv eps4 app}, \ref{tab: bbox abdomen eps4 app}, \ref{tab: bbox hecktor eps4 app}, and \ref{tab: bbox acdc eps4 app}, we report robustness of the volumetric segmentation models on \emph{black box} attacks across \texttt{BTCV}, \texttt{Abdomen-CT}, \texttt{Hecktor}, and \texttt{ACDC} datasets. For pixel-based attacks we craft adversarial examples at $l_{\infty}$ perturbation budget $\epsilon=\frac{4}{255}$ for \texttt{FGSM}, \texttt{PGD}, 
 and \texttt{CosPGD}. For frequency-based attack \texttt{VAFA} we craft adversarial examples with $q_{max}=20$. We report DSC and LPIPS score on generated adversarial examples.
 Similar to our observations for $\epsilon=\frac{8}{255}$ and $q_max=30$ in Section \ref{sec:bbox}, we observe frequency-based attack \texttt{VAFA} results in significant transferability of adversarial examples to target models at $\epsilon=\frac{4}{255}$ and $q_max=20$ as well. Further, we also report the DSC and HD95 score for results reported in Section \ref{sec:bbox} of the main paper in Table  \ref{tab: bbox btcv app},  \ref{tab: bb abdomen app}, \ref{tab: bb hecktor app}, and \ref{tab: bb acdc app}. FInally, in Table \ref{tab:bbox sam}, we report performance of SAM-Med3D on adversarial examples crafted on surrogate models trained on \texttt{BTCV}, \texttt{Abdomen-CT}, \texttt{Hecktor}, and \texttt{ACDC} datasets.
 
We believe our empirical results showing higher transferability obtained from frequency domain attack can be attributed to the ability of frequency domain perturbations to affect a wider range of model architectures and training datasets compared to spatial domain perturbations. One possible reason for the higher transferability obtained from frequency domain adversarial attacks is that frequency domain perturbations can capture more abstract and general features of the input data. These perturbations may affect the underlying patterns and structures that are common across different models and datasets, making them more transferable across various scenarios. Additionally, frequency domain transformations can be less sensitive to small changes in pixel values, leading to more robust and consistent adversarial examples across different scenarios.

\label{appendix: bb_results_eps_4}
\input{tables/black_box_btcv_eps4}
\input{tables/black_box_abdomen_eps4}
\input{tables/black_box_hecktor_eps4}
\input{tables/black_box_acdc_eps4}

\input{tables/black_box_btcv}
\input{tables/black_box_abdomen}
\input{tables/black_box_hecktor}
\input{tables/black_box_acdc}


\input{tables/medsam}



\end{document}

%% file: tables/black_box_btcv_eps4.tex
\begin{table}[!t]
\centering\small
   \scalebox{0.65}[0.65]{
    \begin{tabular}{l|c|cc|cc|cc|cc|cc|cc}
        \toprule
        \rowcolor{LightCyan} 
        Surrogate & Attack  & \multicolumn{2}{c|}{\textbf{UNet}} & \multicolumn{2}{c|}{\textbf{SegResNet}}  & \multicolumn{2}{c|}{\textbf{UNETR}}  & \multicolumn{2}{c|}{\textbf{SwinUNETR}}  &  \multicolumn{2}{c|}{\textbf{UMamba-B}} & \multicolumn{2}{c}{\textbf{UMamba-E}}  \\
                \rowcolor{LightCyan} 
          &  & ~$\mathrm{DSC}\hspace{-0.3em}\downarrow$~ & $\mathrm{HD95}\hspace{-0.3em}\uparrow$~  &  $\mathrm{DSC}\hspace{-0.3em}\downarrow$~ & ~$\mathrm{HD95}\hspace{-0.3em}\uparrow$~ & $\mathrm{DSC}\hspace{-0.3em}\downarrow$~  &  $\mathrm{HD95}\hspace{-0.3em}\uparrow$ & $\mathrm{DSC}\hspace{-0.3em}\downarrow$~  &  $\mathrm{HD95}\hspace{-0.3em}\uparrow$ & $\mathrm{DSC}\hspace{-0.3em}\downarrow$~  &  $\mathrm{HD95}\hspace{-0.3em}\uparrow$ & $\mathrm{DSC}\hspace{-0.3em}\downarrow$~  &  $\mathrm{HD95}\hspace{-0.3em}\uparrow$ \\
          \midrule
\rowcolor{Gray}  & Clean  & 75.72 & 9.15 & 80.84 & 8.14 & 72.53 & 15.08 & 78.07 & 10.01 & 78.37 & 8.12 & 77.06 & 11.11\\
\rowcolor{Gray} & GN  & 75.65 & 9.69 & 80.72 & 8.37 & 72.66 & 14.72 & 77.88 & 9.98 & 78.24 & 8.18 & 76.68 & 16.04\\
\midrule
\multirow{4}{*}{\rotatebox[origin=c]{0}{\parbox[c]{1.5cm}{\centering\texttt{UNet}}}} 
& \texttt{FGSM-4}  & \cellcolor{green!25}58.77 & 37.71\cellcolor{green!25} & 79.6 & 8.36 & 71.72 & 14.72 & 76.7 & 10.39 & 76.85 & 9.4 & 75.18 & 15.44 \\
& \texttt{PGD-4}  & \cellcolor{green!25}50.53 & 53.9\cellcolor{green!25} & 80.42 & 8.51 & 72.42 & 14.68 & 77.57 & 10.21 & 77.7 & 8.69 & 76.22 & 14.28\\
& \texttt{CosPGD-4}  & \cellcolor{green!25}50.97 & 48.99\cellcolor{green!25} & 80.4 & 8.49 & 72.37 & 14.78 & 77.57 & 10.18 & 77.66 & 8.57 & 76.26 & 14.53
\\
& \texttt{VAFA-20}  & \cellcolor{green!25}56.13 & 39.73\cellcolor{green!25} & 69.69 & 18.68 & 53.19 & 31.66 & 66.8 & 28.34 & 72.53 & 14.38 & 69.63 & 15.5
\\

    \midrule
    \multirow{4}{*}{\rotatebox[origin=c]{0}{\parbox[c]{1.5cm}{\centering\texttt{SegResNet}}}} & \texttt{FGSM-4}  & 74.0 & 10.75 & \cellcolor{green!25}66.68 & \cellcolor{green!25}32.41 & 71.45 & 15.21 & 75.27 & 10.69 & 73.33 & 11.1 & 72.95 & 15.6\\
& \texttt{PGD-4}  &74.57 & 13.03 & \cellcolor{green!25}49.49 & \cellcolor{green!25}65.31 & 71.83 & 15.25 & 76.67 & 10.45 & 74.96 & 11.63 & 73.97 & 16.1
\\
& \texttt{CosPGD-4}  & 74.51 & 13.02 & \cellcolor{green!25}49.89 & \cellcolor{green!25}60.42 & 71.78 & 15.45 & 76.64 & 10.47 & 74.99 & 11.41 & 73.85 & 16.77

\\
& \texttt{VAFA-20}  & 61.59 & 20.73 & \cellcolor{green!25}51.44 & \cellcolor{green!25}37.74 & 43.33 & 41.75 & 58.62 & 28.87 & 68.92 & 13.88 & 66.03 & 17.45
\\
    \midrule
    \multirow{4}{*}{\rotatebox[origin=c]{0}{\parbox[c]{1.5cm}{\centering\texttt{UNETR}}}} & \texttt{FGSM-4}  & 73.68 & 9.95 & 79.08 & 8.72 & \cellcolor{green!25}60.52 & \cellcolor{green!25}32.3 & 75.12 & 10.79 & 76.02 & 8.82 & 74.91 & 12.88
\\
& \texttt{PGD-4}  & 74.34 & 11.04 & 79.8 & 8.54 & \cellcolor{green!25}59.27 & \cellcolor{green!25}36.6 & 76.27 & 10.48 & 76.83 & 9.08 & 75.47 & 11.66
\\
& \texttt{CosPGD-4}  & 74.29 & 10.62 & 79.79 & 8.43 & \cellcolor{green!25}59.17 & \cellcolor{green!25}38.39 & 76.29 & 10.48 & 76.83 & 9.05 & 75.35 & 12.85
\\
& \texttt{VAFA-20}  & 55.31 & 26.62 & 57.5 & 23.07 & \cellcolor{green!25}35.09 & \cellcolor{green!25}45.42 & 50.44 & 30.85 & 63.53 & 17.54 & 59.1 & 21.81
\\
    \midrule
    \multirow{4}{*}{\rotatebox[origin=c]{0}{\parbox[c]{1.5cm}{\centering\texttt{SwinUNETR}}}} & \texttt{FGSM-4}  & 73.91 & 11.86 & 78.36 & 9.34 & 70.89 & 15.55 & \cellcolor{green!25}66.8 & \cellcolor{green!25}24.71 & 75.41 & 9.41 & 74.37 & 13.14
\\
& \texttt{PGD-4}  & 74.22 & 12.24 & 79.47 & 9.24 & 71.45 & 15.96 & \cellcolor{green!25}66.37 & \cellcolor{green!25}32.63 & 76.36 & 10.24 & 75.21 & 16.84
\\
& \texttt{CosPGD-4}  & 74.24 & 12.22 & 79.4 & 9.24 & 71.43 & 15.96 & \cellcolor{green!25}65.82 & \cellcolor{green!25}32.86 & 76.37 & 10.22 & 75.09 & 19.19
\\
& \texttt{VAFA-20}  & 59.36 & 24.77 & 61.67 & 26.59 & 47.41 & 40.06 & \cellcolor{green!25}54.22 & \cellcolor{green!25}45.64 & 65.13 & 17.85 & 62.44 & 21.08
\\
\midrule
    \multirow{4}{*}{\rotatebox[origin=c]{0}{\parbox[c]{1.5cm}{\centering\texttt{UMamba-B}}}} & \texttt{FGSM-4}  & 73.81 & 13.17 & 76.57 & 11.88 & 71.52 & 14.8 & 75.39 & 10.84 & \cellcolor{green!25}65.17 & \cellcolor{green!25}24.69 & 71.63 & 20.33
\\
& \texttt{PGD-4}  &74.69 & 11.6 & 78.78 & 10.44 & 71.92 & 15.03 & 76.88 & 10.58 & \cellcolor{green!25}56.03 & \cellcolor{green!25}47.41 & 73.43 & 19.04
 \\
& \texttt{CosPGD-4}  & 74.56 & 11.52 & 78.7 & 11.24 & 71.88 & 15.13 & 76.85 & 10.54 & \cellcolor{green!25}56.53 & \cellcolor{green!25}45.15 & 73.18 & 19.13
\\
& \texttt{VAFA-20}  & 73.3 & 14.16 & 76.82 & 10.92 & 67.21 & 16.77 & 74.83 & 14.62 & \cellcolor{green!25}68.22 & \cellcolor{green!25}19.06 & 73.58 & 15.0
\\
\midrule
    \multirow{4}{*}{\rotatebox[origin=c]{0}{\parbox[c]{1.5cm}{\centering\texttt{UMamba-E}}}} & \texttt{FGSM-4}  & 74.34 & 10.36 & 78.49 & 10.11 & 71.69 & 15.08 & 76.37 & 10.56 & 75.04 & 11.29 & \cellcolor{green!25}59.47 & \cellcolor{green!25}36.77
\\
& \texttt{PGD-4}  & 75.2 & 10.54 & 80.34 & 8.6 & 72.31 & 14.89 & 77.71 & 10.29 & 77.41 & 10.21 & \cellcolor{green!25}51.3 & \cellcolor{green!25}57.83
\\
& \texttt{CosPGD-4}  & 75.23 & 12.35 & 80.28 & 9.36 & 72.31 & 14.74 & 77.62 & 10.26 & 77.33 & 10.1 & \cellcolor{green!25}51.67 & \cellcolor{green!25}54.16
\\
& \texttt{VAFA-20}  & 73.74 & 13.94 & 78.67 & 10.6 & 68.15 & 13.88 & 75.36 & 16.89 & 77.09 & 8.56 & \cellcolor{green!25}62.15 & \cellcolor{green!25}38.97
\\
    \midrule
    \end{tabular}
    }
    \vspace{0.3em}
    \caption{\small Performance of models against transfer-based \emph{black box} attacks on  \texttt{BTCV} dataset. For pixel-based attacks results are reported for $\epsilon=\frac{4}{255}$ indicated by attack names followed by the suffixes $-4$, respectively. Regarding frequency-based attack \texttt{VAFA}, the results are reported with a constraint on $q_{\text{max}}$ set to  $20$, denoted as \texttt{VAFA-20}. DSC score \emph{(lower is better)} is  reported on the generated adversarial examples. }
    \label{tab: bbox btcv eps4 app}
\end{table}

%% file: tables/black_box_abdomen_eps4.tex
\begin{table}[!t]
\centering\small
   \scalebox{0.65}[0.65]{
    \begin{tabular}{l|c|cc|cc|cc|cc|cc|cc}
        \toprule
        \rowcolor{LightCyan} 
        Surrogate & Attack  & \multicolumn{2}{c|}{\textbf{UNet}} & \multicolumn{2}{c|}{\textbf{SegResNet}}  & \multicolumn{2}{c|}{\textbf{UNETR}}  & \multicolumn{2}{c|}{\textbf{SwinUNETR}}  &  \multicolumn{2}{c|}{\textbf{UMamba-B}} & \multicolumn{2}{c}{\textbf{UMamba-E}}  \\
                \rowcolor{LightCyan} 
          &  & ~$\mathrm{DSC}\hspace{-0.3em}\downarrow$~ & $\mathrm{HD95}\hspace{-0.3em}\uparrow$~  &  $\mathrm{DSC}\hspace{-0.3em}\downarrow$~ & ~$\mathrm{HD95}\hspace{-0.3em}\uparrow$~ & $\mathrm{DSC}\hspace{-0.3em}\downarrow$~  &  $\mathrm{HD95}\hspace{-0.3em}\uparrow$ & $\mathrm{DSC}\hspace{-0.3em}\downarrow$~  &  $\mathrm{HD95}\hspace{-0.3em}\uparrow$ & $\mathrm{DSC}\hspace{-0.3em}\downarrow$~  &  $\mathrm{HD95}\hspace{-0.3em}\uparrow$ & $\mathrm{DSC}\hspace{-0.3em}\downarrow$~  &  $\mathrm{HD95}\hspace{-0.3em}\uparrow$ \\
          \midrule
\rowcolor{Gray}  & Clean  & 76.79 & 19.72 & 80.89 & 13.30 & 71.35 & 27.73 & 79.33 & 25.79 & 81.08 & 15.51 & 78.05 & 18.31\\
\rowcolor{Gray} & GN  & - & - & - & - & - & - & - & - & - & - & - & -\\
\midrule
\multirow{4}{*}{\rotatebox[origin=c]{0}{\parbox[c]{1.5cm}{\centering\texttt{UNet}}}} 
& \texttt{FGSM-4}  & \cellcolor{green!25}55.49 & \cellcolor{green!25}49.07 & 77.63 & 14.51 & 70.19 & 31.9 & 77.31 & 28.89 & 77.51 & 18.02 & 74.69 & 18.88
\\
& \texttt{PGD-4}  & \cellcolor{green!25}45.11 & \cellcolor{green!25}60.73 & 78.86 & 14.2 & 70.78 & 32.96 & 78.49 & 27.38 & 79.26 & 18.67 & 76.37 & 17.3
\\
& \texttt{CosPGD-4}  & \cellcolor{green!25}46.15 & \cellcolor{green!25}56.43 & 78.92 & 13.65 & 70.78 & 32.49 & 78.48 & 27.06 & 79.38 & 16.8 & 76.49 & 17.09
\\
& \texttt{VAFA-20}  &\cellcolor{green!25}20.98 & \cellcolor{green!25}79.82 & 25.59 & 64.99 & 39.1 & 65.7 & 38.01 & 60.38 & 22.87 & 69.81 & 17.97 & 72.59
\\

    \midrule
    \multirow{4}{*}{\rotatebox[origin=c]{0}{\parbox[c]{1.5cm}{\centering\texttt{SegResNet}}}} & \texttt{FGSM-4}  & 73.99 & 21.52 & \cellcolor{green!25}58.49 & \cellcolor{green!25}40.26 & 70.12 & 32.88 & 76.92 & 26.18 & 73.01 & 22.27 & 71.63 & 19.58
\\
& \texttt{PGD-4}  & 75.32 & 21.71 & \cellcolor{green!25}38.38 & \cellcolor{green!25}76.59 & 70.74 & 32.11 & 78.1 & 27.63 & 76.2 & 18.0 & 74.31 & 19.26
\\
& \texttt{CosPGD-4}  & 75.28 & 21.9 & \cellcolor{green!25}39.99 & \cellcolor{green!25}71.41 & 70.73 & 31.93 & 78.09 & 28.11 & 76.09 & 20.27 & 74.31 & 18.12
\\
& \texttt{VAFA-20}  & 37.16 & 48.6 & \cellcolor{green!25}\cellcolor{green!25}17.07 & \cellcolor{green!25}86.27 & 45.66 & 48.14 & 43.38 & 50.99 & 22.17 & 66.08 & 17.25 & 72.17
\\
    \midrule
    \multirow{4}{*}{\rotatebox[origin=c]{0}{\parbox[c]{1.5cm}{\centering\texttt{UNETR}}}} & \texttt{FGSM-4}  & 74.07 & 21.03 & 77.61 & 14.12 & \cellcolor{green!25}57.17 & \cellcolor{green!25}47.62 & 75.8 & 27.04 & 77.75 & 19.74 & 74.62 & 18.03
\\
& \texttt{PGD-4}  & 74.88 & 21.39 & 78.53 & 14.16 & \cellcolor{green!25}56.09 & \cellcolor{green!25}54.17 & 77.06 & 27.34 & 78.7 & 18.36 & 75.46 & 17.04
\\
& \texttt{CosPGD-4}  & 74.9 & 21.2 & 78.62 & 14.25 & \cellcolor{green!25}56.47 & \cellcolor{green!25}51.35 & 77.14 & 26.5 & 78.76 & 17.46 & 75.59 & 16.86
\\
& \texttt{VAFA-20}  & 41.79 & 51.35 & 33.3 & 54.93 & \cellcolor{green!25}25.07 & \cellcolor{green!25}81.29 & 39.73 & 54.54 & 29.44 & 55.98 & 27.03 & 55.56
\\
    \midrule
    \multirow{4}{*}{\rotatebox[origin=c]{0}{\parbox[c]{1.5cm}{\centering\texttt{SwinUNETR}}}} & \texttt{FGSM-4}  & 73.67 & 21.14 & 76.4 & 15.09 & 69.15 & 32.35 & \cellcolor{green!25}62.36 & \cellcolor{green!25}53.75 & 76.46 & 18.36 & 73.79 & 19.52
\\
& \texttt{PGD-4}  & 74.7 & 22.26 & 77.75 & 14.97 & 69.93 & 35.14 & \cellcolor{green!25}61.12 & \cellcolor{green!25}62.08 & 78.21 & 17.08 & 75.29 & 18.76
\\
& \texttt{CosPGD-4}  & 74.74 & 21.24 & 77.81 & 15.17 & 69.95 & 33.62 & \cellcolor{green!25}61.11 & \cellcolor{green!25}59.67 & 78.14 & 16.96 & 75.3 & 18.39
\\
& \texttt{VAFA-20}  & 31.55 & 58.31 & 27.55 & 59.83 & 35.04 & 61.92 & \cellcolor{green!25}27.41 & \cellcolor{green!25}72.71 & 24.01 & 65.52 & 21.14 & 68.78
\\
\midrule
    \multirow{4}{*}{\rotatebox[origin=c]{0}{\parbox[c]{1.5cm}{\centering\texttt{UMamba-B}}}} & \texttt{FGSM-4}  & 73.88 & 23.47 & 73.57 & 17.24 & 70.06 & 33.54 & 76.92 & 27.25 & \cellcolor{green!25}59.06 & \cellcolor{green!25}45.82 & 70.01 & 25.71
\\
& \texttt{PGD-4}  & 75.23 & 21.66 & 76.18 & 16.31 & 70.71 & 32.94 & 78.15 & 27.58 & \cellcolor{green!25}41.0 & \cellcolor{green!25}72.28 & 72.81 & 21.73
\\
& \texttt{CosPGD-4}  & 75.14 & 22.07 & 76.19 & 16.21 & 70.71 & 32.76 & 78.13 & 26.97 & \cellcolor{green!25}41.31 & \cellcolor{green!25}73.11 & 72.87 & 21.47
\\
& \texttt{VAFA-20}  & 37.29 & 55.72 & 22.86 & 65.32 & 45.02 & 57.95 & 43.66 & 52.85 & \cellcolor{green!25}14.45 & \cellcolor{green!25}91.51 & 17.12 & 76.75
\\
\midrule
    \multirow{4}{*}{\rotatebox[origin=c]{0}{\parbox[c]{1.5cm}{\centering\texttt{UMamba-E}}}} & \texttt{FGSM-4}  & 74.1 & 21.44 & 75.43 & 15.89 & 70.08 & 33.24 & 77.17 & 27.36 & 73.23 & 21.73 & \cellcolor{green!25}55.75 & \cellcolor{green!25}46.17
 \\
& \texttt{PGD-4}  & 75.58 & 21.91 & 77.85 & 15.16 & 70.86 & 32.12 & 78.45 & 27.14 & 76.53 & 18.44 & \cellcolor{green!25}45.08 & \cellcolor{green!25}64.08
\\
& \texttt{CosPGD-4}  & 75.54 & 21.46 & 77.77 & 15.09 & 70.82 & 32.04 & 78.44 & 27.76 & 76.72 & 18.25 & \cellcolor{green!25}44.7 & \cellcolor{green!25}59.92
 \\
& \texttt{VAFA-20}  & 38.54 & 54.64 & 25.12 & 64.86 & 45.39 & 60.19 & 45.1 & 53.93 & 22.68 & 72.02 & \cellcolor{green!25}12.76 & \cellcolor{green!25}93.07
\\
    \midrule 
    \end{tabular}
    }
    \vspace{0.3em}
    \caption{\small Performance of models against transfer-based \emph{black box} attacks on  \texttt{Abdomen-CT} dataset.  For pixel-based attacks results are reported for $\epsilon=\frac{4}{255}$ indicated by attack names followed by the suffixes $-4$, respectively. Regarding frequency-based attack \texttt{VAFA}, the results are reported with a constraint on $q_{\text{max}}$ set to  $20$, denoted as \texttt{VAFA-20}. DSC score \emph{(lower is better)} is  reported on the generated adversarial examples. }
    \label{tab: bbox abdomen eps4 app}
\end{table}

%% file: tables/black_box_hecktor_eps4.tex
\begin{table}[!t]
\centering\small
   \scalebox{0.65}[0.65]{
    \begin{tabular}{l|c|cc|cc|cc|cc|cc|cc}
        \toprule
        \rowcolor{LightCyan} 
        Surrogate & Attack  & \multicolumn{2}{c|}{\textbf{UNet}} & \multicolumn{2}{c|}{\textbf{SegResNet}}  & \multicolumn{2}{c|}{\textbf{UNETR}}  & \multicolumn{2}{c|}{\textbf{SwinUNETR}}  &  \multicolumn{2}{c|}{\textbf{UMamba-B}} & \multicolumn{2}{c}{\textbf{UMamba-E}}  \\
                \rowcolor{LightCyan} 
          &  & ~$\mathrm{DSC}\hspace{-0.3em}\downarrow$~ & $\mathrm{HD95}\hspace{-0.3em}\uparrow$~  &  $\mathrm{DSC}\hspace{-0.3em}\downarrow$~ & ~$\mathrm{HD95}\hspace{-0.3em}\uparrow$~ & $\mathrm{DSC}\hspace{-0.3em}\downarrow$~  &  $\mathrm{HD95}\hspace{-0.3em}\uparrow$ & $\mathrm{DSC}\hspace{-0.3em}\downarrow$~  &  $\mathrm{HD95}\hspace{-0.3em}\uparrow$ & $\mathrm{DSC}\hspace{-0.3em}\downarrow$~  &  $\mathrm{HD95}\hspace{-0.3em}\uparrow$ & $\mathrm{DSC}\hspace{-0.3em}\downarrow$~  &  $\mathrm{HD95}\hspace{-0.3em}\uparrow$ \\
          \midrule
\rowcolor{Gray}  & Clean  & 73.91 & 11.36 & 74.73 & 11.08 & 72.36 & 14.61 &71.61 & 22.07 & 73.50 & 10.89 & 72.19 & 13.29\\
\rowcolor{Gray} & GN  & - & - & - & - & - & - & - & - & - & - & - & -\\
\midrule
\multirow{4}{*}{\rotatebox[origin=c]{0}{\parbox[c]{1.5cm}{\centering\texttt{UNet}}}} 
& \texttt{FGSM-4}  & \cellcolor{green!25}47.09 & \cellcolor{green!25}34.93 & 66.43 & 14.93 & 69.56 & 15.36 & 64.99 & 27.04 & 66.4 & 13.95 & 64.54 & 15.78
\\
& \texttt{PGD-4}  & \cellcolor{green!25}35.45 & \cellcolor{green!25}74.93 & 70.95 & 12.63 & 71.06 & 15.5 & 68.1 & 25.57 & 69.56 & 12.38 & 68.93 & 14.33
 \\
& \texttt{CosPGD-4}  & \cellcolor{green!25}35.35 & \cellcolor{green!25}75.02 & 70.94 & 12.44 & 71.12 & 15.48 & 68.26 & 24.87 & 69.76 & 12.26 & 69.11 & 15.95
\\
& \texttt{VAFA-20}  & \cellcolor{green!25}47.18 & \cellcolor{green!25}37.67 & 71.81 & 13.39 & 65.29 & 18.38 & 66.4 & 19.96 & 68.94 & 14.23 & 62.6 & 19.23
\\

    \midrule
    \multirow{4}{*}{\rotatebox[origin=c]{0}{\parbox[c]{1.5cm}{\centering\texttt{SegResNet}}}} & \texttt{FGSM-4}  & 65.95 & 19.65 & \cellcolor{green!25}51.08 & \cellcolor{green!25}29.89 & 70.11 & 16.15 & 63.28 & 29.1 & 60.19 & 19.86 & 61.2 & 21.39
\\
& \texttt{PGD-4}  & 67.57 & 16.92 & \cellcolor{green!25}34.53 & \cellcolor{green!25}76.44 & 70.65 & 14.89 & 64.93 & 27.82 & 60.05 & 25.77 & 62.3 & 20.78
\\
& \texttt{CosPGD-4}  & 68.18 & 16.59 & \cellcolor{green!25}34.44 & \cellcolor{green!25}76.67 & 70.78 & 14.84 & 65.21 & 27.15 & 60.25 & 24.63 & 62.28 & 18.18
 \\
& \texttt{VAFA-20}  & 70.06 & 17.43 & \cellcolor{green!25}53.74 & \cellcolor{green!25}26.07 & 66.36 & 17.73 & 65.13 & 23.35 & 67.99 & 14.94 & 63.06 & 21.2
\\
    \midrule
    \multirow{4}{*}{\rotatebox[origin=c]{0}{\parbox[c]{1.5cm}{\centering\texttt{UNETR}}}} & \texttt{FGSM-4}  & 65.86 & 16.76 & 67.22 & 13.6 & \cellcolor{green!25}49.69 & \cellcolor{green!25}28.51 & 62.87 & 28.55 & 66.27 & 13.45 & 64.41 & 15.66
\\
& \texttt{PGD-4}  & 67.24 & 18.42 & 68.89 & 14.18 & \cellcolor{green!25}39.98 & \cellcolor{green!25}57.26 & 64.09 & 28.24 & 68.12 & 14.02 & 66.81 & 14.71
\\
& \texttt{CosPGD-4}  & 67.62 & 18.51 & 68.9 & 14.62 & \cellcolor{green!25}39.97 & \cellcolor{green!25}58.22 & 64.41 & 27.93 & 68.09 & 14.1 & 66.77 & 15.9\\
& \texttt{VAFA-20}  & 66.46 & 16.01 & 71.08 & 13.66 & \cellcolor{green!25}50.4 & \cellcolor{green!25}30.36 & 58.58 & 26.82 & 66.81 & 15.32 & 59.47 & 22.95
\\
    \midrule
    \multirow{4}{*}{\rotatebox[origin=c]{0}{\parbox[c]{1.5cm}{\centering\texttt{SwinUNETR}}}} & \texttt{FGSM-4}  & 64.77 & 17.07 & 62.78 & 15.66 & 68.79 & 16.25 & \cellcolor{green!25}47.62 & \cellcolor{green!25}50.52 & 61.94 & 17.02 & 61.6 & 19.59
\\
& \texttt{PGD-4}  & 64.77 & 20.54 & 63.71 & 17.34 & 69.26 & 17.86 & \cellcolor{green!25}35.71 & \cellcolor{green!25}71.66 & 63.03 & 19.62 & 62.32 & 19.43
\\
& \texttt{CosPGD-4}  & 64.89 & 20.28 & 64.0 & 17.41 & 69.3 & 17.63 & \cellcolor{green!25}35.67 & \cellcolor{green!25}71.17 & 63.62 & 18.24 & 62.63 & 19.32
\\
& \texttt{VAFA-20}  & 67.49 & 19.81 & 70.16 & 14.05 & 63.04 & 20.19 & \cellcolor{green!25}45.55 & \cellcolor{green!25}39.63 & 68.31 & 16.17 & 58.99 & 21.62
\\
\midrule
    \multirow{4}{*}{\rotatebox[origin=c]{0}{\parbox[c]{1.5cm}{\centering\texttt{UMamba-B}}}} & \texttt{FGSM-4}  & 65.9 & 15.64 & 60.75 & 17.1 & 70.0 & 15.18 & 62.54 & 29.27 & \cellcolor{green!25}54.27 & \cellcolor{green!25}25.87 & 60.26 & 19.26
\\
& \texttt{PGD-4}  & 68.38 & 15.01 & 57.51 & 27.59 & 71.01 & 15.12 & 64.41 & 28.32 & \cellcolor{green!25}36.54 & \cellcolor{green!25}73.08 & 59.32 & 23.49
\\
& \texttt{CosPGD-4}  & 68.57 & 15.83 & 57.11 & 29.09 & 71.06 & 15.1 & 64.85 & 28.94 & \cellcolor{green!25}36.2 & \cellcolor{green!25}72.71 & 60.1 & 24.75
\\
& \texttt{VAFA-20}  & 67.38 & 19.59 & 68.45 & 14.79 & 65.08 & 17.12 & 65.41 & 22.1 & \cellcolor{green!25}53.55 & \cellcolor{green!25}31.11 & 60.21 & 21.25
\\
\midrule
    \multirow{4}{*}{\rotatebox[origin=c]{0}{\parbox[c]{1.5cm}{\centering\texttt{UMamba-E}}}} & \texttt{FGSM-4}  & 67.51 & 15.97 & 64.71 & 14.8 & 70.57 & 15.6 & 65.7 & 27.37 & 63.32 & 17.41 & \cellcolor{green!25}50.06 & \cellcolor{green!25}31.45
 \\
& \texttt{PGD-4}  & 71.99 & 12.47 & 71.35 & 12.6 & 71.57 & 14.44 & 69.5 & 23.35 & 70.11 & 13.39 & \cellcolor{green!25}35.94 & \cellcolor{green!25}74.19
\\
& \texttt{CosPGD-4}  & 71.84 & 13.13 & 71.61 & 12.14 & 71.58 & 14.44 & 69.47 & 23.63 & 70.48 & 14.36 & \cellcolor{green!25}36.07 & \cellcolor{green!25}73.97
\\
& \texttt{VAFA-20}  & 68.52 & 19.79 & 71.56 & 13.43 & 64.93 & 18.85 & 66.36 & 22.05 & 68.19 & 21.2 & \cellcolor{green!25}45.28 & \cellcolor{green!25}31.42
\\
    \midrule
    \end{tabular}
    }
    \vspace{0.3em}
    \caption{\small Performance of models against transfer-based \emph{black box} attacks on  \texttt{Hecktor} dataset. For pixel-based attacks results are reported for $\epsilon=\frac{4}{255}$ indicated by attack names followed by the suffixes $-4$, respectively. Regarding frequency-based attack \texttt{VAFA}, the results are reported with a constraint on $q_{\text{max}}$ set to  $20$, denoted as \texttt{VAFA-20}. DSC score \emph{(lower is better)} is  reported on the generated adversarial examples. }
    \label{tab: bbox hecktor eps4 app}
\end{table}

%% file: tables/black_box_acdc_eps4.tex
\begin{table}[!t]
\centering\small
   \scalebox{0.65}[0.65]{
    \begin{tabular}{l|c|cc|cc|cc|cc|cc|cc}
        \toprule
        \rowcolor{LightCyan} 
        Surrogate & Attack  & \multicolumn{2}{c|}{\textbf{UNet}} & \multicolumn{2}{c|}{\textbf{SegResNet}}  & \multicolumn{2}{c|}{\textbf{UNETR}}  & \multicolumn{2}{c|}{\textbf{SwinUNETR}}  &  \multicolumn{2}{c|}{\textbf{UMamba-B}} & \multicolumn{2}{c}{\textbf{UMamba-E}}  \\
                \rowcolor{LightCyan} 
          &  & ~$\mathrm{DSC}\hspace{-0.3em}\downarrow$~ & $\mathrm{HD95}\hspace{-0.3em}\uparrow$~  &  $\mathrm{DSC}\hspace{-0.3em}\downarrow$~ & ~$\mathrm{HD95}\hspace{-0.3em}\uparrow$~ & $\mathrm{DSC}\hspace{-0.3em}\downarrow$~  &  $\mathrm{HD95}\hspace{-0.3em}\uparrow$ & $\mathrm{DSC}\hspace{-0.3em}\downarrow$~  &  $\mathrm{HD95}\hspace{-0.3em}\uparrow$ & $\mathrm{DSC}\hspace{-0.3em}\downarrow$~  &  $\mathrm{HD95}\hspace{-0.3em}\uparrow$ & $\mathrm{DSC}\hspace{-0.3em}\downarrow$~  &  $\mathrm{HD95}\hspace{-0.3em}\uparrow$ \\
          \midrule
\rowcolor{Gray}  & Clean  & 85.52 & 5.75 & 89.65 & 2.56 & 76.37 & 16.31 & 84.19 & 7.93 & 88.22 & 6.01 & 80.91 & 8.48\\
\rowcolor{Gray} & GN  & - & - & - & - & - & - & - & - & - & - & - & -\\
\midrule
\multirow{4}{*}{\rotatebox[origin=c]{0}{\parbox[c]{1.5cm}{\centering\texttt{UNet}}}} 
& \texttt{FGSM-4}  & \cellcolor{green!25}59.57 & \cellcolor{green!25}18.93 & 88.54 & 3.74 & 75.15 & 17.32 & 82.86 & 8.27 & 86.93 & 7.05 & 79.17 & 8.92
\\
& \texttt{PGD-4}  & \cellcolor{green!25}22.75 & \cellcolor{green!25}38.22 & 89.38 & 3.18 & 75.73 & 17.29 & 83.75 & 8.63 & 87.72 & 6.8 & 80.02 & 8.73
\\
& \texttt{CosPGD-4}  & \cellcolor{green!25}23.52 & \cellcolor{green!25}35.54 & 89.24 & 3.29 & 75.46 & 17.34 & 83.5 & 8.48 & 87.72 & 6.4 & 79.95 & 8.55
\\
& \texttt{VAFA-20}  & \cellcolor{green!25}49.69 & \cellcolor{green!25}27.47 & 74.0 & 17.66 & 55.71 & 22.52 & 58.27 & 22.96 & 60.81 & 24.13 & 52.7 & 25.48
\\

    \midrule
    \multirow{4}{*}{\rotatebox[origin=c]{0}{\parbox[c]{1.5cm}{\centering\texttt{SegResNet}}}} & \texttt{FGSM-4}  & 84.67 & 5.92 & \cellcolor{green!25}69.85 & \cellcolor{green!25}11.22 & 75.08 & 17.52 & 82.49 & 8.16 & 85.2 & 10.12 & 78.95 & 8.57
\\
& \texttt{PGD-4}  & 85.04 & 6.09 & \cellcolor{green!25}21.96 & \cellcolor{green!25}38.44 & 75.62 & 17.08 & 83.62 & 8.24 & 87.42 & 6.84 & 80.03 & 8.77
\\
& \texttt{CosPGD-4}  & 85.08 & 5.85 & \cellcolor{green!25}22.68 & \cellcolor{green!25}38.34 & 75.67 & 17.15 & 83.62 & 8.16 & 87.31 & 6.77 & 79.3 & 8.73
\\
& \texttt{VAFA-20}  & 51.79 & 27.04 & \cellcolor{green!25}54.8 & \cellcolor{green!25}23.06 & 54.41 & 22.32 & 56.64 & 22.84 & 59.04 & 24.27 & 51.82 & 26.04
\\
    \midrule
    \multirow{4}{*}{\rotatebox[origin=c]{0}{\parbox[c]{1.5cm}{\centering\texttt{UNETR}}}} & \texttt{FGSM-4}  & 83.39 & 8.43 & 87.74 & 4.14 & \cellcolor{green!25}39.45 & \cellcolor{green!25}26.49 & 80.67 & 10.34 & 85.09 & 12.68 & 78.52 & 9.45
\\
& \texttt{PGD-4}  & 85.01 & 7.95 & 88.31 & 4.25 & \cellcolor{green!25}24.3 & \cellcolor{green!25}33.77 & 82.93 & 9.62 & 86.46 & 11.63 & 78.72 & 9.19
\\
& \texttt{CosPGD-4}  & 85.09 & 7.79 & 88.16 & 4.37 & \cellcolor{green!25}24.79 & \cellcolor{green!25}33.27 & 82.63 & 10.53 & 86.22 & 12.23 & 78.15 & 9.53
\\
& \texttt{VAFA-20}  & 52.38 & 26.55 & 73.19 & 16.56 & \cellcolor{green!25}48.99 & \cellcolor{green!25}23.31 & 55.29 & 24.52 & 57.86 & 24.69 & 50.38 & 26.23
\\
    \midrule
    \multirow{4}{*}{\rotatebox[origin=c]{0}{\parbox[c]{1.5cm}{\centering\texttt{SwinUNETR}}}} & \texttt{FGSM-4}  & 84.44 & 6.38 & 85.64 & 4.41 & 74.37 & 16.69 & \cellcolor{green!25}60.68 & \cellcolor{green!25}19.26 & 84.64 & 8.6 & 78.99 & 8.58
 \\
& \texttt{PGD-4}  & 84.95 & 5.85 & 87.83 & 4.6 & 74.53 & 17.5 & \cellcolor{green!25}23.47 & \cellcolor{green!25}34.53 & 86.65 & 8.74 & 77.56 & 9.95
\\
& \texttt{CosPGD-4}  & 85.0 & 6.39 & 87.92 & 3.96 & 74.46 & 17.59 & \cellcolor{green!25}23.54 & \cellcolor{green!25}34.97 & 86.78 & 8.84 & 77.86 & 9.68
\\
& \texttt{VAFA-20}  & 51.47 & 26.55 & 70.59 & 16.1 & 53.55 & 22.66 & \cellcolor{green!25}48.42 & \cellcolor{green!25}25.2 & 57.92 & 24.72 & 51.46 & 26.11
\\
\midrule
    \multirow{4}{*}{\rotatebox[origin=c]{0}{\parbox[c]{1.5cm}{\centering\texttt{UMamba-B}}}} & \texttt{FGSM-4}  & 83.68 & 8.65 & 86.22 & 4.41 & 74.71 & 17.35 & 82.2 & 8.95 & \cellcolor{green!25}76.9 & \cellcolor{green!25}14.19 & 77.72 & 8.86
\\
& \texttt{PGD-4}  & 84.22 & 8.86 & 86.58 & 5.62 & 74.53 & 17.49 & 82.45 & 9.33 & \cellcolor{green!25}29.55 & \cellcolor{green!25}31.7 & 74.43 & 10.37
\\
& \texttt{CosPGD-4}  & 84.26 & 8.03 & 86.54 & 5.24 & 74.66 & 17.27 & 82.6 & 9.51 & \cellcolor{green!25}30.5 & \cellcolor{green!25}29.28 & 73.78 & 11.24
\\
& \texttt{VAFA-20}  & 50.83 & 27.17 & 69.36 & 17.54 & 53.71 & 22.46 & 55.54 & 23.19 & \cellcolor{green!25}54.22 & \cellcolor{green!25}24.92 & 50.69 & 26.11
\\

\midrule
    \multirow{4}{*}{\rotatebox[origin=c]{0}{\parbox[c]{1.5cm}{\centering\texttt{UMamba-E}}}} & \texttt{FGSM-4}  &  84.28 & 7.03 & 87.74 & 3.43 & 74.41 & 17.51 & 83.5 & 8.09 & 85.73 & 8.73 & \cellcolor{green!25}54.11 & \cellcolor{green!25}19.96
\\
& \texttt{PGD-4}  & 85.13 & 6.02 & 89.17 & 3.39 & 75.56 & 17.04 & 83.65 & 8.09 & 87.36 & 6.35 & \cellcolor{green!25}28.09 & \cellcolor{green!25}29.77
\\
& \texttt{CosPGD-4}  & 85.24 & 5.88 & 88.91 & 3.09 & 75.54 & 17.16 & 83.72 & 8.2 & 87.39 & 6.67 & \cellcolor{green!25}27.24 & \cellcolor{green!25}30.24
\\
& \texttt{VAFA-20}  & 54.27 & 26.67 & 75.58 & 15.64 & 57.17 & 22.19 & 59.84 & 22.92 & 61.11 & 24.42 & \cellcolor{green!25}50.47 & \cellcolor{green!25}25.7
\\
    \midrule
    \end{tabular}
    }
    \vspace{0.3em}
    \caption{\small Performance of models against transfer-based \emph{black box} attacks on  \texttt{ACDC} dataset. For pixel-based attacks results are reported for $\epsilon=\frac{4}{255}$ indicated by attack names followed by the suffixes $-4$, respectively. Regarding frequency-based attack \texttt{VAFA}, the results are reported with a constraint on $q_{\text{max}}$ set to  $20$, denoted as \texttt{VAFA-20}. DSC score \emph{(lower is better)} is  reported on the generated adversarial examples. }
    \label{tab: bbox acdc eps4 app}
\end{table}

%% file: tables/black_box_btcv.tex
\begin{table}[!t]
\centering\small
   \scalebox{0.65}[0.65]{
    \begin{tabular}{l|c|cc|cc|cc|cc|cc|cc}
        \toprule
        \rowcolor{LightCyan} 
        Surrogate & Attack  & \multicolumn{2}{c|}{\textbf{UNet}} & \multicolumn{2}{c|}{\textbf{SegResNet}}  & \multicolumn{2}{c|}{\textbf{UNETR}}  & \multicolumn{2}{c|}{\textbf{SwinUNETR}}  &  \multicolumn{2}{c|}{\textbf{UMamba-B}} & \multicolumn{2}{c}{\textbf{UMamba-E}}  \\
                \rowcolor{LightCyan} 
          &  & ~$\mathrm{DSC}\hspace{-0.3em}\downarrow$~ & $\mathrm{HD95}\hspace{-0.3em}\uparrow$~  &  $\mathrm{DSC}\hspace{-0.3em}\downarrow$~ & ~$\mathrm{HD95}\hspace{-0.3em}\uparrow$~ & $\mathrm{DSC}\hspace{-0.3em}\downarrow$~  &  $\mathrm{HD95}\hspace{-0.3em}\uparrow$ & $\mathrm{DSC}\hspace{-0.3em}\downarrow$~  &  $\mathrm{HD95}\hspace{-0.3em}\uparrow$ & $\mathrm{DSC}\hspace{-0.3em}\downarrow$~  &  $\mathrm{HD95}\hspace{-0.3em}\uparrow$ & $\mathrm{DSC}\hspace{-0.3em}\downarrow$~  &  $\mathrm{HD95}\hspace{-0.3em}\uparrow$ \\
          \midrule
\rowcolor{Gray}  & Clean  & 75.72 & 9.15 & 80.84 & 8.14 & 72.53 & 15.08 & 78.07 & 10.01 & 78.37 & 8.12 & 77.06 & 11.11\\
\rowcolor{Gray} & GN  & 75.40 & 10.31 & 80.34 & 8.44 & 72.33 & 14.45 & 77.44 & 9.73 & 77.69 & 9.44 & 75.34 & 21.80\\
\midrule
\multirow{4}{*}{\rotatebox[origin=c]{0}{\parbox[c]{1.5cm}{\centering\texttt{UNet}}}} 
& \texttt{FGSM-8}  & \cellcolor{green!25}49.49 & \cellcolor{green!25}48.23 & 77.98 & 9.49 & 70.51 & 15.99 & 74.97 & 12.26 & 74.64 & 10.03 & 71.84 & 25.95\\
& \texttt{PGD-8}  & \cellcolor{green!25}32.06 & \cellcolor{green!25}89.45 & 79.78 & 8.73 &  72.02 & 15.28 & 76.89 & 10.69 & 76.92 & 9.02 & 74.74 & 13.86\\
& \texttt{CosPGD-8}  & \cellcolor{green!25}32.51 & \cellcolor{green!25}75.89 & 79.78 & 8.71 &  71.99 & 14.94 & 76.87 & 11.02 & 76.83 & 10.36 & 74.79 & 16.50\\
& \texttt{VAFA-30}  & \cellcolor{green!25}19.49 & \cellcolor{green!25}86.75 & 34.39 & 53.31 &  45.49 & 34.68 & 41.48 & 37.00 & 29.00 & 57.25 & 19.38 & 71.95\\

    \midrule
    \multirow{4}{*}{\rotatebox[origin=c]{0}{\parbox[c]{1.5cm}{\centering\texttt{SegResNet}}}} & \texttt{FGSM-8}  & 71.70 & 12.01 & \cellcolor{green!25} 59.97 & \cellcolor{green!25}37.36 & 69.93 & 15.34 & 72.48 & 11.81 & 68.73 & 14.77 & 67.95 & 22.74\\
& \texttt{PGD-8}  & 73.34 & 13.94 & \cellcolor{green!25}20.11 & \cellcolor{green!25}100.13 & 70.74 & 15.15 & 74.95 & 12.29 & 70.79 & 14.25 & 70.08 & 21.67\\
& \texttt{CosPGD-8}  & 73.29 & 13.76 & \cellcolor{green!25}19.68 & \cellcolor{green!25}96.74 & 70.75 & 15.12 & 75.04 & 11.75 & 70.50 & 14.16 & 69.79 & 21.31\\
& \texttt{VAFA-30}  & 36.43 & 44.43 & \cellcolor{green!25}22.00 & \cellcolor{green!25}72.16 & 50.16 & 28.19 & 45.54 & 36.83 & 28.39 & 55.87 & 19.53 & 69.45\\
    \midrule
    \multirow{4}{*}{\rotatebox[origin=c]{0}{\parbox[c]{1.5cm}{\centering\texttt{UNETR}}}} & \texttt{FGSM-8}  & 70.89 & 12.07 & 76.13 & 10.51 & \cellcolor{green!25}50.98 & \cellcolor{green!25}44.86 & 71.48 & 12.97 & 72.47 & 10.35 & 71.21 & 15.05\\
& \texttt{PGD-8}  & 72.24 & 13.49 & 78.46 & 9.57 & \cellcolor{green!25}{48.01} & \cellcolor{green!25}{53.78} & 74.64 & 12.62 & 74.62 & 10.29 & 72.73 & 15.29\\
& \texttt{CosPGD-8}  & 72.31 & 13.61 & 78.73 & 9.29 & \cellcolor{green!25}{47.46} & \cellcolor{green!25}{51.14} & 74.83 & 13.44 & 74.78 & 9.80 & 72.81 & 15.43\\
& \texttt{VAFA-30}  & 40.27 & 35.24 & 42.55 & 34.84 & \cellcolor{green!25}{26.68} & \cellcolor{green!25}{58.83} & 41.90 & 39.54 & 36.16 & 43.97 & 31.47 & 45.45\\
    \midrule
    \multirow{4}{*}{\rotatebox[origin=c]{0}{\parbox[c]{1.5cm}{\centering\texttt{SwinUNETR}}}} & \texttt{FGSM-8}  & 71.47 & 12.58 & 75.09 & 10.86 & 68.94 & 15.74 & \cellcolor{green!25} 58.80 & \cellcolor{green!25}40.84 & 71.88 & 11.46 & 70.15 & 19.12\\
& \texttt{PGD-8}  & 72.28 & 13.68 & 77.92 & 9.61 & 70.17 & 17.17 & \cellcolor{green!25} 54.09 & \cellcolor{green!25} 64.43 & 74.62 & 11.56 & 72.19 & 20.70\\
& \texttt{CosPGD-8}  & 72.40 & 14.59 & 77.87 & 9.71 & 70.20 & 18.13 & \cellcolor{green!25} 52.93 & \cellcolor{green!25} 58.46 & 74.72 & 12.03 & 72.19 & 21.12\\
& \texttt{VAFA-30}  & 29.22 & 54.11 & 30.30 & 44.20 & 38.69 & 38.22 & \cellcolor{green!25} 24.30 & \cellcolor{green!25}62.38 & 25.69 & 49.59 & 21.96 & 56.46\\
\midrule
    \multirow{4}{*}{\rotatebox[origin=c]{0}{\parbox[c]{1.5cm}{\centering\texttt{UMamba-B}}}} & \texttt{FGSM-8}  & 71.39 & 14.14 & 71.96 & 13.29 & 70.13 & 15.97 & 72.63 & 12.73 & \cellcolor{green!25}58.63 & \cellcolor{green!25}35.45& 65.75 & 23.13\\
& \texttt{PGD-8}  & 73.27 & 12.02 & 76.11 & 12.59 & 71.06 & 15.61 & 75.59 & 12.31 & \cellcolor{green!25}30.08 & \cellcolor{green!25}98.04 & 67.72 & 23.63\\
& \texttt{CosPGD-8}  & 73.30 & 12.92 & 76.41 & 12.78 & 71.11 & 14.97 & 75.74 & 11.72 & \cellcolor{green!25}30.96 & \cellcolor{green!25}82.75 & 68.26 & 25.44\\
& \texttt{VAFA-30}  & 35.18 & 42.78 & 30.45 & 49.72 & 46.93 & 34.42 & 42.89 & 38.19 & \cellcolor{green!25}16.65 & \cellcolor{green!25}79.04 & 17.01 & 73.73\\
\midrule
    \multirow{4}{*}{\rotatebox[origin=c]{0}{\parbox[c]{1.5cm}{\centering\texttt{UMamba-E}}}} & \texttt{FGSM-8}  & 72.59 & 12.33 & 75.69 & 11.32 & 70.49 & 16.27 & 74.48 & 11.02 & 71.36 & 14.58 & \cellcolor{green!25}53.59 & \cellcolor{green!25}45.54 \\
& \texttt{PGD-8}  & 74.39 & 13.66 & 79.09 & 9.55 & 71.69 & 15.04 & 76.91 & 11.23 & 75.58 & 10.74 & \cellcolor{green!25}29.51 & \cellcolor{green!25}104.51\\
& \texttt{CosPGD-8}  & 74.41 & 13.73 & 79.14 & 9.94 & 71.75 & 15.38 & 76.95 & 11.24 & 75.83 & 11.52 & \cellcolor{green!25}31.10 & \cellcolor{green!25}79.69\\
& \texttt{VAFA-30}  & 44.08 & 36.78 & 42.47 & 36.74 & 53.61 & 25.57 & 54.41 & 25.99 & 32.90 & 47.13 & \cellcolor{green!25}16.45 & \cellcolor{green!25}91.35\\
    \midrule
    \end{tabular}
    }
    \vspace{0.3em}
    \caption{\small Performance of models against transfer-based \emph{black box} attacks on  \texttt{BTCV} dataset.  For pixel-based attacks results are reported for $\epsilon=\frac{8}{255}$ indicated by attack names followed by the suffixes $-8$, respectively. Regarding frequency-based attack \texttt{VAFA}, the results are reported with a constraint on $q_{\text{max}}$ set to  $30$, denoted as \texttt{VAFA-30}. DSC score \emph{(lower is better)} is  reported on the generated adversarial examples.  }
    \label{tab: bbox btcv app}
\end{table}

%% file: tables/black_box_abdomen.tex
\begin{table}[!t]
\centering\small
   \scalebox{0.65}[0.65]{
    \begin{tabular}{l|c|cc|cc|cc|cc|cc|cc}
        \toprule
        \rowcolor{LightCyan} 
        Surrogate & Attack  & \multicolumn{2}{c|}{\textbf{UNet}} & \multicolumn{2}{c|}{\textbf{SegResNet}}  & \multicolumn{2}{c|}{\textbf{UNETR}}  & \multicolumn{2}{c|}{\textbf{SwinUNETR}}  &  \multicolumn{2}{c|}{\textbf{UMamba-B}} & \multicolumn{2}{c}{\textbf{UMamba-E}}  \\
                \rowcolor{LightCyan} 
          &  & ~$\mathrm{DSC}\hspace{-0.3em}\downarrow$~ & $\mathrm{HD95}\hspace{-0.3em}\uparrow$~  &  $\mathrm{DSC}\hspace{-0.3em}\downarrow$~ & ~$\mathrm{HD95}\hspace{-0.3em}\uparrow$~ & $\mathrm{DSC}\hspace{-0.3em}\downarrow$~  &  $\mathrm{HD95}\hspace{-0.3em}\uparrow$ & $\mathrm{DSC}\hspace{-0.3em}\downarrow$~  &  $\mathrm{HD95}\hspace{-0.3em}\uparrow$ & $\mathrm{DSC}\hspace{-0.3em}\downarrow$~  &  $\mathrm{HD95}\hspace{-0.3em}\uparrow$ & $\mathrm{DSC}\hspace{-0.3em}\downarrow$~  &  $\mathrm{HD95}\hspace{-0.3em}\uparrow$ \\
          \midrule
\rowcolor{Gray}  & Clean  & 76.79 & 19.72 & 80.89 & 13.30 & 71.35 & 27.73 & 79.33 & 25.79 & 81.08 & 15.51 & 78.05 & 18.31\\
\rowcolor{Gray} & GN  & 76.13 & 21.97 & 77.59 & 18.16 & 71.13 & 34.37 & 78.95 & 29.09 & 79.33 & 16.78 & 76.27 & 18.16\\
\midrule
\multirow{4}{*}{\rotatebox[origin=c]{0}{\parbox[c]{1.5cm}{\centering\texttt{UNet}}}} 
& \texttt{FGSM-8}  & \cellcolor{green!25}46.85 & \cellcolor{green!25}60.19 & 71.90 & 18.50  & 68.90 & 35.00 & 74.99 & 31.74 & 72.55 & 22.68 & 70.24 & 24.43\\
& \texttt{PGD-8}  & \cellcolor{green!25}29.14 & \cellcolor{green!25}90.85  &  76.35 & 15.56 & 70.19 & 32.92 & 77.73 & 28.77 & 77.22 & 19.29 & 74.47 & 20.15\\
& \texttt{CosPGD-8}  & \cellcolor{green!25}30.80 & \cellcolor{green!25}77.47  &  76.44 & 15.66 & 70.24 & 33.87 & 77.86 & 28.52 & 77.43 & 18.27 & 74.58 & 20.25\\
& \texttt{VAFA-30}  & \cellcolor{green!25}20.92 & \cellcolor{green!25}88.19  &  26.62 & 66.39 & 38.36 & 63.57 & 38.75 & 57.28 & 22.36 & 71.25 & 18.89 & 75.11\\

    \midrule
    \multirow{4}{*}{\rotatebox[origin=c]{0}{\parbox[c]{1.5cm}{\centering\texttt{SegResNet}}}} & \texttt{FGSM-8}  & 70.50 & 23.59 & \cellcolor{green!25}50.73 & \cellcolor{green!25}49.07 & 68.71 & 34.66 & 74.29 & 27.96 & 64.91 & 29.14 & 63.94 & 26.72\\
& \texttt{PGD-8}  & 73.76 & 22.77 & \cellcolor{green!25}17.42 & \cellcolor{green!25}106.73 & 69.90 & 32.58 & 76.90 & 28.83 & 70.87 & 23.57 & 69.81 & 24.80\\
& \texttt{CosPGD-8}  & 74.07 & 24.07 & \cellcolor{green!25}18.88 & \cellcolor{green!25}102.25 & 69.93 & 32.26 & 77.00 & 29.17 & 71.59 & 23.13 & 70.36 & 23.56\\
& \texttt{VAFA-30}  & 38.47 & 51.36 & \cellcolor{green!25} 17.68 & \cellcolor{green!25} 84.91 & 43.33 & 54.59 & 45.00 & 47.85 & 22.43 & 63.84 & 19.75 & 64.63\\
    \midrule
    \multirow{4}{*}{\rotatebox[origin=c]{0}{\parbox[c]{1.5cm}{\centering\texttt{UNETR}}}} & \texttt{FGSM-8}  & 70.38 & 23.55& 71.68 & 16.66 & \cellcolor{green!25}48.48 & \cellcolor{green!25}58.85 & 71.54 & 31.48 & 71.70 & 25.19 & 69.28 & 21.42\\
& \texttt{PGD-8}  & 72.57 & 22.99 & 74.85 & 17.19 & \cellcolor{green!25}{44.83} & \cellcolor{green!25}{76.08} & 75.28 & 30.96 & 74.82 & 18.84 & 70.82 & 23.81\\
& \texttt{CosPGD-8}  & 72.67 & 22.73 & 75.11 & 17.05 & \cellcolor{green!25}{44.64} & \cellcolor{green!25}{74.34} & 75.46 & 30.61 & 74.82 & 18.47 & 71.00 & 23.05\\
& \texttt{VAFA-30}  & 39.68 & 48.60 & 32.63 & 56.05 & \cellcolor{green!25}{22.34} & \cellcolor{green!25}{88.08} & 38.11 & 54.49 & 29.60 & 54.23 & 26.77 & 58.62\\
    \midrule
    \multirow{4}{*}{\rotatebox[origin=c]{0}{\parbox[c]{1.5cm}{\centering\texttt{SwinUNETR}}}} & \texttt{FGSM-8}  & 69.99 & 24.44 & 69.39 & 18.48 & 66.81 & 35.05 & \cellcolor{green!25} 53.46 & \cellcolor{green!25}71.03 & 70.21 & 23.45 & 68.12 & 23.85\\
& \texttt{PGD-8}  & 72.67 & 24.14 & 73.19 & 18.20 & 68.77 & 38.35 & \cellcolor{green!25}46.95 & \cellcolor{green!25}88.88 & 74.39 & 23.43 & 71.40 & 24.23\\
& \texttt{CosPGD-8}  & 72.88 & 23.67 & 73.56 & 18.34 & 68.85 & 38.09 & \cellcolor{green!25}46.79 & \cellcolor{green!25}82.92 & 74.59 & 22.30 & 71.67 & 23.91\\
& \texttt{VAFA-30}  & 30.47 & 60.63 & 27.51 & 60.51 & 33.10 & 64.36 & \cellcolor{green!25}25.43 & \cellcolor{green!25}76.41 & 24.11 & 65.63 & 22.06 & 65.39\\
\midrule
    \multirow{4}{*}{\rotatebox[origin=c]{0}{\parbox[c]{1.5cm}{\centering\texttt{UMamba-B}}}} & \texttt{FGSM-8}  & 70.60 & 24.91 & 65.06 & 23.98 & 68.69 & 35.94 & 74.25 & 31.40 & \cellcolor{green!25}51.36 & \cellcolor{green!25}53.89& 62.19 & 30.01\\
& \texttt{PGD-8}  & 73.49 & 22.17 & 69.75 & 21.84 & 69.93 & 32.71 & 76.89 & 28.43 & \cellcolor{green!25}18.20 & \cellcolor{green!25}111.79 & 65.72 & 28.54\\
& \texttt{CosPGD-8}  & 73.46 & 22.99 & 70.35 & 20.94 & 69.94 & 33.08 & 76.92 & 28.69 & \cellcolor{green!25}18.99 & \cellcolor{green!25}101.89 & 66.36 & 30.04\\
& \texttt{VAFA-30}  & 38.49 & 55.47 & 24.20 & 65.44 & 43.80 & 58.37 & 44.61 & 52.18 & \cellcolor{green!25}14.99 & \cellcolor{green!25}92.09 & 18.46 & 69.64\\
\midrule
    \multirow{4}{*}{\rotatebox[origin=c]{0}{\parbox[c]{1.5cm}{\centering\texttt{UMamba-E}}}} & \texttt{FGSM-8}  & 70.93 & 23.48 & 68.27 & 20.87 & 68.69 & 35.07 & 74.85 & 29.52 & 65.66 & 30.35 & \cellcolor{green!25}48.81 & \cellcolor{green!25}52.13 \\
& \texttt{PGD-8}  & 74.20 & 21.71 & 73.82 & 17.58 & 70.25 & 32.12 & 77.55 & 28.89 & 71.18 & 23.11 & \cellcolor{green!25}23.59 & \cellcolor{green!25}96.84\\
& \texttt{CosPGD-8}  & 74.41 & 21.16 & 74.15 & 17.06 & 70.30 & 32.56 & 77.65 & 28.78 & 71.86 & 22.90 & \cellcolor{green!25}24.99 & \cellcolor{green!25}82.01\\
& \texttt{VAFA-30}  & 39.41 & 53.15 & 27.39 & 61.14 & 43.32 & 58.63 & 46.53 & 50.97 & 23.31 & 72.51 & \cellcolor{green!25}13.88 & \cellcolor{green!25}92.91\\
    \midrule
    \end{tabular}
    }
    \vspace{0.3em}
    \caption{\small Performance of models against transfer-based \emph{black box} attacks on  \texttt{Abdomen-CT} dataset.  For pixel-based attacks results are reported for $\epsilon=\frac{8}{255}$ indicated by attack names followed by the suffixes $-8$, respectively. Regarding frequency-based attack \texttt{VAFA}, the results are reported with a constraint on $q_{\text{max}}$ set to  $30$, denoted as \texttt{VAFA-30}. DSC score \emph{(lower is better)} is  reported on the generated adversarial examples.  }
    \label{tab: bb abdomen app}
\end{table}

%% file: tables/black_box_hecktor.tex
\begin{table}[!t]
\centering\small
   \scalebox{0.65}[0.65]{
    \begin{tabular}{l|c|cc|cc|cc|cc|cc|cc}
        \toprule
        \rowcolor{LightCyan} 
        Surrogate & Attack  & \multicolumn{2}{c|}{\textbf{UNet}} & \multicolumn{2}{c|}{\textbf{SegResNet}}  & \multicolumn{2}{c|}{\textbf{UNETR}}  & \multicolumn{2}{c|}{\textbf{SwinUNETR}}  &  \multicolumn{2}{c|}{\textbf{UMamba-B}} & \multicolumn{2}{c}{\textbf{UMamba-E}}  \\
                \rowcolor{LightCyan} 
          &  & ~$\mathrm{DSC}\hspace{-0.3em}\downarrow$~ & $\mathrm{HD95}\hspace{-0.3em}\uparrow$~  &  $\mathrm{DSC}\hspace{-0.3em}\downarrow$~ & ~$\mathrm{HD95}\hspace{-0.3em}\uparrow$~ & $\mathrm{DSC}\hspace{-0.3em}\downarrow$~  &  $\mathrm{HD95}\hspace{-0.3em}\uparrow$ & $\mathrm{DSC}\hspace{-0.3em}\downarrow$~  &  $\mathrm{HD95}\hspace{-0.3em}\uparrow$ & $\mathrm{DSC}\hspace{-0.3em}\downarrow$~  &  $\mathrm{HD95}\hspace{-0.3em}\uparrow$ & $\mathrm{DSC}\hspace{-0.3em}\downarrow$~  &  $\mathrm{HD95}\hspace{-0.3em}\uparrow$ \\
          \midrule
\rowcolor{Gray}  & Clean  & 73.91 & 11.36 & 74.73 & 11.08 & 72.36 & 14.61 &71.61 & 22.07 & 73.50 & 10.89 & 72.19 & 13.29\\
\rowcolor{Gray} & GN  & 72.11 & 12.97 & 73.26 & 10.08 & 71.28 & 14.51 &  69.61 & 23.65 & 72.12 & 13.96 & 66.58 & 13.45\\
\midrule
\multirow{4}{*}{\rotatebox[origin=c]{0}{\parbox[c]{1.5cm}{\centering\texttt{UNet}}}} 
& \texttt{FGSM-8}  & \cellcolor{green!25} 43.78 & \cellcolor{green!25}37.79 & 59.11 & 18.23 & 66.49 & 17.23 & 58.76 & 30.61 & 59.26 & 18.63 & 57.47 & 19.08\\
& \texttt{PGD-8}  & \cellcolor{green!25} 34.07 & \cellcolor{green!25}78.61 & 68.24 & 13.80 &  69.87 & 17.54 & 64.99 & 27.33 & 67.38 & 14.62 & 65.55 & 17.80\\
& \texttt{CosPGD-8}  & \cellcolor{green!25} 34.22 & \cellcolor{green!25}78.62 & 69.48 & 12.92 &  70.02 & 16.83 & 66.11 & 26.00 & 68.58 & 13.11 & 66.21 & 16.10\\
& \texttt{VAFA-30}  & \cellcolor{green!25} 44.30 & \cellcolor{green!25}42.19 & 68.81 & 14.80 &  63.10 & 18.37 & 65.10 & 22.05 & 64.99 & 17.93 & 60.37 & 21.85\\

    \midrule
    \multirow{4}{*}{\rotatebox[origin=c]{0}{\parbox[c]{1.5cm}{\centering\texttt{SegResNet}}}} & \texttt{FGSM-8}  & 60.62 & 21.43 & \cellcolor{green!25} 47.70 & \cellcolor{green!25} 32.24& 67.51 & 17.17 & 57.01 & 30.20 & 53.92 & 27.12 & 55.13 & 21.08\\
& \texttt{PGD-8}  & 61.85 & 20.89 & \cellcolor{green!25} 33.02 & \cellcolor{green!25} 81.17 & 68.59 & 17.13 & 58.63 & 29.07 & 51.21 & 31.48 & 54.53 & 25.55\\
& \texttt{CosPGD-8}  & 61.99 & 19.89 & \cellcolor{green!25} 33.41 & \cellcolor{green!25} 81.28 & 68.81 & 16.63 & 58.74 & 29.19 & 50.25 & 32.25 & 54.15 & 23.28\\
& \texttt{VAFA-30}  & 69.07 & 17.77 & \cellcolor{green!25} 47.55 & \cellcolor{green!25} 33.08 & 66.24 & 17.71 & 65.55 & 21.85 & 65.23 & 17.79 & 62.22 & 23.54\\
    \midrule
    \multirow{4}{*}{\rotatebox[origin=c]{0}{\parbox[c]{1.5cm}{\centering\texttt{UNETR}}}} & \texttt{FGSM-8}  & 58.16 & 19.27 & 59.01 & 15.60 & \cellcolor{green!25}44.89 & \cellcolor{green!25}35.08 & 54.29 & 31.27 & 57.88 & 17.11 & 56.95 & 19.62\\
& \texttt{PGD-8}  & 58.33 & 20.35 & 61.02 & 16.89 & \cellcolor{green!25}{34.93} & \cellcolor{green!25}{74.58} & 55.55 & 32.75 & 60.38 & 19.40 & 58.98 & 17.85\\
& \texttt{CosPGD-8}  & 59.02 & 21.18 & 61.97 & 17.29 & \cellcolor{green!25}{34.89} & \cellcolor{green!25}{73.84} & 56.13 & 32.29 & 61.55 & 19.64 & 59.95 & 20.11\\
& \texttt{VAFA-30}  & 64.73 & 19.36 & 69.68 & 13.81 & \cellcolor{green!25}{48.01} & \cellcolor{green!25}{33.82} & 57.85 & 27.53 & 65.38 & 19.19 & 58.59 & 21.28\\
    \midrule
    \multirow{4}{*}{\rotatebox[origin=c]{0}{\parbox[c]{1.5cm}{\centering\texttt{SwinUNETR}}}} & \texttt{FGSM-8}  & 57.96 & 21.82 & 55.59 & 18.97 & 64.66 & 19.39 & \cellcolor{green!25}43.82 & \cellcolor{green!25}52.95 & 54.54 & 24.71 & 53.39 & 21.48\\
& \texttt{PGD-8}  & 55.25 & 25.87 & 56.09 & 20.16 & 66.29 & 19.83 & \cellcolor{green!25}34.12 & \cellcolor{green!25}78.85 & 55.14 & 23.72 & 54.18 & 23.60\\
& \texttt{CosPGD-8}  & 55.83 & 23.88 & 56.61 & 21.35 & 66.53 & 19.43 & \cellcolor{green!25}34.21 & \cellcolor{green!25}77.62 & 55.37 & 25.18 & 54.70 & 23.08\\
& \texttt{VAFA-30}  & 65.58 & 19.76 & 68.38 & 15.46 & 61.97 & 20.25 & \cellcolor{green!25}44.44 & \cellcolor{green!25}42.14 & 65.93 & 16.53 & 58.56 & 20.68\\
\midrule
    \multirow{4}{*}{\rotatebox[origin=c]{0}{\parbox[c]{1.5cm}{\centering\texttt{UMamba-B}}}} & \texttt{FGSM-8}  & 60.34 & 18.63 & 55.50 & 18.78 & 67.23 & 16.73 & 56.29 & 33.02 & \cellcolor{green!25}50.16 & \cellcolor{green!25}30.12& 54.51 & 19.86\\
& \texttt{PGD-8}  & 61.83 & 19.34 & 49.65 & 39.99 & 69.07 & 16.88 & 58.12 & 31.30 & \cellcolor{green!25}34.31 & \cellcolor{green!25}78.39 & 51.03 & 30.70\\
& \texttt{CosPGD-8}  & 61.87 & 20.34 & 48.57 & 40.81 & 69.12 & 17.61 & 59.05 & 31.02 & \cellcolor{green!25}34.41 & \cellcolor{green!25}78.44 & 51.39 & 31.16\\
& \texttt{VAFA-30}  & 66.61 & 20.86 & 64.99 & 16.32 & 64.11 & 18.50 & 64.53 & 21.53 & \cellcolor{green!25}49.27 & \cellcolor{green!25}35.99 & 58.52 & 22.13\\
\midrule
    \multirow{4}{*}{\rotatebox[origin=c]{0}{\parbox[c]{1.5cm}{\centering\texttt{UMamba-E}}}} & \texttt{FGSM-8}  & 62.64 & 16.67 & 59.29 & 16.55 & 68.43 & 17.10 & 60.34 & 29.29 & 57.49 & 22.28 & \cellcolor{green!25}47.15 & \cellcolor{green!25}31.65 \\
& \texttt{PGD-8}  & 69.30 & 14.95 & 68.77 & 12.22 & 70.59 & 15.53 & 66.92 & 25.14 & 66.75 & 16.73 & \cellcolor{green!25}34.29 & \cellcolor{green!25}78.25\\
& \texttt{CosPGD-8}  & 69.55 & 13.86 & 69.49 & 12.88 & 70.67 & 15.07 & 67.34 & 26.04 & 67.43 & 16.99 & \cellcolor{green!25}34.37 & \cellcolor{green!25}77.85\\
& \texttt{VAFA-30}  & 68.09 & 20.51 & 69.84 & 13.90 & 64.26 & 19.12 & 65.86 & 24.17 & 66.03 & 20.12 & \cellcolor{green!25} 44.81 & \cellcolor{green!25}36.84\\
    \midrule
    \end{tabular}
    }
    \vspace{0.3em}
    \caption{\small Performance of models against transfer-based \emph{black box} attacks on  \texttt{Hecktor} dataset.  For pixel-based attacks results are reported for $\epsilon=\frac{8}{255}$ indicated by attack names followed by the suffixes $-8$, respectively. Regarding frequency-based attack \texttt{VAFA}, the results are reported with a constraint on $q_{\text{max}}$ set to  $30$, denoted as \texttt{VAFA-30}. DSC score \emph{(lower is better)} is  reported on the generated adversarial examples.  }
    \label{tab: bb hecktor app}
\end{table}

%% file: tables/black_box_acdc.tex
\begin{table}[!t]
\centering\small
   \scalebox{0.65}[0.65]{
    \begin{tabular}{l|c|cc|cc|cc|cc|cc|cc}
        \toprule
        \rowcolor{LightCyan} 
        Surrogate & Attack  & \multicolumn{2}{c|}{\textbf{UNet}} & \multicolumn{2}{c|}{\textbf{SegResNet}}  & \multicolumn{2}{c|}{\textbf{UNETR}}  & \multicolumn{2}{c|}{\textbf{SwinUNETR}}  &  \multicolumn{2}{c|}{\textbf{UMamba-B}} & \multicolumn{2}{c}{\textbf{UMamba-E}}  \\
                \rowcolor{LightCyan} 
          &  & ~$\mathrm{DSC}\hspace{-0.3em}\downarrow$~ & $\mathrm{HD95}\hspace{-0.3em}\uparrow$~  &  $\mathrm{DSC}\hspace{-0.3em}\downarrow$~ & ~$\mathrm{HD95}\hspace{-0.3em}\uparrow$~ & $\mathrm{DSC}\hspace{-0.3em}\downarrow$~  &  $\mathrm{HD95}\hspace{-0.3em}\uparrow$ & $\mathrm{DSC}\hspace{-0.3em}\downarrow$~  &  $\mathrm{HD95}\hspace{-0.3em}\uparrow$ & $\mathrm{DSC}\hspace{-0.3em}\downarrow$~  &  $\mathrm{HD95}\hspace{-0.3em}\uparrow$ & $\mathrm{DSC}\hspace{-0.3em}\downarrow$~  &  $\mathrm{HD95}\hspace{-0.3em}\uparrow$ \\
          \midrule
\rowcolor{Gray}  & Clean  & 85.52 & 5.75 & 89.65 & 2.56 & 76.37 & 16.31 & 84.19 & 7.93 & 88.22 & 6.01 & 80.91 & 8.48\\
\rowcolor{Gray} & GN  & 85.01 & 5.67 & 89.26 & 2.89 & 74.88 & 18.49 & 83.06 & 10.69 & 86.95 & 6.68 & 78.11 & 11.67\\
\midrule
\multirow{4}{*}{\rotatebox[origin=c]{0}{\parbox[c]{1.5cm}{\centering\texttt{UNet}}}} 
& \texttt{FGSM-8}  & \cellcolor{green!25}55.55 & \cellcolor{green!25}21.99 & 86.65 & 4.55 & 73.32 & 18.87 & 81.12 & 11.22 & 84.58 & 11.19 & 74.96 & 13.42\\
& \texttt{PGD-8}  & \cellcolor{green!25}21.42 & \cellcolor{green!25}39.14 & 88.05 & 5.21 & 74.67 & 18.96 & 82.57 & 10.59 & 86.32 & 10.08 & 77.82 & 10.82\\
& \texttt{CosPGD-8}  & \cellcolor{green!25}22.62 & \cellcolor{green!25}36.84 & 88.18 & 5.12 & 74.36 & 18.73 & 82.92 & 10.12 & 86.50 & 9.92 & 77.33 & 11.97\\
& \texttt{VAFA-30}  & \cellcolor{green!25}49.85 & \cellcolor{green!25}27.44 & 74.15 & 18.17 & 56.09 & 21.77 & 58.72 & 22.95 & 60.27 & 24.46 & 53.68 & 25.98\\

    \midrule
    \multirow{4}{*}{\rotatebox[origin=c]{0}{\parbox[c]{1.5cm}{\centering\texttt{SegResNet}}}} & \texttt{FGSM-8}  & 83.63 & 6.32 & \cellcolor{green!25}66.58 & \cellcolor{green!25}11.88 & 73.03 & 19.72 & 80.41 & 12.75 & 81.15 & 19.95 & 74.57 & 14.46\\
& \texttt{PGD-8}  & 84.38 & 6.34 & \cellcolor{green!25}20.42 & \cellcolor{green!25}37.34 & 74.59 & 17.62 & 82.27 & 9.84 & 85.72 & 10.48 & 76.39 & 11.34\\
& \texttt{CosPGD-8}  & 84.54 & 6.08 & \cellcolor{green!25}22.33 & \cellcolor{green!25}38.61 & 74.79 & 18.05 & 82.44 & 9.58 & 85.69 & 9.58 & 76.41 & 11.32\\
& \texttt{VAFA-30}  & 51.83 & 27.22 & \cellcolor{green!25}54.47 & \cellcolor{green!25}23.38 & 54.96 & 21.60 & 57.00 & 24.20 & 59.24 & 24.27 & 53.12 & 25.38\\
    \midrule
    \multirow{4}{*}{\rotatebox[origin=c]{0}{\parbox[c]{1.5cm}{\centering\texttt{UNETR}}}} & \texttt{FGSM-8}  & 80.15 & 12.78 & 83.33 & 6.48 & \cellcolor{green!25}33.16 & \cellcolor{green!25}29.77 & 75.79 & 16.74 & 77.48 & 25.93 & 72.88 & 14.93\\
& \texttt{PGD-8}  & 83.25 & 13.69 & 85.54 & 7.19 & \cellcolor{green!25}{22.31} & \cellcolor{green!25}{35.44} & 80.12 & 15.19 & 80.67 & 22.48 & 68.37 & 16.98\\
& \texttt{CosPGD-8}  & 82.98 & 13.48 & 86.08 & 7.09 & \cellcolor{green!25}{23.08} & \cellcolor{green!25}{34.80} & 80.12 & 15.41 & 81.37 & 21.11 & 70.52 & 17.51\\
& \texttt{VAFA-30}  & 52.67 & 26.66 & 74.18 & 15.78 & \cellcolor{green!25}{49.39} & \cellcolor{green!25}{22.63} & 55.90 & 24.47 & 58.27 & 24.85 & 52.16 & 25.09\\
    \midrule
    \multirow{4}{*}{\rotatebox[origin=c]{0}{\parbox[c]{1.5cm}{\centering\texttt{SwinUNETR}}}} & \texttt{FGSM-8}  & 83.11 & 7.61 & 81.34 & 5.11 & 71.91 & 18.15 & \cellcolor{green!25}56.11 & \cellcolor{green!25}21.83 & 80.69 & 15.61 & 74.72 & 11.74\\
& \texttt{PGD-8}  & 84.04 & 7.49 & 84.37 & 6.75 & 71.73 & 19.22 & \cellcolor{green!25}21.69 & \cellcolor{green!25}36.99 & 82.42 & 15.23 & 72.15 & 12.77\\
& \texttt{CosPGD-8}  & 83.92 & 7.75 & 84.38 & 7.25 & 71.91 & 20.24 & \cellcolor{green!25}22.50 & \cellcolor{green!25}36.20 & 83.34 & 19.98 & 70.49 & 13.23\\
& \texttt{VAFA-30}  & 51.50 & 26.94 & 70.18 & 17.16 & 53.50 & 23.25 & \cellcolor{green!25}48.44 & \cellcolor{green!25}25.57 & 57.68 & 25.01 & 52.52 & 25.27\\
\midrule
    \multirow{4}{*}{\rotatebox[origin=c]{0}{\parbox[c]{1.5cm}{\centering\texttt{UMamba-B}}}} & \texttt{FGSM-8}  & 81.55 & 14.53 & 83.27 & 5.11 & 71.83 & 19.99 & 79.61 & 13.74 & \cellcolor{green!25}72.55 & \cellcolor{green!25}21.19& 71.54 & 14.86\\
& \texttt{PGD-8}  & 82.82 & 12.81 & 82.49 & 6.18 & 72.14 & 19.13 & 79.95 & 12.33 & \cellcolor{green!25}24.95 & \cellcolor{green!25}34.66 & 69.66 & 15.29\\
& \texttt{CosPGD-8}  & 82.58 & 10.26 & 82.69 & 7.37 & 72.50 & 18.78 & 80.14 & 11.97 & \cellcolor{green!25}25.54 & \cellcolor{green!25}31.32 & 68.93 & 14.86\\
& \texttt{VAFA-30}  & 51.46 & 27.35 & 71.36 & 17.79 & 54.18 & 22.82 & 57.14 & 13.39 & \cellcolor{green!25}53.38 & \cellcolor{green!25}25.22 & 51.83 & 26.27\\
\midrule
    \multirow{4}{*}{\rotatebox[origin=c]{0}{\parbox[c]{1.5cm}{\centering\texttt{UMamba-E}}}} & \texttt{FGSM-8}  & 82.44 & 13.41 & 84.96 & 4.75 & 71.23 & 20.59 & 81.82 & 12.16 & 81.97 & 15.77 & \cellcolor{green!25}51.65 & \cellcolor{green!25}24.22 \\
& \texttt{PGD-8}  & 84.66 & 6.11 & 87.71 & 3.43 & 74.51 & 18.16 & 82.95 & 9.17 & 86.12 & 7.46 & \cellcolor{green!25}25.01 & \cellcolor{green!25}32.43\\
& \texttt{CosPGD-8}  & 84.79 & 6.32 & 87.31 & 3.84 & 74.86 & 17.58 & 83.01 & 9.52 & 86.19 & 7.14 & \cellcolor{green!25}24.44 & \cellcolor{green!25}31.55\\
& \texttt{VAFA-30}  & 55.01 & 27.08 & 75.57 & 15.77 & 57.40 & 20.80 & 60.22 & 23.85 & 61.26 & 24.54 & \cellcolor{green!25}51.61 & \cellcolor{green!25}25.72\\
    \midrule
    \end{tabular}
    }
    \vspace{0.3em}
    \caption{\small Performance of models against transfer-based \emph{black box} attacks on  \texttt{ACDC} dataset.  For pixel-based attacks results are reported for $\epsilon=\frac{8}{255}$ indicated by attack names followed by the suffixes $-8$, respectively. Regarding frequency-based attack \texttt{VAFA}, the results are reported with a constraint on $q_{\text{max}}$ set to  $30$, denoted as \texttt{VAFA-30}. DSC score \emph{(lower is better)} is  reported on the generated adversarial examples.  }
    \label{tab: bb acdc app}
\end{table}

%% file: tables/medsam.tex
\begin{table}[!t]
\centering\small
   \scalebox{0.65}[0.65]{
    \begin{tabular}{l|c|cc|cc|cc|cc}
        \toprule
        \rowcolor{LightCyan} 
        Surrogate & Attack  & \multicolumn{2}{c|}{\textbf{BTCV}} & \multicolumn{2}{c|}{\textbf{ACDC}}  & \multicolumn{2}{c|}{\textbf{Hecktor}}  & \multicolumn{2}{c}{\textbf{Abdomen-CT}}    \\
                \rowcolor{LightCyan} 
          &  & ~$\mathrm{DSC}\hspace{-0.3em}\downarrow$~ & $\mathrm{IoU}\hspace{-0.3em}\downarrow$~  &  $\mathrm{DSC}\hspace{-0.3em}\downarrow$~ & ~$\mathrm{IoU}\hspace{-0.3em}\downarrow$~ & $\mathrm{DSC}\hspace{-0.3em}\downarrow$~  &  $\mathrm{IoU}\hspace{-0.3em}\downarrow$ & $\mathrm{DSC}\hspace{-0.3em}\downarrow$~  &  $\mathrm{IoU}\hspace{-0.3em}\downarrow$  \\
          \midrule
\multirow{4}{*}{\rotatebox[origin=c]{0}{\parbox[c]{1.5cm}{\centering\texttt{UNet}}}} 
& \texttt{FGSM-8}  & 72.56 & 60.84 & 61.74 & 48.09 & 36.41 & 24.66 & 76.73 & 65.07 \\
& \texttt{PGD-8}  & 73.23 & 61.39 & 60.44 & 46.54 & 36.15 & 24.55 & 77.15 & 65.53 \\
& \texttt{CosPGD-8}  & 73.34 & 61.51 & 60.50 & 46.68 & 36.32 & 24.59 & 77.05 & 65.44 \\

& \texttt{VAFA-30}  & 66.26 & 53.80 & 47.01 & 35.51 & 36.77 & 24.70 & 69.82 & 56.99 \\
    \midrule
    \multirow{4}{*}{\rotatebox[origin=c]{0}{\parbox[c]{1.5cm}{\centering\texttt{SegResNet}}}} & \texttt{FGSM-8}  & 73.16 & 61.34 & 63.15 & 49.86 & 36.64 & 24.73 & 76.82 & 65.21 \\
& \texttt{PGD-8}  & 72.95 & 61.24 & 61.38 & 47.64 & 34.98 & 23.57 & 76.84 & 85.27  \\
& \texttt{CosPGD-8}  & 73.05 & 61.31 & 61.99 & 48.35 & 34.76 & 23.38 & 76.92 & 65.32  \\
& \texttt{VAFA-30}  & 66.28 & 53.76 & 45.13 & 33.64 & 36.27 & 24.39  & 71.57& 58.92 \\
    \midrule
    \multirow{4}{*}{\rotatebox[origin=c]{0}{\parbox[c]{1.5cm}{\centering\texttt{UNETR}}}} & \texttt{FGSM-8}  & 71.59 & 59.82 & 61.18 & 47.65 & 37.52 & 25.49 & 75.49 & 63.66 \\
& \texttt{PGD-8}  & 72.15 & 60.33 & 58.93 & 45.80 & {38.19} & {26.08} &  76.15& 64.44 \\
& \texttt{CosPGD-8}  & 72.34 & 60.46 & 59.74 & 46.55 & {38.32} & {26.37} &  76.16& 64.43 \\
& \texttt{VAFA-30}  & 65.31 & 52.77 & 45.16 & 33.43 & {40.00} & {27.24} & 68.35 & 55.07 \\
    \midrule
    \multirow{4}{*}{\rotatebox[origin=c]{0}{\parbox[c]{1.8cm}{\centering\texttt{SwinUNETR}}}} & \texttt{FGSM-8}  & 73.05 & 61.19 & 62.72 & 48.68 & 35.94 & 24.20 & 76.54 & 64.91 \\
& \texttt{PGD-8}  & 73.06 & 61.23 & 61.59 & 47.82 & 36.35 & 24.81 & 76.67 & 64.99 \\
& \texttt{CosPGD-8}  & 73.04 & 61.26 & 61.32 & 47.65 & 36.21 & 24.62 & 76.82 & 65.22 \\
& \texttt{VAFA-30}  & 65.23 & 52.59 & 43.32 & 31.76 & 38.61 & 26.12 & 68.03 & 54.81 \\
\midrule
    \multirow{4}{*}{\rotatebox[origin=c]{0}{\parbox[c]{1.8cm}{\centering\texttt{UMamba-B}}}} & \texttt{FGSM-8}  & 72.93 & 61.12 & 62.85 & 49.62 & 36.86 & 25.05 & 76.83 & 65.23 \\
& \texttt{PGD-8}  & 72.94 & 61.16 & 60.10 & 46.45 & 35.56 & 24.09 & 77.01 & 65.38 \\
& \texttt{CosPGD-8}  & 73.17 & 61.34 & 60.91 & 46.99 & 36.37 & 24.64 & 77.12 & 65.54 \\
& \texttt{VAFA-30}  & 65.47 & 53.15 & 44.51 & 32.98 & 37.54 & 25.41 & 71.61 & 58.82 \\
\midrule
    \multirow{4}{*}{\rotatebox[origin=c]{0}{\parbox[c]{1.8cm}{\centering\texttt{UMamba-E}}}} & \texttt{FGSM-8}  & 73.03 & 61.19 & 63.56 & 50.06 & 36.95 & 25.04 & 77.01 & 65.38 \\
& \texttt{PGD-8}  & 73.04 & 61.28 & 60.56 & 46.78 & 35.72 & 24.14 & 77.15 & 65.53 \\
& \texttt{CosPGD-8}  & 73.24 & 61.42 & 61.15 & 47.62 & 35.67 & 24.12 & 77.02 & 65.43 \\
& \texttt{VAFA-30}  & 67.43 & 55.23 & 45.47 & 34.04 & 36.55 & 24.63 & 71.47 & 58.75 \\
\midrule
    \end{tabular}
    }
    \vspace{0.3em}
    \caption{\small Evaluating SAM-Med3D on adversarial examples crafted on surrogate models trained on BTCV, ACDC, Hecktor, and Abdomen-CT datasets. }
    \label{tab:bbox sam}
\end{table}